\documentclass[preprint]{JHEP3}
\JHEPspecialurl{http://jhep.sissa.it/JOURNAL/JHEP3.tar.gz}
\usepackage{epsfig,multicol,bbm,bigcenter}
\usepackage{citesort}
\usepackage{enumerate}
\newcommand\fverb{\setbox\fverbbox=\hbox\bgroup\verb}
\newcommand\fverbdo{\egroup\medskip\noindent%
\fbox{\unhbox\fverbbox}\ }
\newcommand\fverbit{\egroup\item[\fbox{\unhbox\fverbbox}]}
\newbox\fverbbox

\newcommand{\lsim}{\raisebox{-0.13cm}{~\shortstack{$<$ \\[-0.07cm] $\sim$}}~} 
\newcommand{\gsim}{\raisebox{-0.13cm}{~\shortstack{$>$ \\[-0.07cm] $\sim$}}~} 
\newcommand{\beq}{\begin{eqnarray}} 
\newcommand{\eeq}{\end{eqnarray}} 
\newcommand{\tb}{\tan \beta}

\newcommand{\lhc}{\rm lHC}
 
\preprint{LPT Orsay 10-89,\\ CERN-PH-TH/2010--289}

\title{Higgs production at the lHC}

\author{Julien Baglio\\
Laboratoire de Physique Th\'eorique, U. Paris-Sud et CNRS, 91405 Orsay Cedex, 
France.\\
E-mail: \email{Julien.Baglio@th.u-psud.fr}}

\author{Abdelhak Djouadi\\
Laboratoire de Physique Th\'eorique, U. Paris-Sud et CNRS, 91405 Orsay Cedex, 
France.\\
Theory Unit, CERN, 1211  Gen\`eve 23, Switzerland.\\
E-mail: \email{Abdelhak.Djouadi@cern.ch}}

\abstract{We analyze the production of Higgs particles at the early stage of
the CERN large Hadron Collider with a 7 TeV center of mass energy (lHC).  We
first consider the case of the Standard Model Higgs boson that is mainly
produced in the gluon--gluon fusion channel and to be detected in its decays
into electroweak gauge bosons, $gg\to H \to WW,ZZ,\gamma\gamma$. The production
cross sections at $\sqrt s=7$ TeV and the decay branching ratios, including all
relevant higher order QCD and electroweak corrections, are evaluated. An
emphasis is put  on the various theoretical uncertainties that affect the
production rates: the significant uncertainties from scale variation and from
the parametrization of the parton distribution functions as well as the
uncertainties which arise due to the use of an effective field theory in the
calculation of the next--to--next--to--leading order corrections. The parametric
uncertainties stemming from the values of the strong coupling constant and the
heavy  quark masses in the Higgs decay branching ratios, which turn out to be
non--negligible, are also discussed. The implications for  different center of
mass energies of the proton collider, $\sqrt s=8$--10 TeV as well as for the
design energy $\sqrt s= 14$ TeV, are briefly summarized. We then discuss the
production of the neutral Higgs particles of the Minimal Supersymmetric
extension of the Standard Model in the two main channels:  gluon--gluon and
bottom quark fusion leading to Higgs bosons which subsequently decay into tau
lepton or $b$--quark pairs, $gg, b\bar b \to {\rm Higgs} \to \tau^+\tau^-,b\bar
b$. The Higgs production cross sections at the $\lhc$  and the decay branching
ratios are analyzed. The  associated theoretical uncertainties are found to be 
rather large and will have a significant impact on the parameter space of the
model   that can be probed.}  


\begin{document} 

\renewcommand{\thefootnote}{\arabic{footnote}}
\setcounter{footnote}{0}
\setcounter{page}{2}

\section{Introduction}  

The Standard Model (SM) of the electroweak and strong interactions crucially
relies on the Higgs mechanism in order to spontaneously break the electroweak
symmetry and generate in a gauge invariant way the elementary particle  masses
\cite{Higgs,Review}. In this minimal realization of electroweak symmetry
breaking, one complex Higgs doublet field is introduced which implies the
existence of a single neutral scalar particle, the Higgs boson $H$.  In
supersymmetric theories \cite{SUSY}, that are widely considered as the most
attractive extensions of the SM as they protect the Higgs boson mass against
large radiative corrections and stabilize the hierarchy between the electroweak
and Planck scales, the Higgs sector is extended to contain at least two Higgs
doublet fields. The Minimal Supersymmetric Standard Model (MSSM) predicts the 
existence of  five Higgs particles: two CP--even Higgs particles $h$ and $H$, a
CP--odd $A$ boson and two charged $H^\pm$ particles
\cite{HHG,Review2,MichaelR,SvenPR}. The search for these new scalar particles
is the main goal of present high--energy colliders.

With its very successful operation in the last years, the Fermilab Tevatron  $p
\bar p$ collider has now collected a substantial amount of integrated
luminosity which allows the CDF and D0 experiments to be sensitive to  Higgs
particles. Exclusion limits beyond  the well established LEP bounds
\cite{LEP-Higgs} are now being set on Higgs masses both in the SM 
\cite{Tevatron} and in the MSSM \cite{Tevatron-MSSM}.  The CERN proton--proton
collider successfully started operation  but at a reduced center of mass
energy of 7 TeV \cite{LHC}. Not to confuse it with the LHC, which is expected
to operate at the design energy close to 14 TeV, we will call this early stage
machine the $\lhc$ for large (or littler) Hadron Collider.  The $\lhc$ will
also be sensitive to the SM Higgs particle and will start to be competitive
with the Tevatron  once it will accumulate the expected 1 fb$^{-1}$ of data
\cite{LHC,LHCXS,ATLAS,CMS}. However, with the present expectations, the
Tevatron and presumably also  the $\lhc$ will only be able to exclude  the
existence of the SM Higgs particle in some given mass range. An undebatable
discovery in the entire Higgs mass range favored by theoretical 
considerations,  115 GeV $\lsim M_H \lsim {\cal O}(1\;{\rm TeV})$, has to await
for the full--fledged LHC with a center of mass energy of 14 TeV and ${\cal
O}(10\;$fb$^{-1})$ of integrated luminosity. Observing  the MSSM Higgs bosons
at the Tevatron and the  $\lhc$ will be more likely as, in some areas of the
supersymmetric parameter  space, the production cross sections are much higher
than in the SM \cite{Review2}.  

The exclusion of Higgs mass regions relies crucially on the theoretical
predictions for the production cross sections for the Higgs signal as well as
for the relevant SM backgrounds. At the Tevatron, the two main search channels
for the SM Higgs boson are the top and bottom quark loops mediated  gluon--gluon
fusion mechanism $gg\to H$ \cite{ggH-LO} with the Higgs decaying into $WW$
pairs which lead to $\ell \nu \ell \bar \nu$ (with $\ell\!=\!e,\mu$) final
states  \cite{HWW-DD} and the Higgs--strahlung processes $q \bar q\! \to\! VH$
(with $V\!=\!W,Z$) \cite{HV-LO} with the subsequent $H\!\to \!b\bar b$ and $V
\to \ell\!+\!X$ decays of the Higgs and the associated gauge bosons.
Analyses performed by the CDF and D0 experiments, under some assumptions for
the production cross sections and their associated uncertainties, have recently
excluded the Higgs mass range $M_{H}\!=\!158$--175 GeV at the 95\% confidence
level (CL) \cite{Tevatron}. In this mass range, the Higgs signal is mainly due
to the $gg\!\to\!H\to\!WW\to\! \ell \nu \ell \bar  \nu$ production and decay
channels.

At the $\lhc$, the Higgs--strahlung processes, as well as as other production
channels such as weak vector--boson fusion and associated Higgs production with
top quark pairs, have too small cross sections and/or are plagued with too
large QCD backgrounds. Thus, in practice,  only the gluon--gluon fusion process
with the Higgs boson decaying into $H\!\to\! WW\!\to\! \ell \ell \nu \bar
\nu$,  $H\!\to\! ZZ\! \to\!2\ell\!+\!X$, where $X$ stands for charged leptons,
neutrinos and eventually also jets including $b$--quark jets, and to a lesser
extent $H \to \gamma \gamma$  final states will be mostly relevant.  At a 
center of mass energy $\sqrt{s}\!=\!7$ TeV and with 1 fb$^{-1}$ of data, recent
studies by the ATLAS and CMS collaborations have shown that the mass  range
$M_H\!\approx\! 150$--190 GeV can be excluded at 95\% CL if no Higgs signal is
observed \cite{ATLAS,CMS}.     

Nevertheless, it is well known  that the production cross sections at hadron
colliders as well as the associated  kinematical distributions are  generally
affected by various and possibly large uncertainties. In a recent study of SM
Higgs production at the Tevatron \cite{Hpaper}, it has been re-emphasized that
while the signal production cross sections are well under control in the
Higgs--strahlung processes, the theoretical uncertainties are rather large in
the case of the gluon--gluon fusion channel. These uncertainties are stemming
mainly from the variation of the energy scale at which the process is evaluated
(a variation that provides a hint of the not yet calculated perturbative higher
order corrections) and the parametrization of the parton distribution functions
or PDFs (in particular, the gluon density at moderate to high Bjorken--$x$
values).  At the Tevatron, these uncertainties can shift the central values of
the $gg\to H$ cross section \cite{ggH-NLO,SDGZ,ggH-NNLO,ggH-resum}, by more
than $\approx 20\%$ in both  the case of the scale and the PDF uncertainty.
Arguing that these are two theoretical uncertainties that have in principle no
statistical ground and should thus not be added in quadrature, we have proposed
a  more adequate procedure to combine them which led to a total theoretical
uncertainty of about $\pm 40\%$ on the $gg\to H\to \ell \nu \ell \bar \nu$
signal. This  overall uncertainty is much larger than the uncertainty assumed
in  the CDF and D0 combined analysis that excluded the SM Higgs boson at the
95\% CL in the mass range $M_H=158$--175 GeV \cite{Tevatron}.   To our opinion,
this exclusion limit should thus be reconsidered.  

In the present paper we extend the analysis of Ref.~\cite{Hpaper} to the case
of the $\lhc$, i.e. for proton--proton collisions with a center of mass  energy
of 7 TeV. We first update the Higgs production cross sections in the $gg\to H$
channel, including all the relevant higher order QCD and electroweak
corrections and then analyze the various uncertainties that affect them. We
show that  the scale and PDF uncertainties, as well as the non--negligible 
uncertainty due to the use of an effective approach in the calculation of QCD
and electroweak higher order corrections beyond next--to--leading order, add up
to $\approx 25$--30\% depending on the considered Higgs mass range.  The total
uncertainty  is significantly smaller than that obtained at the Tevatron,  as a
result of smaller QCD radiative corrections and a better knowledge of the gluon
distribution function at the energies relevant at the $\lhc$. 

We then discuss an additional source of theoretical uncertainties that has not
been considered neither in the experimental analyses \cite{Tevatron,ATLAS,CMS}
nor in Ref.~\cite{Hpaper}: the one affecting the Higgs decay branching ratios.
Indeed, while the Higgs decays into lepton and gauge boson pairs are well under
control (as mainly small electroweak effects are involved),  the partial decays
widths into quark pairs and gluons are plagued with  uncertainties  that are
mainly due to  the imperfect knowledge of the bottom and charm  quark masses
and the value of the strong coupling constant $\alpha_s$.  Updating a previous
analysis \cite{DSZ}, we show that at least in the intermediate mass range, $M_H
\approx 120$--150 GeV, where the SM Higgs decay rates into $b\bar b$ and
$W^+W^-$ final states have the same order of magnitude, the parametric
uncertainties on these two main Higgs decay branching ratios are
non--negligible, being of the order of 3 to 10\% at the $1\sigma$ level. 

For completeness, we will extend our analysis and explore the implications for
SM Higgs searches at center of mass energies beyond  $\sqrt s=7$ TeV. Not only
the full--fledged  LHC with $\sqrt s =14$ TeV will be considered, but also
intermediate energies between  $\sqrt s= 7$ and 10 TeV that are being currently
considered for a  very near future, with integrated luminosities significantly
larger than 1 fb$^{-1}$ \cite{LHC}. The main result that we obtain  is that
while the production cross sections for the SM Higgs particle are higher at
energies $\sqrt s=8$--14 TeV compared to $\sqrt s= 7$ TeV, the overall
theoretical  uncertainties that affect them  are approximately the same.   

In the MSSM, the chances of observing at the $\lhc$ the neutral Higgs
particles, that we will collectively denote by $\Phi$, are significantly higher
than in the  case of the SM Higgs boson. Indeed, for large values of the ratio
of the vacuum expectation values of the two MSSM Higgs fields, $\tan\beta
=v_2/v_1$, the couplings  of some of the Higgs bosons  to isospin down--type
particles are strongly enhanced. For not too heavy Higgs particles, this leads 
to production cross sections that are possibly orders of magnitude larger than
in the SM, in processes in which the MSSM Higgs particles couple to
bottom--quarks \cite{Review2,MichaelR}.   This is the case of the gluon--gluon
fusion mechanism for neutral Higgs production  $gg \to \Phi$ which, in this
context, dominantly  proceeds through $b$--quark triangular loops
\cite{ggH-LO,SDGZ},  as well as  bottom--quark fusion, in which the bottom
quarks are  directly taken from the protons in a five active flavor scheme,
$b\bar b \to \Phi$ \cite{bbH-LO,bbH-NLO,bbH-NNLO}. For the relevant high values
of $\tan\beta$, the $\Phi$ bosons decay almost exclusively into $b\bar b$ and
$\tau^+ \tau^-$ final states with, respectively, $\approx 90\%$ and $\approx
10\%$ branching ratios.  This leads  to the dominant production channels, $gg,
b \bar b \to \Phi \to \tau^+ \tau^-, b\bar b$. Again, besides the uncertainties
due to scale variation and the parametrization  of the PDFs, additional
uncertainties originate from   errors in the extraction of  the bottom quark
mass $m_b$ and the strong coupling constant $\alpha_s$. These uncertainties
affect both the production cross sections, as the amplitudes involve the
Higgs--$b\bar b$ Yukawa coupling $\propto m_b$,  and the  $\Phi \to \tau^+
\tau^-, b\bar b$ decay branching fractions.

A second aim of this paper is to analyze in detail these two main MSSM Higgs 
production  channels at the $\lhc$: the production cross sections, including
the relevant higher order radiative corrections are  updated, the important QCD
effects are summarized and the various theoretical uncertainties are evaluated.
We show that the uncertainties from scale variation and the parametrization of
the PDFs in the cross sections are at least as large as in the SM $gg\to H$
case. Additional   parametric uncertainties in both the production and the
decay rates, as well as a freedom in the choice of the renormalization scheme
that defines the $b$--quark mass,  lead to a  total theoretical uncertainty
that is  even  larger, of the order of 40 to 50\% depending on the Higgs
masses. This uncertainty  will thus have a significant impact on the MSSM
parameter space that can be probed at the $\lhc$.  

The rest of the paper is organized as follows. In the next section we discuss
the production of the SM Higgs boson at the $\lhc$ with $\sqrt s=7$ TeV, mainly
concentrating on the dominant $gg\to H$ production channel; the total inclusive
cross section including the relevant higher order contributions is updated, the
various theoretical  uncertainties evaluated and combined, and the implications
for higher center of mass energies summarized. In section 3, the parametric
uncertainties in the SM Higgs decay branching ratios are summarized and
combined with the uncertainties in the production cross sections. In section 
4, we discuss the production of the MSSM neutral Higgs bosons: the cross
sections in the two production channels, $gg \to \Phi$ and $ b\bar b \to \Phi$
are analyzed  and the associated uncertainties from scale and scheme 
dependence, the PDF uncertainties as well as the parametric uncertainties are
discussed; the errors in the Higgs branching ratios in the main decay modes 
$\Phi \to \tau^+ \tau^-$ and $\Phi \to b\bar b$ are addressed and the way to
combine all these theoretical uncertainties in the production and in the decays
is discussed. Here, we only discuss the case of the lHC with  $\sqrt s=7$ TeV
and simply summarize  the implications at slightly higher energies,  $\sqrt
s=8$--10  TeV. Finally, a conclusion is given in the last section.

\section{The SM Higgs boson at the lHC}

\subsection{The Higgs production cross sections} 

The main production channel for the Higgs particles at hadron colliders is the
gluon--gluon fusion channel, $gg\to H$, which proceeds through triangular heavy
quark loops \cite{ggH-LO}. In the SM, it is dominantly mediated by the top
quark loop contribution, with a bottom quark contribution that does not exceed
the 10\% level at leading order. This process is known to be subject to
extremely  large QCD radiative corrections
\cite{ggH-NLO,SDGZ,ggH-NNLO,ggH-resum,ggH-HO,ggH-radja,ggH-FG} that can be
described by an associated $K$--factor defined as the ratio of the higher
order (HO) to the lowest order (LO) cross  sections, consistently evaluated
with the value of the strong coupling $\alpha_s$ and the parton distribution 
functions (PDF)  taken at the considered order,   
\beq   
K_{\rm HO}= \sigma^{\rm HO} |_{ (\alpha_s^{\rm HO} \, , \, {\rm PDF^{HO} )} } 
\; / \;  \sigma^{\rm LO} |_{ (\alpha_s^{\rm LO} \, , \,  {\rm PDF^{LO}}) }
\label{kfactor}    
\eeq  
The next-to-leading-order (NLO) corrections in QCD are known both  for infinite
\cite{ggH-NLO} and finite \cite{SDGZ} loop quark masses and, at $\sqrt{s}=7 $
TeV, lead to a $K$--factor $K_{\rm NLO}\sim 1.8$ in the low Higgs mass
range, if the central scale of the cross section is chosen to be the Higgs
mass.  It has been shown in Ref.~\cite{SDGZ} that working in an effective field
theory (EFT) approach in which the top quark mass is assumed to be infinite is
a very  good approximation for Higgs mass values below the $t\bar t$
threshold $M_H \lsim 2m_t$, provided that the leading order
cross section contains the full  $m_t$ and $m_b$ dependence. The calculation of
the next-to-next-to-leading-order (NNLO) contribution has been done 
\cite{ggH-NNLO} only in the EFT approach $M_H \ll 2m_t$  and, at $\sqrt s=7$ 
TeV, it leads to a $\approx 25\%$ increase of the cross section\footnote{The 
QCD corrections to $gg\to H$ at $\sqrt s=7$ TeV are thus smaller than the
corresponding ones at the Tevatron as the $K$--factors in this case are $K_{\rm
NLO} \approx 2$ and $K_{\rm NNLO} \approx 3$ (with a central scale equal to
$M_H$). At the LHC with $\sqrt s=14$ TeV, the $K$--factors  are even smaller,
$K_{\rm NLO} \approx 1.7$ and $K_{\rm NNLO} \approx 2$. The perturbative series
shows thus a better (converging) behavior at LHC than at Tevatron energies.}, 
$K_{\rm NNLO} \sim 2.5$. The resummation of soft gluons is known up to
next-to-next-to-leading-logarithm (NNLL) and, again, increases the cross section
by slightly less than 10\% \cite{ggH-resum,ggH-FG}. The effects of soft--gluon
resumation at NNLL can be accounted for in $\sigma^{\rm NNLO}(gg\to H)$ by
lowering the central value of the  renormalization and factorization
scales\footnote{We will include the soft--gluon resummation contributions
in this indirect way since, as discussed in Ref.~\cite{Hpaper}, we would like
to stick to the fixed order NNLO calculation for two main reasons: $i)$   there
are not yet  parton distribution functions which include soft--gluon 
resummation  and it  appears inconsistent to fold a partonic cross section with
PDFs that are not  at the same order of perturbation theory  (although the
effects might be  small in practice \cite{Magnea}) and $ii)$ the soft--gluon
resummation is not available for the Higgs+jet production cross sections and/or
the  Higgs cross sections including kinematical cuts (both are known only at
NNLO \cite{ggH-cuts,ggH-ADGSW}) which, ultimately, are the basic experimental
inputs.}, from  $\mu_0=M_H$ to $\mu_0=\frac12 M_H$. We will thus choose for
definiteness the value   
\begin{equation}
\mu_0=\mu_R=\mu_F=\frac12 M_H \label{central-scale}  
\end{equation}
for the central scale of the process which, in passing,  also improves the
convergence of the perturbative series and is more appropriate to describe the
kinematics of the process \cite{BabisHH}. The electroweak corrections are known
exactly up to NLO \cite{ggH-actis,ggH-ew} and contribute at the level of a few
percent; there  are also small mixed NNLO QCD--electroweak effects which have 
been calculated \cite{ggH-radja} in an effective approach valid for $M_{H}\ll
M_W$. 

\begin{figure}[!t]
\begin{center}
\includegraphics[scale=0.75]{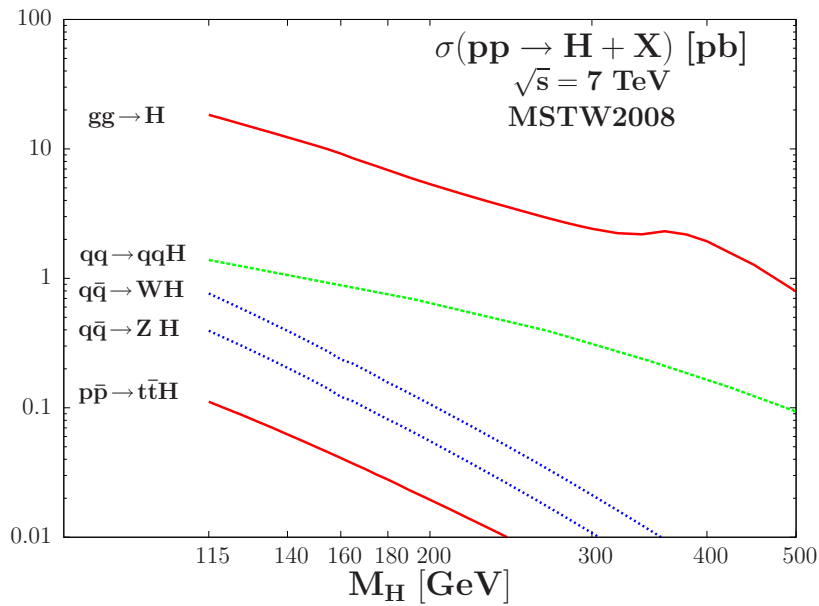}
\end{center}
\vspace*{-4mm}
\caption[]{The total cross sections for Higgs production at the $\lhc$ 
with $\sqrt s=7$ TeV as a function of the Higgs mass. The MSTW set of PDFs 
has been used and the higher order corrections are included as discussed in 
the text.}
\vspace*{-2mm}
\label{pp-H-TeV}
\end{figure}

Our calculation\footnote{Other recent updates of the $gg\to H$ cross section 
can be found in Ref.~\cite{ggH-radja,ggH-FG,other-updates}.} of the $gg\to H$
production cross section at the $\lhc$,  including these higher order
corrections strictly follows the one performed in Ref.~\cite{Hpaper} for the
Tevatron: the starting point is the Fortran code {\tt HIGLU} \cite{Michael}
which evaluates the $gg \to H$ cross section at exact NLO in QCD (i.e. with  the
exact contributions of the top and bottom quark loops) to which we add   the
NNLO QCD contribution in the infinite top--quark mass limit, but with the LO
normalisation containing  the exact $m_t$ and $m_b$ dependence;  grids  for the
exact NLO electroweak  and the mixed QCD--electroweak  corrections are then
implemented within the code (they have been added in the partial factorization
scheme \cite{ggH-actis}, see section 2.2.2). The production cross sections are
shown at the $\lhc$ with $\sqrt s=7$ TeV in Fig.~\ref{pp-H-TeV} for an
updated value of the top quark mass and when the partonic cross
section is folded with the latest NNLO MSTW2008 public set of PDFs
\cite{PDF-MSTW}. The renormalization and factorization scales are
fixed to the central values $\mu_F=\mu_R=\frac12 M_H$. Our results
agree with those given in Refs.~\cite{ggH-radja,ggH-FG} and updated
in Ref.~\cite{LHCXS} within a few percent.

For completeness, we also display in Fig.~\ref{pp-H-TeV} the cross
sections for the three other Higgs production channels at hadron
colliders that we evaluate using the programs of Ref.~\cite{Michael}:
\vspace{-2mm}
\begin{enumerate}[{\it i)}]
\itemsep-3pt
\item{the Higgs--strahlung processes  $q\bar q \to HV$ with $V=W,Z$ that are
known exactly up to NNLO in QCD \cite{HV-NLO,MichaelR,HV-NNLO} and up to NLO
for the electroweak corrections \cite{HV-EW}; they are evaluated at 
$\mu_0= M_{HV}$ (the invariant mass of the $HV$ system) for the central
scale\footnote{Here, the QCD $K$--factors are moderate, $K_{\rm NNLO} \sim 1.5$
and the electroweak  corrections reduce the cross section by an amount of 
$\approx 3-8\%$. For the evaluation of the cross section, we have used the NLO
code {\tt V2HV} \cite{Michael} in which we implemented  these higher order 
contributions. Our results agree with those of Ref.~\cite{LHCXS}. };} 

\item{the weak vector boson fusion channel $qq \to Hqq$ evaluated at the scale
$\mu_0=Q_V$ (the momentum transfer at the gauge boson leg) in which only the 
NLO QCD corrections \cite{VVH-NLO,MichaelR} have been included; the NNLO
corrections have  been found to be very small \cite{VVH-NNLO} and we omit the
electroweak corrections \cite{VVH-EW};}  

\item{associated Higgs production with top quark pairs in which only the
leading order order contribution  is implemented but at a central scale
$\mu_0\!=\!\frac12(M_H+2m_t)$, which is a good approximation  at these energies
as the NLO $K$--factor is very close to unity \cite{ttH-NLO,MichaelP}.} 
\end{enumerate}

\begin{table}[!h]{\small%
\let\lbr\{\def\{{\char'173}%
\let\rbr\}\def\}{\char'175}%
\renewcommand{\arraystretch}{1.41}
\vspace*{2mm}
\begin{center}
\begin{tabular}{|c||ccccc|}\hline
$M_H$ & $\sigma^{\rm NNLO}_{g g\to H}~$ & $\sigma^{\rm NLO}_{q q\to H q q}~$ &
$\sigma^{\rm NNLO}_{q \bar q\to HW}~$ & $\sigma^{\rm NNLO}_{q \bar q\to HZ}~$ 
& $\sigma^{\rm LO}_{pp\to t\bar t H}~$ \\ \hline
$115$ & $18347.4$ & $1386.1$ & $764.1$ & $394.0$ & $111.5$ \\ \hline
$120$ & $16844.6$ & $1313.2$ & $664.8$ & $343.2$ & $98.7$ \\ \hline
$125$ & $15509.2$ & $1259.5$ & $580.0$ & $300.3$ & $87.9$ \\ \hline
$130$ & $14322.6$ & $1192.1$ & $507.5$ & $263.2$ & $78.1$ \\ \hline
$135$ & $13260.6$ & $1148.6$ & $446.0$ & $231.4$ & $69.8$ \\ \hline
$140$ & $12305.6$ & $1087.0$ & $392.9$ & $203.6$ & $62.3$ \\ \hline
$145$ & $11446.4$ & $1051.5$ & $347.3$ & $180.0$ & $56.0$ \\ \hline
$150$ & $10665.9$ & $1006.2$ & $307.1$ & $159.1$ & $50.2$ \\ \hline
$155$ & $9936.4$ & $964.6$ & $272.0$ & $140.4$ & $45.3$ \\ \hline
$160$ & $9205.9$ & $908.4$ & $236.8$ & $121.7$ & $40.9$ \\ \hline
$165$ & $8470.8$ & $875.3$ & $218.9$ & $112.3$ & $37.0$ \\ \hline
$170$ & $7872.3$ & $842.5$ & $196.0$ & $100.6$ & $33.7$ \\ \hline
$175$ & $7345.3$ & $796.6$ & $175.6$ & $90.2$ & $30.5$ \\ \hline
$180$ & $6861.2$ & $768.0$ & $156.8$ & $81.2$ & $27.9$ \\ \hline
$185$ & $6416.3$ & $732.8$ & $142.7$ & $74.0$ & $25.4$ \\ \hline
$190$ & $6010.0$ & $705.0$ & $129.3$ & $67.1$ & $23.1$ \\ \hline
$195$ & $5654.8$ & $683.8$ & $117.5$ & $60.9$ & $21.2$ \\ \hline
$200$ & $5344.1$ & $651.5$ & $106.8$ & $55.3$ & $19.5$ \\ \hline
$220$ & $4357.0$ & $556.7$ & $74.4$ & $38.4$ & $14.0$ \\ \hline
$240$ & $3646.4$ & $479.6$ & $52.9$ & $27.2$ & $10.4$ \\ \hline
$260$ & $3110.7$ & $411.6$ & $38.4$ & $19.7$ & $7.8$ \\ \hline
$280$ & $2706.4$ & $360.7$ & $28.3$ & $14.5$ & $6.1$ \\ \hline
$300$ & $2415.4$ & $312.6$ & $21.2$ & $10.8$ & $4.8$ \\ \hline
\end{tabular}
\vspace*{-2mm}
\end{center}
\caption{The total Higgs production cross sections (in fb) in the processes 
$\protect{g g\to H}$, vector--boson fusion  $\protect{q q\to H q q}$,
Higgs--strahlung  $\protect{q\bar q\to HW.HZ}$  and  associated 
production $\protect{pp \to t\bar t H}$  at the $\lhc$  with $\sqrt s=7$ TeV 
for given Higgs mass values (in GeV) with the corresponding central scales 
described in the main text. The MSTW sets of PDFs have been used at the 
relevant  order.} 
\label{lhc7_allchannels}
\vspace*{-1mm}
}
\end{table}

As can be seen, the $gg\to H$ process is by far dominating in the entire Higgs
mass range, with a cross section that is one to two orders of magnitudes larger
than in the other production channels. For low Higgs masses, $M_H \lsim
2M_W$, it leads to more than 10.000 events for a luminosity of order 1
fb$^{-1}$.  Table \ref{lhc7_allchannels} displays the values of the
cross sections for the five Higgs production processes for a selection
of Higgs masses relevant at the $\lhc$.

\subsection{The theoretical uncertainties}

\subsubsection{Higher order contributions and scale variation }

In the calculation of production cross sections and kinematical distributions
at hadron colliders, as the perturbative series are truncated and the results
are available only at a given perturbative order, there is a residual
dependence of the observables on  the renormalization scale $\mu_{R}$ which
defines the strong coupling constant $\alpha_{s}$ and on the factorization
scale $\mu_{F}$ at which the matching between the perturbative matrix elements
calculation and the non--perturbative parton distribution functions is
performed. The  uncertainty due to the variation of these two scales is viewed
as an estimate of the unknown (not yet calculated) higher--order terms and is
rather often the dominant source of theoretical uncertainties. 

In general, starting  with the median scale $\mu_F=\mu_R=\mu_{0}$ for which the
central value of the cross section  is obtained, the two scales $\mu_{R}$ and
$\mu_{F}$ are varied within the interval ${\mu_0}/{\kappa} \le \mu_R, \mu_F \le
{\kappa}\! \times \! {\mu_0}$ with the value of the constant $\kappa=2,3,4,$
etc... to be chosen.  The additional restriction $1/\kappa \le \mu_F/\mu_R \leq
\kappa$ is  also often applied\footnote{In the case of the $gg\to H$ process,
the  maximal (minimal) cross sections at a given fixed order in perturbation
theory (in particular at NNLO) is obtained for the scale choices $\mu_R=\mu_F= 
\mu_0/\kappa~(\kappa \mu_0)$ and, hence, this restriction is irrelevant in
practice.}. The choice of the scale variation domain and hence the constant
factor $\kappa$, is rather subjective and a possibility would be  to adopt the 
$\kappa$ value which allows the uncertainty band due to the scale variation of the
lower order cross section to incorporate the central value of the cross section
(i.e. with $\mu_R=\mu_F=\mu_0$) calculated at the highest available order. 

\begin{figure}[!h]
\vspace*{3mm}
\begin{center}
\mbox{
\epsfig{file=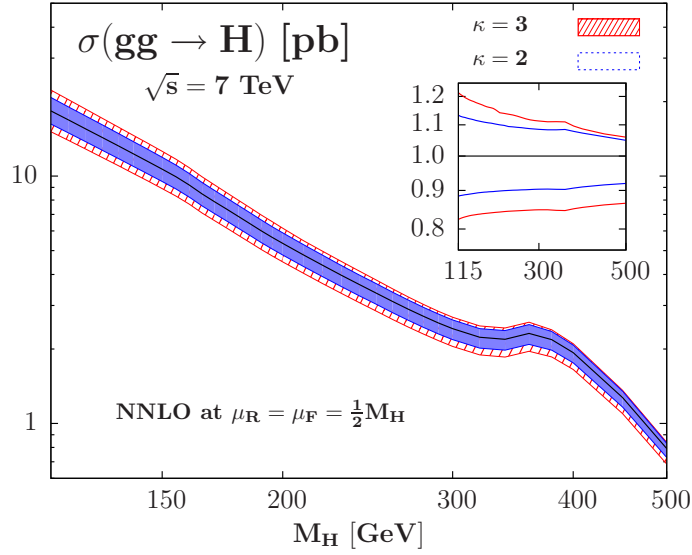,width=9cm} }
\end{center}
\vspace*{-4mm}
\caption[]{
The scale uncertainty band of $\sigma^{\rm NNLO} (gg\to H)$ at the $\lhc$
as a function of $M_H$ for a scale variation in the domain $M_H/(2\kappa) 
\le \mu_R, \mu_F  \le \kappa\!\times\! \frac12 M_H$ for $\kappa=2$ and 3; 
in the insert, shown are the relative deviations from the central value with a 
scale $\mu_R=\mu_F=\frac12 M_H$.}
\label{scale-SM}
\vspace*{-.1mm}
\end{figure}

In our analysis, we have adopted the central scale value $\mu_R=\mu_F= \mu_0=
\frac12 M_H$ as discussed earlier. At $\sqrt s=7$ TeV, for the scale uncertainty
band of $\sigma^{\rm NLO} (gg\to H)$ to catch the central value of $\sigma^{\rm
NNLO}( gg\to H)$ for $\mu_0=\frac12M_H$, a value $\kappa=2$ is sufficient.
Adopting the range $\frac12 \mu_0 \le \mu_R,\mu_F \le 2\mu_0$  with
$\mu_0=\frac12 M_H$ for the cross section $\sigma^{\rm NNLO}(gg\to H)$, one
obtains the scale variation  displayed in  Fig.~\ref{scale-SM} as a function of
$M_H$; it is compared to a variation with a factor $\kappa=3$.  In the insert,
shown  are the maximal and minimal variations of $\sigma^{\rm NNLO} (gg\to H)$ 
compared to the central value. One sees that  for $\kappa=2$, a scale
uncertainty of  $\approx \pm 10 \%$ is obtained in the low mass range, $M_H
\approx 120$ GeV, which decreases to the level of $\approx -8\%, +4\%$ at high
masses, $M_H \approx 500$ GeV. Note that if the domain for scale  variation were
extended to $\kappa=3$, the uncertainty would have  increased to $\approx \pm
17\%$ in the low $M_H$ range\footnote{In the analysis  of Ref.~\cite{Hpaper} for
the case of the Tevatron,  the value $\kappa=3$ was adopted instead of 
$\kappa=2$  and, hence, a larger domain for the scale  variation  was assumed as
the higher order QCD corrections in $\sigma(gg\to H)$ are larger at the Tevatron
compared to  the $\lhc$.} as also shown in the figure.

\subsubsection{The use of an effective theory approach}

Another source  of theoretical uncertainties is specific to the gluon--gluon
fusion mechanism in which the cross section beyond the NLO approximation is
evaluated  in an  effective theory (EFT) approach where the particles that are
running in the loops are assumed to be much heavier than the produced Higgs 
boson. At NLO, the approximation  $m_t \gg M_{H}$ for the contribution of the
top quark in the loop is rather good for Higgs masses below the $t \bar t$
threshold,  $M_H \lsim 340$ GeV, in particular when the full quark mass
dependence  of the leading order cross section $\sigma^{\rm LO}_{\rm exact}$ is
taken into account \cite{SDGZ}. At NNLO,  this approximation for the top quark
contribution  seems  also to be accurate as studies of the effect of a finite
$m_t$ value in expansions  of $M_H/(2m_t)$ have shown a difference below the
percent level with respect to the EFT calculation for $M_H \lesssim 300$ GeV
\cite{ggH-NNLO-mt}.

Nevertheless, the EFT approach should definitely not be valid for Higgs masses
beyond the $t\bar t$ threshold, where the $gg\to H$ amplitude develops imaginary
parts. This can be seen at NLO where both the exact and the approximate results
are known. We will thus include an uncertainty related to the use of the EFT
approach for $M_H \gsim 2m_t$ which is taken as  the difference between
$\sigma^{\rm NLO}_{\rm exact}$ and $\sigma^{\rm NLO}_{m_t \to  \infty}$ when the
exact top quark mass dependence is included in the LO cross section and when one
rescales this difference with the relative magnitudes of the NLO and NNLO 
$K$--factors, i.e. $K^{\rm NLO}_{m_t \to \infty}/K^{\rm NNLO}_{m_t \to
\infty}$. 

\begin{figure}[!h]
\vspace*{3mm}
\begin{center}
\mbox{
\epsfig{file=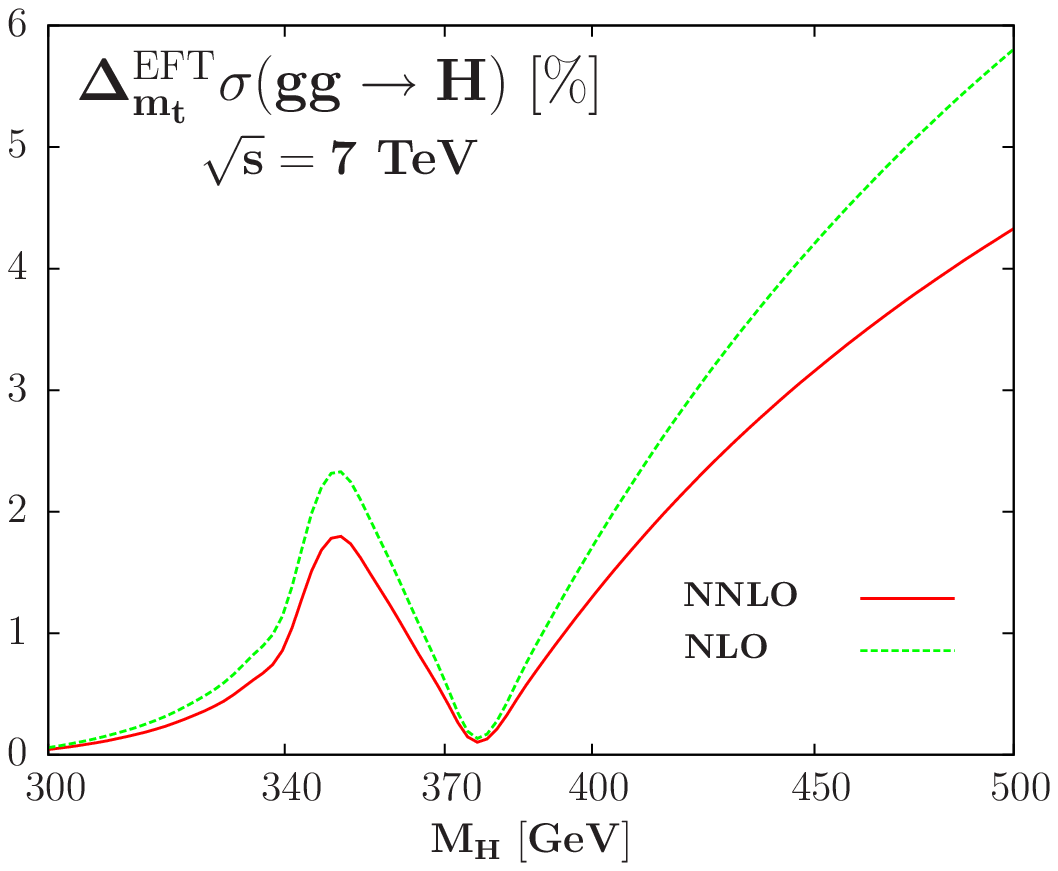, scale=0.7}\hspace*{2mm} 
\epsfig{file=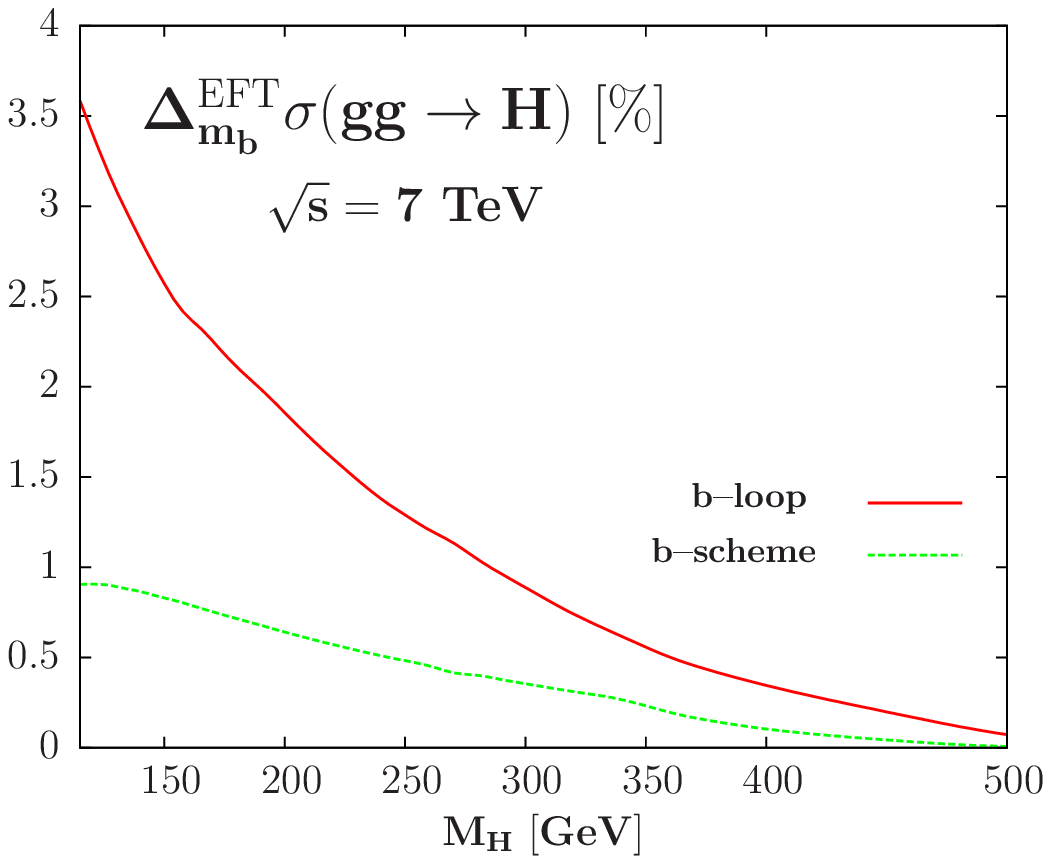, scale=0.7} 
}
\end{center}
\vspace*{-3mm}
\caption[]{The estimated uncertainties (in \%) due to the use of the EFT
approach in the evaluation of $\sigma(gg \to H)$ at NNLO for the top quark 
loop contribution for Higgs masses beyond  $2m_t$ (left) and for the 
bottom--quark loop contributions (right).}
\label{EFT-mq}
\vspace*{-3mm}
\end{figure}

In the left--hand side of Fig.~\ref{EFT-mq}, shown is this difference in
percent  for $\sqrt s=7$ TeV. It is small for Higgs masses below $M_H=400$ GeV,
at most  2\%, it increases very rapidly with increasing $M_H$ values and, for
$M_H\gsim 500$ GeV, it is larger than $\approx 5\%$. This uncertainty is thus
not negligible and it should be accounted for (at the Tevatron, this uncertainty
has not been discussed \cite{Hpaper} as only Higgs masses below 200 GeV were
considered).

Furthermore, the EFT approach is certainly also not valid for the $b$--quark
loop  and, for instance, the omission of this contribution (at LO) leads to a
$\approx 10 \%$ difference compared to the exact case. Furthermore, the NLO
$K$--factor for the $b$--quark loop, $K^{\rm NNLO}_{\rm b-loop} \approx
1.2$--1.4, is much smaller than that for the top--quark loop. In fact, the
problem arises mainly because of the significant negative interference  between
the top and bottom loop contributions (the bottom quark contribution itself is
rather small) up to NLO and the fact that the bottom contribution has been
factorized out in the LO cross section. Thus, including the top quark
contribution only using the EFT approach might overestimate the NNLO correction
as this interference component is missing (and, since it is resulting from an
approximation it is not, in principle, accounted for by  e.g. the scale
variation that is performed at NNLO).

In order to estimate the uncertainty due the missing $b$--loop contribution at
NNLO, we simply follow the previous procedure for the top quark: we  assign an
error on the NNLO QCD result  which is  approximately the difference between the
exact result $\sigma_{\rm exact}^{\rm NLO}$ and the approximate result
$\sigma_{m_t \! \to \! \infty}^{\rm NLO}$  obtained at NLO  but rescaled with
the relative magnitude of the $K$--factors that one obtains at NLO and NNLO for
the top--loop, i.e. $K^{\rm NLO}_{m_t \to \infty}/K^{\rm NNLO}_{m_t \to
\infty}$. This leads to the uncertainty on the $\sigma^{\rm NNLO}(gg \to H)$ at
$\sqrt  s=7$ TeV that  is shown in the right--hand side of Fig.~\ref{EFT-mq}.
It   ranges from $\sim \pm 3\%$  for low Higgs values $M_H  \sim 120$ GeV where
the $b$--loop is relatively important to  less than $\sim \pm 1\%$ for
Higgs masses above $M_H \sim 300$ GeV. Note that this correction is larger than
that obtained at the Tevatron \cite{Hpaper} where a different  central scale,
$\mu_0=M_H$, has been adopted for its evaluation.

In addition, there is some freedom in the choice of the renormalization scheme
for the $b$--quark  mass in the $gg\to H$ amplitude: for instance, one can use
the on--shell scheme in which the pole mass is $m_b \approx  4.7$ GeV or the
$\overline{\rm MS}$ scheme in which the mass $\overline{m}_{b}(\overline{m}_{b}) \approx
4.2$ GeV is adopted. This leads to a difference of $\approx 1\%$ in the 
$b$--quark loop contribution at NLO that we will take as an additional 
uncertainty due to the scheme dependence (this scheme dependence will be
discussed in more details in the MSSM case).

Finally,  in the  corrections to the $gg\to H$ amplitude, the mixed
QCD--electroweak corrections have been calculated at NNLO  \cite{ggH-radja} in
an EFT approach with $M_{W/Z} \gg M_{H}$, a limit that  is obviously not valid
in practice. We thus should be cautious when using this correction and assign an
uncertainty which is of the same size as the contribution itself \cite{Hpaper}. 
This uncertainty is comparable in size to the difference between the electroweak
correction calculated exactly at NLO \cite{ggH-actis} evaluated in the  partial
factorization scheme (where  the correction $\sigma^{\rm LO} \Delta_{\rm EW}$ is
simply added to the QCD corrected cross section)  compared to the result in the
complete factorization  scheme (where the NNLO cross section is multiplied by
$1+\Delta_{\rm EW}$). This generates  an additional uncertainty of $\approx
3\%$ at most, which is not shown as it is the same as the one discussed in
Ref.~\cite{Hpaper} for the Tevatron.

\begin{figure}[!h]
\vspace*{-1mm}
\begin{center}
\mbox{
\epsfig{file=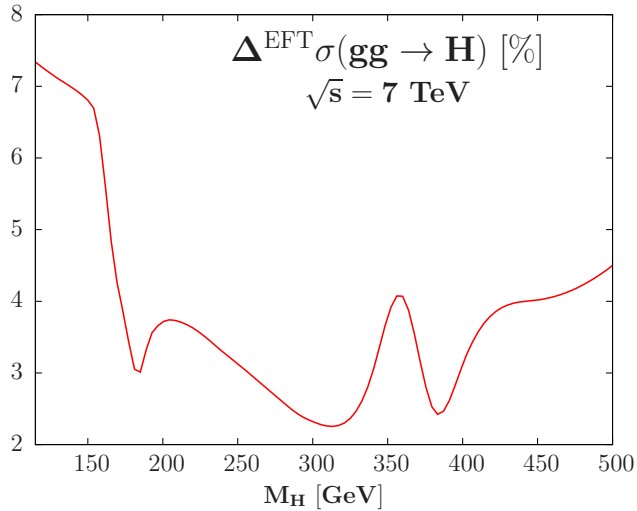, scale=0.78} 
}
\end{center}
\vspace*{-4mm}
\caption[]{The estimated total uncertainty (in \%)  at $\lhc$ energies 
and as a function of $M_H$ from  the use of the EFT approach for the 
calculation of the $gg\to H$ amplitude at NNLO and from the scheme dependence
of the $b$--quark mass.}
\label{ggH-eft}
\vspace*{-2mm}
\end{figure}

Adding the uncertainties from these three sources linearly, the resulting
overall uncertainty is displayed in  Fig.~\ref{ggH-eft}  as a function of $M_H$.
It amounts to  $\approx 7\%$ in the low Higgs mass range, drops to the level of
$\approx 4\%$ in the mass range $M_H \approx 200$--400 GeV, and then increases
to reach the level of $\approx 6\%$ at  600 GeV, as a result of the bad $M_H
\ll 2 m_t$ approximation for the top loop contribution. Since this is a pure
theoretical uncertainty, it should be added linearly to the uncertainty from
scale variation. 

\subsubsection{The PDFs and $\alpha_{s}$ uncertainties}

One of the most important sources of theoretical uncertainties in the $gg\to H$
production cross section is the one stemming from the parametrization of the
gluon densities. There are two quite different ways to estimate this
uncertainty.  

The first one is the so--called Hessian method where, besides the best fit PDF
with which the central values of the cross sections are evaluated,  a set of
$2N_{\rm PDF}$ PDF parameterizations  is provided that reflect the $\pm
1\sigma$ variation of all ($N_F$)  parameters that enter into the global fit.
These uncertainties are thus mostly due to the experimental errors in the
various data that are used in the fits.

\begin{figure}[!h]
\vspace*{-1mm}
\begin{center}
\mbox{
\epsfig{file=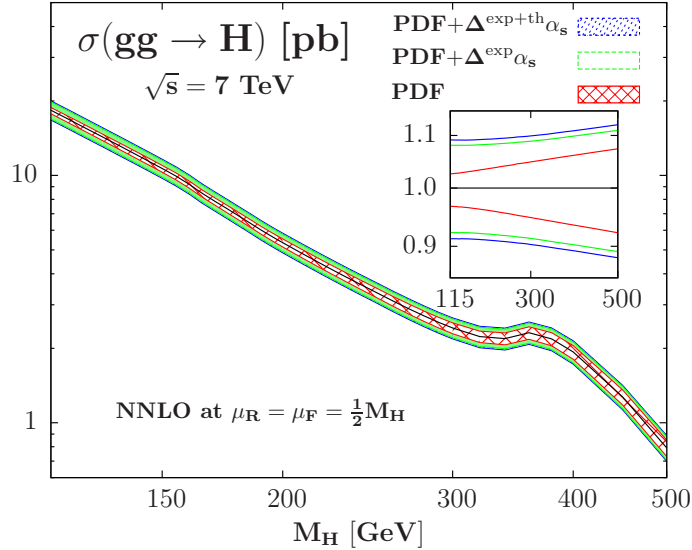,width=9cm} 
}
\end{center}
\vspace*{-4mm}
\caption[]{The central values and the 90\% CL PDF, PDF+$\Delta^{\rm
exp}\alpha_s$ and PDF+$\Delta^{\rm exp}\alpha_s+\Delta^{\rm th}\alpha_s$
uncertainty bands in  $\sigma^{\rm NNLO} (gg \to H+X)$ at the 
$\lhc$ when evaluated within the MSTW scheme. In the inserts, shown are the 
same but with the cross sections normalized to the central cross section.}
\label{ggH-pdf1}
\end{figure}

If one takes the latest NNLO public set provided by the MSTW collaboration 
\cite{PDF-MSTW,PDF-as} for instance, the PDF uncertainty at the 90\% CL that is
obtained at the $\lhc$ for the NNLO $gg\to H$ cross section is shown (by the
red lines and red band)  in  Fig.~\ref{ggH-pdf1} as a function of $M_H$.
For low Higgs masses, the uncertainty is at the level of 5\% but it increases
to reach the level of $\approx 8\%$ at high Higgs masses, $M_H\approx 500$ GeV.
If only the (more optimistic) 68\%CL errors are to be considered, the previous
numbers have to be divided by $\approx 1.6$.

\begin{figure}[!h]
\vspace*{-1mm}
\begin{center}
\mbox{
\epsfig{file=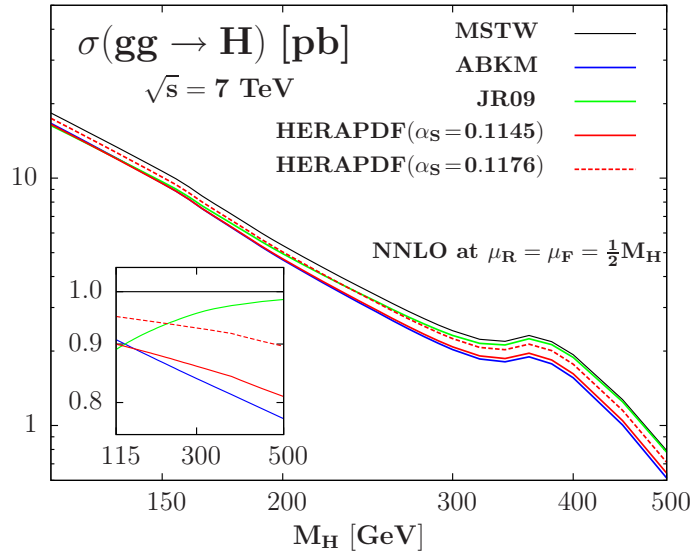,width=9cm}
}
\end{center}
\vspace*{-4mm}
\caption[]{The central values of the  NNLO cross section $\sigma(gg \to H+X)$ 
at the  $\lhc$ as a function of $M_H$ when evaluated in the four schemes 
which provide NNLO PDFs. In the inserts, shown are the same but with the 
cross sections normalized to the MSTW central cross section.}
\label{ggH-pdf2}
\end{figure}

The previous Hessian errors do not account for the theoretical assumptions that
enter into the parametrization of the PDFs and which may come from many
different sources. These assumptions lead to an uncertainty in the
parametrization that  is difficult to quantify in a given scheme, say the MSTW 
scheme. A way to access this theoretical uncertainty is to compare the results
for the central values of the cross section (and hence, using the best fit PDFs)
when using  different sets of PDFs which involve, in principle, different
assumptions. In  Fig.~\ref{ggH-pdf2}, we display the values of $\sigma^{\rm
NNLO}(gg\to H)$ that are obtained when folding the partonic cross section with
the gluon  densities that are predicted by the four PDF sets\footnote{As we 
make the choice of  using parton cross sections and PDFs that are consistently
taken at the same perturbative order, we will not consider two other
parameterizations, CTEQ \cite{PDF-CTEQ} and NNPDF \cite{PDF-NN}, which provide
PDF sets only at NLO.} that have parameterizations at NNLO: MSTW
\cite{PDF-MSTW},  JR \cite{PDF-JR}, ABKM \cite{PDF-ABKM} and HERAPDF
\cite{PDF-HERA}.  In the later case, two sets are provided: one with the value 
$\alpha_s=0.1176$ that is close to the  world average value \cite{PDG} and the
MSTW best--fit $\alpha_s=0.1171$ \cite{PDF-MSTW}, and one with  the value
$\alpha_s=0.1145$ used also in the ABKM set \cite{PDF-ABKM}, that is close to
the preferred values that one obtains using deep--inelastic scattering data
alone. 

As can be seen, while the differences in the cross sections are moderate in the
low Higgs mass range, being of the order of 10\% or less, there is a significant
difference at higher Higgs masses, ${\cal O}(25\%$) for $M_H\approx 500$ GeV.
Even with the same $\alpha_s$ input or best--fit values, HERAPDF  with
$\alpha_s=0.1176$ and MSTW results are strikingly different. For the large $M_H$
values, this discrepancy is mainly due to the gluon densities at moderate to
high Bjorken--$x$ values  which are less constrained by the data (in particular
if the Tevatron high $E_T$ jet data are not included as is the case of the ABKM
and HERAPDF sets).

We thus conclude when comparing Fig.~\ref{ggH-pdf1} and Fig.~\ref{ggH-pdf2}
that, depending on how the  uncertainties are estimated, through the $2 \times
20$ grids given in the MSTW scheme (other PDF sets such as ABKM for instance
give approximately the same error) or by comparing the cross sections
evaluated  using PDFs from different schemes, a rather different
uncertainty is obtained. To partly reconcile these two different results, we
will follow the analysis made in Ref.~\cite{Hpaper} and include the additional
errors on the  coupling $\alpha_s$.  

As a matter of fact, in addition to the PDF uncertainties, one should also
consider the ones coming from the uncertainties in the determination of the
value of the strong coupling constant\footnote{One can see that the input value
of  $\alpha_s$ is very important by comparing the two HERAPDF results in 
Fig.~\ref{ggH-pdf2}. Nevertheless, the difference in $\alpha_s$ is not
sufficient to explain the behavior of the cross sections in the MSTW and HERAPDF
cases for instance, and  there is some part that is presumably also due to the
different shapes of the  gluon densities.} $\alpha_{s}$ which, in general, is
fitted together with the  PDFs and this point is now well taken by the
experimental collaborations.  Indeed, since already at leading order one has 
$\sigma^{\rm LO}(gg\to H) \propto  \mathcal{O}(\alpha_s^2)$ and the $K$--factors
are large,  the uncertainty in the determination of $\alpha_s$ may induce a
non--negligible error on the final cross section. In the MSTW scheme,  the value
of   $\alpha_s$  together with its experimental accuracy, is  \cite{PDF-as}
\beq
\label{alphas}
\alpha_s(M_Z^2)&=& 0.1171~^{+0.0014}_{-0.0014}~{\rm (68\%~CL)}~{\rm or}~
                      ^{+0.0032}_{-0.0032}~{\rm (90\%~CL)}~{\rm at~NNLO} 
\eeq
We have evaluated  the 90\% CL correlated PDF+$\Delta^{\rm exp} \alpha_{s}$
uncertainties using the new  set--up  provided by the MSTW  collaboration
\cite{PDF-as} and the result is shown in  Fig.~\ref{ggH-pdf1} (green band and
green lines) as a function of $M_H$. As one can see, the obtained uncertainties
are much larger than the pure PDF uncertainties alone. This is particularly
true at low Higgs masses where the uncertainty is doubled.  Nevertheless,  even
with this additional uncertainty, one still cannot reconcile the MSTW and  the
ABKM/HERAPDF predictions for instance, in particular at high Higgs masses.

However, it has been argued in Ref.~\cite{Hpaper} that one also needs to
consider the theoretical uncertainty on $\alpha_{s}$ that is due to the
truncation of the perturbative series, the different heavy flavor schemes,
etc... This is particularly true as one has a significant difference between the
$\alpha_s$ world average value and the value obtained from deep--inelastic 
scattering data alone, a difference which could be reduced if this theory
uncertainty is included. The MSTW collaboration estimates the theoretical
uncertainty to be  $\Delta^{\rm th} \alpha_{s}  =  0.003~{\rm at~NLO}$
\cite{PDF-as}, which gives $\Delta^{\rm th} \alpha_{s} = 0.002$ at most at NNLO.
Using a fixed $\alpha_s$ NNLO grid with central PDFs given by the MSTW
collaboration, with $\alpha_s$ values different from the best--fit value (in the
range 0.107--0.127 with steps of 0.001 and which thus include the values
$\alpha_s(M_Z^2) =  0.1171\pm 0.002$ at NNLO), we have evaluated this
uncertainty. Adding it in quadrature to the PDF+$\Delta^{\rm exp} \alpha_{s}$
uncertainty, we obtain a total  PDF+$\Delta^{\rm exp}\alpha_s+\Delta^{\rm
th}\alpha_s$ uncertainty of  $\approx 11\%-15\%$ depending on the Higgs mass. At
least for not too heavy  Higgs bosons, this larger uncertainty reconciles the
MSTW and ABKM/HERAPDF predictions\footnote{Note that we could also take into
account the effects due to the bottom and charm quark masses on the PDFs.
Indeed, a change in the fitted masses for these quarks may affect the gluon
splitting and hence alter the shape of the gluon--gluon luminosity in $gg\to
H$. We have estimated quantitatively this effect by using the $m_b$ and $m_c$
dependent MSTW PDFs \cite{MSTW-mb} with $m_c = 1.40 \pm 0.15$ GeV and $m_b=4.75
\pm 0.05$ GeV. In the case of the charm quark, we obtain a $\approx 0.2\%$ 
change at $M_H=115$ GeV and at most $\approx 1.5\%$ change at $M_H=500$ GeV. In
the case of the $b$--quark, the change is below the percent level (see also
section 4.3). These uncertainties can be neglected as we adopted the attitude
of adding them in quadrature with the ones due to the PDF and $\alpha_s$ which
are much larger.}.  

Note that for $M_H \lsim 200$ GeV, we obtain smaller uncertainties compared to
the $\approx \pm 15$--20\% uncertainty of $\sigma^{\rm NNLO}(gg\to H)$ at the
Tevatron, a reflection of  the better control on the  behavior of the gluon
density at moderate--$x$ values (relevant for the lHC) compared to high--$x$
values (relevant for the Tevatron).

We note also that the result that we obtain here for the PDF+$\Delta^{\rm
exp}\alpha_s+\Delta^{\rm th}\alpha_s$ uncertainty is  larger than what one
obtains using the recommendation of the PDF4LHC group \cite{PDF4LHC}, that is,
to take the envelope of the PDF+$\Delta^{\rm exp}\alpha_s$ values provided by
the MSTW, CTEQ and NNPDF sets or more simply (and more consistently at NNLO),
take the 68\%CL MSTW PDF+$\Delta^{\rm exp}\alpha_s$ error and multiply it by a
factor of two. To our opinion, this estimate of the uncertainty is rather
optimistic and, as can be seen from Fig.~\ref{ggH-pdf2}, cannot account for the
difference between  the predictions using the various NNLO parameterizations. 

In the rest of our analysis, we will take the 90\% CL PDF+$\Delta^{\rm
exp}\alpha_s+ \Delta^{\rm th}\alpha_s$ uncertainty evaluated as above as the
total PDF uncertainty in the determination of the $gg\to H$ production cross
section. However, we will view this uncertainty  as a measure of the difference
between the various possible parameterizations of the PDFs and, hence, consider
it as a true theoretical uncertainty with no statistical ground as will be
discussed shortly.

\subsection{The combined uncertainty}

A very important issue that remains is the way to combine the various
theoretical uncertainties on the cross section discussed in the previous
subsection. Let us first reiterate an important comment: the uncertainties
associated  to the PDFs in a given scheme should be viewed as purely theoretical
uncertainties despite of the fact that  they are presented as the $1\sigma$ or
more departure from the central values of the data included in the PDF fits.
Indeed, this uncertainty should be interpreted as being due to the
parametrization of the quark and gluon densities and, hence,  to the 
theoretical assumptions in determining these densities and equivalent to the
spread that one observes when comparing different parameterizations. Thus, the
PDF uncertainties should be considered as having no statistical
ground\footnote{In statistical language, the PDF uncertainties should be
considered as having a flat prior, exactly like the scale uncertainty. A more
elaborated discussion can be found in \cite{LHCXS} where the
recommended combination is that of the linear type.}  and they cannot be
combined in quadrature with the two other
purely theoretical errors, namely the scale and the scheme (EFT) uncertainties.
This should be contrasted to what has been done, for instance, by  the CDF and
D0 collaborations in their combined Higgs analysis at the Tevatron
\cite{Tevatron}, where the scale and PDF  uncertainties have been added in
quadrature. 

Here, we discuss three possible ways to combine the various uncertainties. 
\vspace{-2mm}
\begin{enumerate}[A)]
\itemsep-3pt
\item{A reasonable procedure to combine the uncertainty from scale and scheme 
dependence with the PDF uncertainties has been presented in Ref.~\cite{Hpaper},
which also takes into account the correlation between the scale and the PDFs
(which are evaluated at a given factorization scale): one calculates the
maximal/minimal cross sections with respect to the scale variation and apply on
these cross sections the PDF+$\Delta^{\rm exp}  \alpha_{s}$+$\Delta^{\rm th}
\alpha_{s}$ uncertainty.  This procedure has been in fact already proposed,
together with other possibilities which give similar results, in
Ref.~\cite{PDF-ttpaper} where  top quark pair production at hadron colliders
was discussed. An argument for the approach  put forward in
Ref.~\cite{PDF-ttpaper} is that unknown high order effects also enter in the
PDFs determination and one needs to use the full information  on the PDF
uncertainties in the determination of the scale dependence. To this combined
scale+PDF uncertainty, one can linearly add the scheme/EFT uncertainty  to
obtain the overall theoretical error.} 

\item{In fact, a much simpler way to combine the scale, EFT/scheme and the PDF
uncertainties which, as mentioned  before, should all be considered as pure
theoretical errors,  is to add them linearly.  Doing so, one in general obtains
uncertainties that are slightly larger than with the procedure A described above.} 

\item{A third possibility is to take the different predictions of the cross
sections when evaluated with  the various sets of NNLO PDF sets as a measure of
the PDF uncertainty.  Since the maximal cross section for $gg\to H$ is obtained
with the MSTW parametrization and the minimal one with the ABKM set (see
Fig.~\ref{ggH-pdf2}), the total uncertainty on the production  cross section will be
simply the  sum of the scale and scheme/EFT uncertainties when evaluated with
the MSTW PDF set (for the + uncertainty)  and when evaluated  with the ABKM PDF
set (for the -- uncertainty) which maximizes/minimizes the cross section.} 
\end{enumerate}

\begin{figure}[!h]
\vspace*{-1mm}
\begin{center}
\mbox{
\epsfig{file=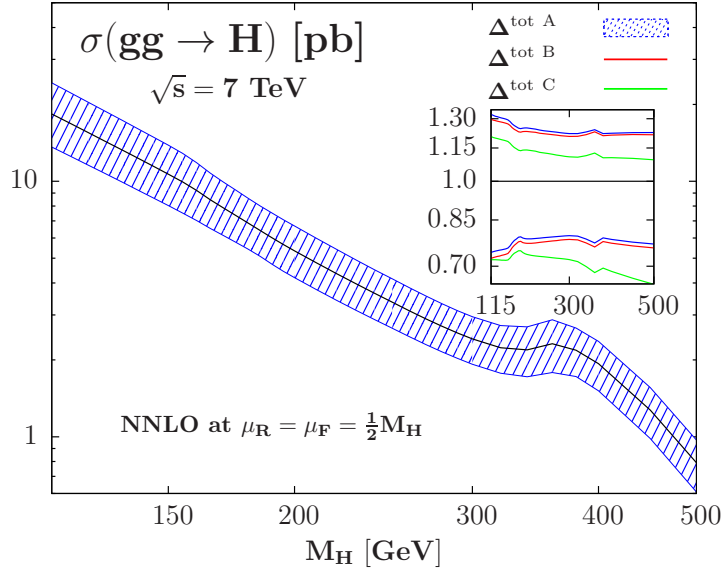, scale=.89} 
}
\end{center}
\vspace*{-4mm}
\caption[]{The production cross section $\sigma(gg\to H)$ at NNLO at the
$\lhc$ with $\sqrt s=7$ TeV,  including the total theoretical uncertainty band 
when all the individual uncertainties are combined using the three procedures
A, B and C described in the text. In the inserts, the relative deviations from 
the central cross section value are shown.}
\label{ggH_all_lhc7}
\vspace*{-1mm}
\end{figure}

Using the three procedures A, B, C  described above, one obtains the overall
theoretical uncertainties on $\sigma^{\rm NNLO}(gg\to H)$  at the 7 TeV lHC that
are shown in Fig.~\ref{ggH_all_lhc7} as a function of the Higgs mass. With the
procedure A, the uncertainty amounts\footnote{ Note that the overall uncertainty
on $\sigma^{\rm NNLO} (gg\to H)$ at the $\lhc$ is  significantly smaller than 
what has been obtained at the  Tevatron  using the  procedure A for the
combination , $\approx -35\%, +40\%$, as a result of the reduction of both the
scale and the PDF uncertainties at $\sqrt s=7$ TeV compared to $\sqrt s=1.96$
TeV. The main reasons are that, at the $\lhc$, the  QCD  $K$--factors are
smaller and the gluon density has a less uncertain behavior  for not too heavy
Higgs bosons.} to $\approx -23\%,+25\%$ for $M_H \lsim 160$ GeV, reduces to
$\approx -21\%, +22\%$ for $M_H \gsim 200$ GeV to reach the value $\approx \pm
22\%$ at $M_H \approx 500$ GeV.  With the procedure B, i.e. when  the various
sources of uncertainties are added linearly, one obtains  approximately the same
total uncertainty as above; they are a few percent (2 to 3\%) smaller for the +
uncertainty and higher for the -- uncertainty.  In the case of procedure C, as
we use for the central prediction of $\sigma^{\rm NNLO} (gg\to H)$ the MSTW PDF
set, the + uncertainty is simply the sum of  the scale and scheme uncertainties
which add to less than 20\% in the entire Higgs mass and are thus much smaller
(in particular in the low  Higgs mass range) than in procedures A and B, while
for the -- uncertainty, one has to add the difference between the MSTW and ABKM
predictions. In this case, the overall uncertainty is comparable to what is
obtained using the procedures A and B in the low Higgs mass range where the
difference between MSTW and ABKM is of the order of 10\% and, hence, the Hessian
MSTW PDF+$\alpha_s$ uncertainty. It becomes much larger at higher $M_H$ values
where the difference between the MSTW and ABKM predictions becomes significant:
at $M_H=500$ GeV, one obtains an uncertainty of $\approx -35\%$, compared to
$\approx -23\%$ using procedures A and B.

\begin{table}[!t]{\small%
\renewcommand{\arraystretch}{1.1}
\vspace*{1mm}
\begin{bigcenter}
\begin{tabular}{|c|cccc|}\hline
$\!M_H\!$ & $\hspace*{-8mm} \sigma_{gg\!\to\! H}^{~~~~~\pm\! \Delta_{\mu}\! \pm \! \Delta^{\rm PDF}
\! \pm\! \Delta^{\rm EFT}} \hspace*{-12mm} $ &  $\hspace*{-12mm}  {\rm A} \hspace*{-12mm}$ & B & C 
\\ \hline 
$115$ & $18.35^{+9.1\%+9.1\%+7.3\%}_{-10.2\%-8.8\%-7.3\%} $ & $^{+27.8\%}_{-24.6\%} $ & $^{+25.6\%}_{-26.3\%} $ & $^{+16.4\%}_{-26.8\%}$ \\ \hline
$120$ & $16.84^{+8.9\%+9.1\%+7.3\%}_{-10.2\%-8.8\%-7.3\%} $ & $^{+27.5\%}_{-24.5\%} $ & $^{+25.3\%}_{-26.2\%} $ & $^{+16.2\%}_{-26.8\%}$ \\ \hline
$125$ & $15.51^{+8.8\%+9.1\%+7.2\%}_{-9.9\%-8.8\%-7.2\%} $ & $^{+27.1\%}_{-24.1\%} $ & $^{+25.1\%}_{-25.8\%} $ & $^{+16.0\%}_{-26.7\%}$ \\ \hline
$130$ & $14.32^{+8.5\%+9.1\%+7.1\%}_{-9.8\%-8.8\%-7.1\%} $ & $^{+26.7\%}_{-24.0\%} $ & $^{+24.7\%}_{-25.6\%} $ & $^{+15.6\%}_{-26.7\%}$ \\ \hline
$135$ & $13.26^{+8.4\%+9.1\%+7.0\%}_{-9.6\%-8.8\%-7.0\%} $ & $^{+26.4\%}_{-23.7\%} $ & $^{+24.5\%}_{-25.4\%} $ & $^{+15.4\%}_{-26.6\%}$ \\ \hline
$140$ & $12.31^{+8.3\%+9.1\%+7.0\%}_{-9.5\%-8.8\%-7.0\%} $ & $^{+26.2\%}_{-23.6\%} $ & $^{+24.3\%}_{-25.2\%} $ & $^{+15.3\%}_{-26.7\%}$ \\ \hline
$145$ & $11.45^{+8.2\%+9.1\%+6.9\%}_{-9.5\%-8.8\%-6.9\%} $ & $^{+26.0\%}_{-23.6\%} $ & $^{+24.2\%}_{-25.2\%} $ & $^{+15.1\%}_{-26.8\%}$ \\ \hline
$150$ & $10.67^{+8.1\%+9.1\%+6.8\%}_{-9.5\%-8.8\%-6.8\%} $ & $^{+25.8\%}_{-23.5\%} $ & $^{+24\%}_{-25.1\%} $ & $^{+14.9\%}_{-26.9\%}$ \\ \hline
$155$ & $9.94^{+7.9\%+9.1\%+6.6\%}_{-9.4\%-8.8\%-6.6\%} $ & $^{+25.4\%}_{-23.2\%} $ & $^{+23.6\%}_{-24.8\%} $ & $^{+14.5\%}_{-26.8\%}$ \\ \hline
$160$ & $9.21^{+7.8\%+9.1\%+5.9\%}_{-9.4\%-8.8\%-5.9\%} $ & $^{+24.4\%}_{-22.7\%} $ & $^{+22.8\%}_{-24.2\%} $ & $^{+13.7\%}_{-26.4\%}$ \\ \hline
$165$ & $8.47^{+7.7\%+9.1\%+4.9\%}_{-9.4\%-8.8\%-4.9\%} $ & $^{+23.4\%}_{-21.7\%} $ & $^{+21.7\%}_{-23.1\%} $ & $^{+12.6\%}_{-25.5\%}$ \\ \hline
$170$ & $7.87^{+7.7\%+9.1\%+4.2\%}_{-9.4\%-8.8\%-4.2\%} $ & $^{+22.7\%}_{-21.0\%} $ & $^{+21.1\%}_{-22.5\%} $ & $^{+11.9\%}_{-25.0\%}$ \\ \hline
$175$ & $7.35^{+7.6\%+9.1\%+3.7\%}_{-9.4\%-8.9\%-3.7\%} $ & $^{+22.0\%}_{-20.5\%} $ & $^{+20.4\%}_{-21.9\%} $ & $^{+11.3\%}_{-24.6\%}$ \\ \hline
$180$ & $6.86^{+7.5\%+9.2\%+3.1\%}_{-9.3\%-8.9\%-3.1\%} $ & $^{+21.4\%}_{-19.9\%} $ & $^{+19.8\%}_{-21.3\%} $ & $^{+10.6\%}_{-24.2\%}$ \\ \hline
$185$ & $6.42^{+7.4\%+9.2\%+3.0\%}_{-9.3\%-8.9\%-3.0\%} $ & $^{+21.1\%}_{-19.8\%} $ & $^{+19.6\%}_{-21.2\%} $ & $^{+10.4\%}_{-24.2\%}$ \\ \hline
$190$ & $6.01^{+7.4\%+9.2\%+3.4\%}_{-9.3\%-8.9\%-3.4\%} $ & $^{+21.5\%}_{-20.3\%} $ & $^{+20.0\%}_{-21.7\%} $ & $^{+10.8\%}_{-24.9\%}$ \\ \hline
$195$ & $5.65^{+7.4\%+9.2\%+3.6\%}_{-9.3\%-8.9\%-3.6\%} $ & $^{+21.8\%}_{-20.5\%} $ & $^{+20.3\%}_{-21.9\%} $ & $^{+11.0\%}_{-25.2\%}$ \\ \hline
$200$ & $5.34^{+7.3\%+9.3\%+3.7\%}_{-9.3\%-9.0\%-3.7\%} $ & $^{+21.8\%}_{-20.6\%} $ & $^{+20.3\%}_{-21.9\%} $ & $^{+11.0\%}_{-25.4\%}$ \\ \hline
 
$210$ & $4.81^{+7.2\%+9.3\%+3.7\%}_{-9.3\%-9.0\%-3.7\%} $ & $^{+21.7\%}_{-20.6\%} $ & $^{+20.2\%}_{-22.0\%} $ & $^{+10.9\%}_{-25.8\%}$ \\ \hline
$220$ & $4.36^{+7.2\%+9.3\%+3.6\%}_{-9.2\%-9.1\%-3.6\%} $ & $^{+21.6\%}_{-20.6\%} $ & $^{+20.2\%}_{-21.9\%} $ & $^{+10.9\%}_{-26.0\%}$ \\ \hline
$230$ & $3.97^{+7.0\%+9.4\%+3.5\%}_{-9.2\%-9.2\%-3.5\%} $ & $^{+21.4\%}_{-20.5\%} $ & $^{+19.9\%}_{-21.8\%} $ & $^{+10.5\%}_{-26.2\%}$ \\ \hline
$240$ & $3.65^{+7.0\%+9.5\%+3.3\%}_{-9.2\%-9.2\%-3.3\%} $ & $^{+21.1\%}_{-20.5\%} $ & $^{+19.8\%}_{-21.8\%} $ & $^{+10.3\%}_{-26.5\%}$ \\ \hline
$250$ & $3.37^{+6.9\%+9.5\%+3.1\%}_{-9.2\%-9.3\%-3.1\%} $ & $^{+20.9\%}_{-20.4\%} $ & $^{+19.5\%}_{-21.6\%} $ & $^{+10.0\%}_{-26.6\%}$ \\ \hline
$260$ & $3.11^{+6.8\%+9.6\%+3.0\%}_{-9.2\%-9.4\%-3.0\%} $ & $^{+20.6\%}_{-20.3\%} $ & $^{+19.3\%}_{-21.6\%} $ & $^{+9.7\%}_{-26.8\%}$ \\ \hline
$270$ & $2.89^{+6.7\%+9.7\%+2.8\%}_{-9.2\%-9.5\%-2.8\%} $ & $^{+20.5\%}_{-20.2\%} $ & $^{+19.1\%}_{-21.4\%} $ & $^{+9.4\%}_{-27.0\%}$ \\ \hline
$280$ & $2.71^{+6.8\%+9.8\%+2.6\%}_{-9.2\%-9.5\%-2.6\%} $ & $^{+20.5\%}_{-20.1\%} $ & $^{+19.2\%}_{-21.4\%} $ & $^{+9.4\%}_{-27.2\%}$ \\ \hline
$290$ & $2.55^{+6.8\%+9.8\%+2.4\%}_{-9.1\%-9.6\%-2.4\%} $ & $^{+20.3\%}_{-20.0\%} $ & $^{+19.1\%}_{-21.1\%} $ & $^{+9.2\%}_{-27.2\%}$ \\ \hline
$300$ & $2.42^{+6.7\%+9.9\%+2.3\%}_{-9.1\%-9.7\%-2.3\%} $ & $^{+20.2\%}_{-19.9\%} $ & $^{+18.9\%}_{-21.1\%} $ & $^{+9.0\%}_{-27.4\%}$ \\ \hline
$320$ & $2.23^{+6.7\%+10.1\%+2.3\%}_{-9.2\%-9.9\%-2.3\%} $ & $^{+20.2\%}_{-20.1\%} $ & $^{+19.0\%}_{-21.4\%} $ & $^{+9.0\%}_{-28.2\%}$ \\ \hline
$340$ & $2.19^{+6.9\%+10.3\%+3.0\%}_{-9.2\%-10.1\%-3.0\%} $ & $^{+21.4\%}_{-21.1\%} $ & $^{+20.2\%}_{-22.3\%} $ & $^{+10.0\%}_{-29.6\%}$ \\ \hline
$360$ & $2.31^{+7.0\%+10.5\%+4.1\%}_{-9.2\%-10.3\%-4.1\%} $ & $^{+22.5\%}_{-22.3\%} $ & $^{+21.6\%}_{-23.6\%} $ & $^{+11.0\%}_{-31.3\%}$ \\ \hline
$380$ & $2.18^{+6.3\%+10.7\%+2.5\%}_{-9.1\%-10.5\%-2.5\%} $ & $^{+20.4\%}_{-20.9\%} $ & $^{+19.6\%}_{-22.2\%} $ & $^{+8.8\%}_{-30.3\%}$ \\ \hline
$400$ & $1.93^{+5.9\%+11.0\%+3.1\%}_{-8.8\%-10.7\%-3.1\%} $ & $^{+20.7\%}_{-21.5\%} $ & $^{+20.0\%}_{-22.7\%} $ & $^{+9.0\%}_{-31.3\%}$ \\ \hline
$450$ & $1.27^{+5.0\%+11.6\%+4.0\%}_{-8.4\%-11.3\%-4.0\%} $ & $^{+21.4\%}_{-22.6\%} $ & $^{+20.5\%}_{-23.8\%} $ & $^{+9.0\%}_{-33.4\%}$ \\ \hline
$500$ & $0.79^{+4.4\%+12.2\%+4.5\%}_{-8.1\%-11.9\%-4.5\%} $ & $^{+22.1\%}_{-23.3\%} $ & $^{+21.1\%}_{-24.5\%} $ & $^{+8.9\%}_{-35.2\%}$ \\ \hline
$600$ & $0.31^{+3.7\%+13.3\%+6.6\%}_{-7.7\%-13.0\%-6.6\%} $ & $^{+24.5\%}_{-26.1\%} $ & $^{+23.7\%}_{-27.3\%} $ & $^{+10.4\%}_{-40.0\%}$ \\ \hline

\end{tabular}
\end{bigcenter} 
\caption{The NNLO total Higgs production cross sections in the $\protect{gg\to
H}$ process at the $\lhc$ with $\sqrt s=7$ TeV  (in pb) for given Higgs mass
values (in GeV) at a central scale $\mu_F=\mu_R=\frac12 M_H$. Shown also are 
the corresponding shifts due to the theoretical uncertainties from the various
sources discussed (first from scale, then from PDF+$\Delta^{\rm \exp+th}\alpha_s$ 
at 90\%CL and from EFT), as well as the total uncertainty when all errors are
added using the procedures A, B and C  described in the text. } 
\label{NF-tab}
\vspace*{-1mm}
}
\end{table}
\clearpage

In Table \ref{NF-tab}, we summarize the results obtained in this
section\footnote{An extended table for Higgs masses up to 1 TeV can be found in
the $gg$--fusion section of Ref.~\cite{LHCXS}.}. The  $gg\to H$ production
cross section for values of the Higgs mass relevant at the  $\lhc$ with $\sqrt
s=7$ TeV are given  together with the uncertainties from scale variations, the
PDF+$\Delta^{\rm exp} \alpha_{s}$+$\Delta^{\rm th} \alpha_{s}$ uncertainty in
the MSTW scheme and  the uncertainty due the use of the EFT approach beyond the
NLO approximation.  The combined uncertainties obtained using the three 
procedures A, B and C proposed above are also given.

\subsection{The cross sections at $\sqrt s=8$--10 TeV and 14 TeV}

\begin{figure}[!h]
\vspace*{-1mm}
\begin{center}
\mbox{
\epsfig{file=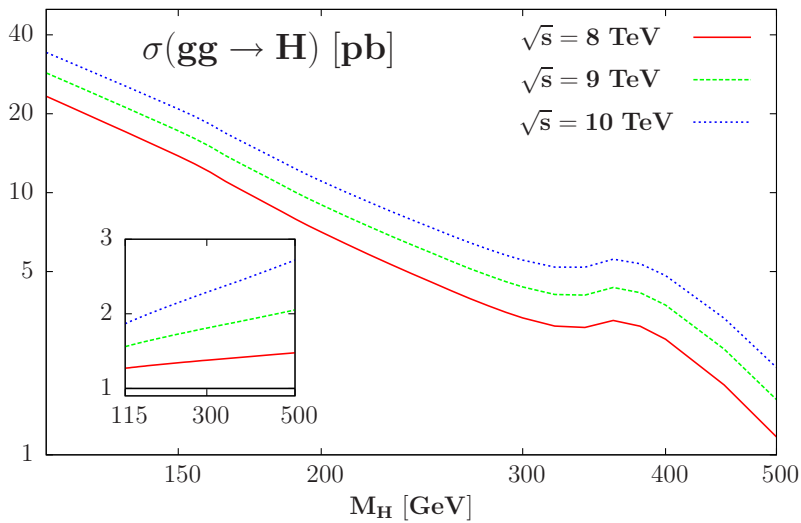, scale=.8} 
}
\end{center}
\vspace*{-4mm}
\caption[]{The production cross section $\sigma(gg\to H)$ at NNLO as a function
of $M_H$ at the $\lhc$ with center of mass energies of $\sqrt s=8,9$ and
10 TeV. In the inserts, the relative increase compared to the cross section
at $\sqrt s=7$ TeV are shown.}
\label{fig:7-10}
\vspace*{-2mm}
\end{figure}

As already stated, intermediate center of mass energies between  $\sqrt s=7$
and 10 TeV are being currently considered for the $\lhc$ in a  very near
future, with the possibility of collecting integrated luminosities that are
significantly larger than 1 fb$^{-1}$ \cite{LHC}. We therefore also display the
theoretical predictions for the Higgs production rates in the $gg\to H$ channel
at these energies, following exactly the same steps as the ones discussed in 
section 2.1. The results for $\sigma^{\rm NNLO} (gg\to H)$, again for a central
scale $\mu_R=\mu_F=\frac12 M_H$ and using the MSTW2008 set of NNLO PDFs, are
displayed for $\sqrt s=8,9$ and 10 TeV in Fig.~\ref{fig:7-10} as a function of
$M_H$ and in  Table \ref{table_lhc8} for the relevant Higgs mass  values .

We see that in the low Higgs mass range, $M_H \approx 120$ GeV, the cross
section $\sigma(gg\to H)$ is $\approx 20\%, 40\%$ and  $\approx 100\%$ higher
at, respectively, $\sqrt s=8,9$ and 10 TeV, compared  to $\sqrt s=7$ TeV. At
higher Higgs masses, $M_H \gsim 300$ GeV, the increase of the cross section is
slightly larger as a result  of the  reduction  of the phase  space at
lower energies.

As far as the theoretical uncertainties are concerned, we have verified that
they differ only by a small amount, less than one to two percent, when
increasing the center of mass energy  from $\sqrt s= 7$ to $\sqrt s=10$ TeV. To
a good approximation, one can  therefore use the values of the scale,
scheme/EFT and  PDF+$\Delta^{\rm exp} \alpha_{s}$+$\Delta^{\rm th} \alpha_{s}$
uncertainties,  as well as the combined uncertainties, that are given in Table
\ref{NF-tab} for the $\sqrt s=7$ TeV case. 

\begin{table}[!t]{\small%
\let\lbr\{\def\{{\char'173}%
\let\rbr\}\def\}{\char'175}%
\renewcommand{\arraystretch}{1.2}
\vspace*{-1mm}
\begin{center}
\begin{tabular}{|c|ccc||c|ccc|}\hline
$M_H$ & $\sigma^{\rm 8~TeV}_{g g\to H}$ & $\sigma^{\rm 9~TeV}_{g g\to H}$ 
& $\sigma^{\rm 10~TeV}_{g g\to H}$
& $M_H$ & $\sigma^{\rm 8~TeV}_{g g\to H}$ & $\sigma^{\rm 9~TeV}_{g g\to H}$ 
& $\sigma^{\rm 10~TeV}_{g g\to H}$ \\ \hline
$115$ & $23.31$ & $28.63$ & $34.26$ &
$210$ & $6.39$ & $8.15$ & $10.07$ \\ \hline 
$120$ & $21.46$ & $26.42$ & $31.68$ &
$220$ & $5.82$ & $7.44$ & $9.22$ \\ \hline 
$125$ & $19.81$ & $24.44$ & $29.37$ &
$230$ & $5.33$ & $6.84$ & $8.50$ \\ \hline 
$130$ & $18.34$ & $22.68$ & $27.30$ &
$240$ & $4.91$ & $6.32$ & $7.88$ \\ \hline 
$135$ & $17.03$ & $21.10$ & $25.44$ &
$250$ & $4.55$ & $5.88$ & $7.34$ \\ \hline 
$140$ & $15.84$ & $19.67$ & $23.76$ &
$260$ & $4.22$ & $5.47$ & $6.85$ \\ \hline 
$145$ & $14.77$ & $18.38$ & $22.24$ &
$270$ & $3.94$ & $5.13$ & $6.44$ \\ \hline 
$150$ & $13.80$ & $17.21$ & $20.86$ &
$280$ & $3.70$ & $4.83$ & $6.08$ \\ \hline 
$155$ & $12.89$ & $16.10$ & $19.55$ &
$290$ & $3.50$ & $4.58$ & $5.78$ \\ \hline 
$160$ & $11.97$ & $14.98$ & $18.22$ &
$300$ & $3.33$ & $4.37$ & $5.53$ \\ \hline 
$165$ & $11.04$ & $13.85$ & $16.87$ &
$320$ & $3.10$ & $4.09$ & $5.20$ \\ \hline 
$170$ & $10.28$ & $12.92$ & $15.77$ &
$340$ & $3.06$ & $4.07$ & $5.19$ \\ \hline 
$175$ & $9.62$ & $12.11$ & $14.80$ &
$360$ & $3.26$ & $4.35$ & $5.58$ \\ \hline 
$180$ & $9.00$ & $11.36$ & $13.90$ &
$380$ & $3.09$ & $4.15$ & $5.36$ \\ \hline 
$185$ & $8.44$ & $10.66$ & $13.08$ &
$400$ & $2.76$ & $3.73$ & $4.83$ \\ \hline 
$190$ & $7.92$ & $10.03$ & $12.32$ &
$450$ & $1.85$ & $2.53$ & $3.32$ \\ \hline 
$195$ & $7.47$ & $9.47$ & $11.65$ &
$500$ & $1.17$ & $1.62$ & $2.15$ \\ \hline 
$200$ & $7.07$ & $8.99$ & $11.07$ &
$600$ & $0.47$ & $0.68$ & $0.92$ \\ \hline
\end{tabular}
\vspace*{-2mm}
\end{center}
\caption{The cross sections in the $\protect{ \sigma^{\rm NNLO}( gg\to H)}$  
at the LHC with $\sqrt s=8,9,10$ TeV  (in pb) for given Higgs mass values (in
GeV) at a central scale $\mu_F=\mu_R=\frac12 M_H$ using  the MSTW PDF set.}
\label{table_lhc8}
\vspace*{-4mm}
}
\end{table}

\begin{figure}[!h]
\begin{center}
\vspace*{-.1mm}
\epsfig{file=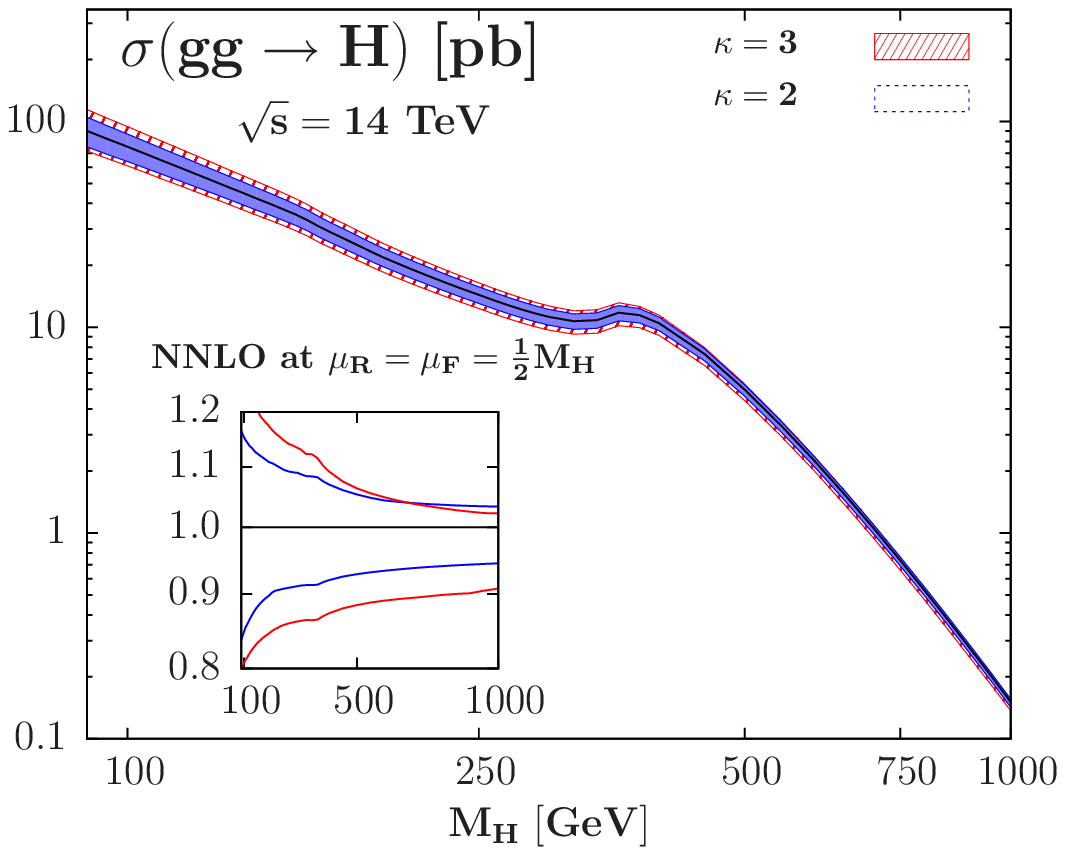,scale=0.67}
\epsfig{file=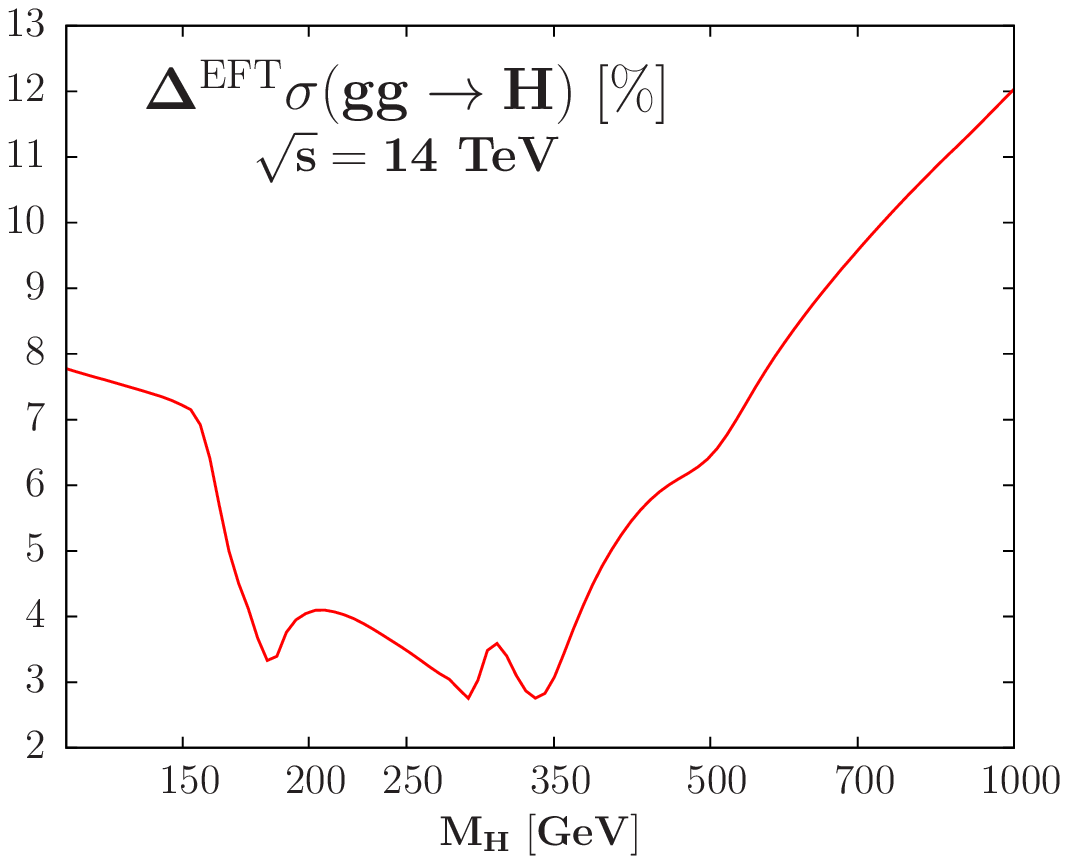,scale=0.67}
\end{center}
\vspace*{-5mm}
\caption[]{The uncertainty bands of $\sigma^{\rm NNLO}_{gg \to H}$ from
scale variation with $\kappa=2$ (left) and the total EFT uncertainty (right) 
at $\sqrt s=14$ TeV  as a function of $M_H$.}
\label{scale_ggH_lhc14}
\end{figure}

\begin{figure}[!h]
\begin{bigcenter}
\vspace*{-.1mm}
\mbox{
\epsfig{file=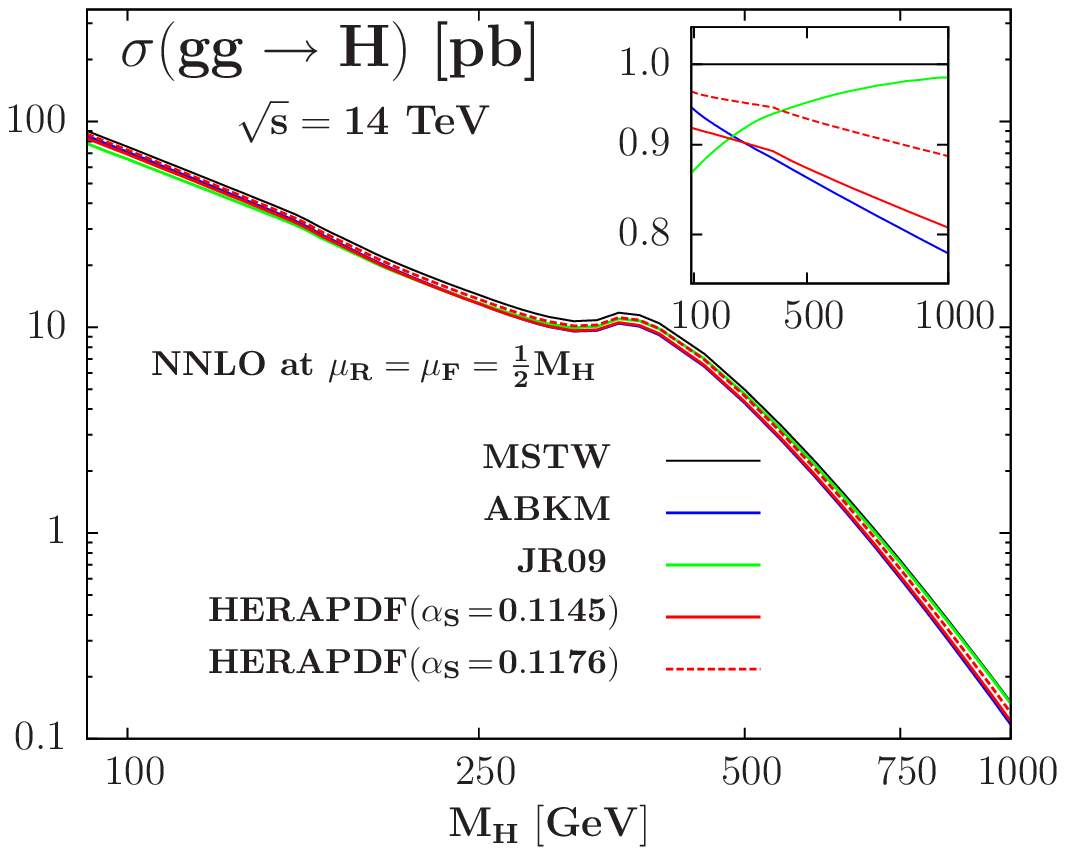,scale=0.67}\hspace*{-0.1cm}
\epsfig{file=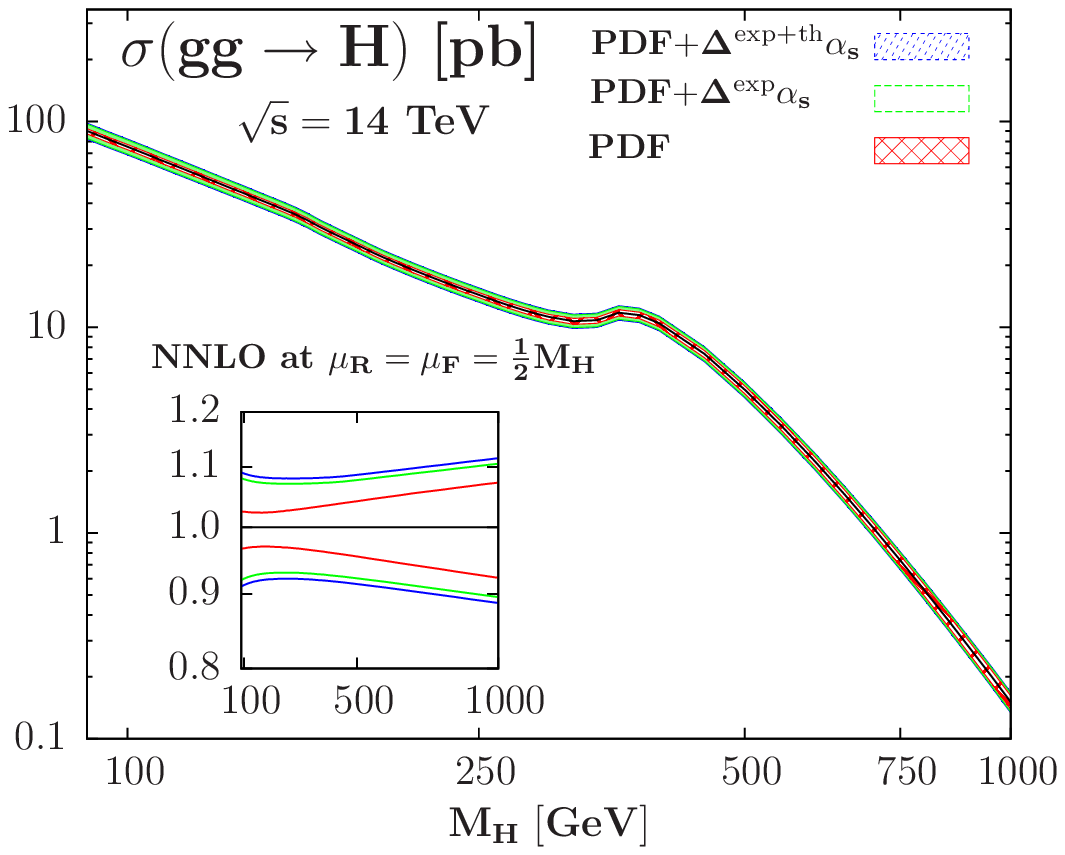,scale=0.67}
}
\end{bigcenter}
\vspace*{-5mm}
\caption[]{The PDF uncertainties in $\sigma^{\rm NNLO}(g g\to H)$ at the LHC
with $\sqrt s= 14$ TeV as a function of $M_H$. Left: the central  values when
using the four NNLO PDFs and right:  the 90\% CL PDF, PDF+$\Delta^{\rm exp}
\alpha_s$ and PDF+$\Delta^{\rm exp}\alpha_s +\Delta^{\rm th}\alpha_s$ 
uncertainties in the MSTW scheme.}
\label{pdf_ggH_lhc14}
\end{figure}

\begin{figure}[!h]
\begin{center}
\vspace*{-.01mm}
\epsfig{file=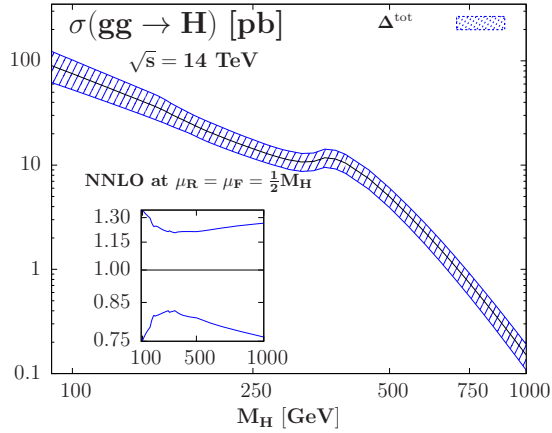,scale=0.67}
\end{center}
\vspace*{-5mm}
\caption[]{The cross section $\sigma^{\rm NNLO} (gg\to H)$ at the LHC 
with the uncertainty band when all theoretical uncertainties are added
using our procedure A described in the text. } 
\label{ggH_all_lhc14}
\vspace*{-5mm}
\end{figure}

Finally, let us briefly summarize the expectations for the LHC with the design 
center of mass energy of $\sqrt s\!=\!14$ TeV, which is expected to collect at
least $\sim 30$ fb$^{-1}$ of data that should allow  to discover the Higgs 
boson in its entire mass range $115\; {\rm GeV} \! \leq\! M_H \!\lsim \!  1$
TeV. The results for the $gg\to H$ cross section at  $\sqrt s=$ 14 TeV are
summarized in Figs.~\ref{scale_ggH_lhc14}, \ref{pdf_ggH_lhc14},
\ref{ggH_all_lhc14} and in Table \ref{table_lhc14}, following exactly the same
lines and procedure as at $\sqrt s\!=\!7$ TeV. 

We will simply make a few comments to highlight the main differences: 

-- The scale uncertainty, estimated by varying $\mu_R$ and $\mu_F$ again in
the  domain $\frac14 M_H \le \mu_R=\mu_F \le M_H$, does not significantly
change when comparing to the $\lhc$ case: one obtains a variation of  $\approx
\pm 10\%$ at low  and $\approx \pm 5\%$ at high Higgs masses; see
Fig.~\ref{scale_ggH_lhc14} (left).

-- The EFT/scheme uncertainty is almost exactly the same than at $\sqrt s=7$
TeV,  as its most important component enters as a multiplicative factor in the
$gg\!\to\! H$ amplitude but is larger starting from $M_H \gsim 500$ 
GeV; Fig.~\ref{scale_ggH_lhc14} (right).

-- For the PDF uncertainties,  one can see in Fig.~\ref{pdf_ggH_lhc14} (left)
that one still has significant differences when folding the partonic $gg\to H$
cross section with the gluon luminosities given by the four NNLO PDFs sets. 
Fig.~\ref{pdf_ggH_lhc14} (right) displays the PDF, PDF$+\Delta^{\rm
exp}\alpha_s$ and the combined PDF$+\Delta^{\rm exp}\alpha_s$+$\Delta^{\rm
th}\alpha_s$ uncertainties at the LHC. One  obtains a slightly smaller 
PDF+$\alpha_s$  uncertainty than at the $\lhc$, 1 to $2\%$ , since  lower $x$
values are probed.

-- For the overall uncertainty on $\sigma^{\rm NNLO} (gg\to H)$ shown  in
Fig.~\ref{ggH_all_lhc14}, where we have restricted ourselves to procedure A only,
it is more or less the same at 14 TeV than at 7 TeV in the low Higgs mass range,
but is slightly  smaller for heavier Higgs bosons.  As an example, we obtain for
$M_H=500$ GeV a total $\approx \pm 21\%$ uncertainty at 14 TeV compared to 
$\approx \pm 23\%$ at 7 TeV.

Table~\ref{table_lhc14} which displays the  cross sections together with the
individual and overall theoretical uncertainties for the Higgs masses relevant
at the LHC, summarizes our results.

\begin{table}[!h]{\small%
\let\lbr\{\def\{{\char'173}%
\let\rbr\}\def\}{\char'175}%
\renewcommand{\arraystretch}{1.42}
\vspace*{1mm}
\begin{bigcenter}
\begin{tabular}{|c|ccc||c|ccc|}\hline
$\!M_H\!$ & $\sigma_{g g\to H}^{\pm \Delta_{\mu}\pm \Delta_{\rm PDF}\pm
\Delta_{\rm EFT}}$ & A & B &
$\!M_H\!$ & $\sigma_{g g\to H}^{\pm \Delta_{\mu}\pm \Delta_{\rm PDF}\pm 
\Delta_{\rm EFT}}$ & A & B\\ \hline 
$115$ & $59.37^{+9.4\% +8.7\% +7.8\%}_{-12.2\% -8.5\% -7.8\%} $ & $^{+28.1\%}_{-26.6\%} $ & $^{+25.8\%}_{-28.5\%}$ &
$240$ & $15.30^{+7.0\% +8.0\% +3.7\%}_{-8.2\% -7.8\% -3.7\%} $ & $^{+20.0\%}_{-18.6\%} $ & $^{+18.7\%}_{-19.6\%}$ \\ \hline
$120$ & $55.20^{+9.2\% +8.6\% +7.7\%}_{-11.9\% -8.4\% -7.7\%} $ & $^{+27.8\%}_{-26.1\%} $ & $^{+25.5\%}_{-28.0\%}$ &
$250$ & $14.38^{+6.9\% +8.0\% +3.5\%}_{-8.2\% -7.8\% -3.5\%} $ & $^{+19.7\%}_{-18.4\%} $ & $^{+18.4\%}_{-19.5\%}$ \\ \hline
$125$ & $51.45^{+9.0\% +8.5\% +7.6\%}_{-11.6\% -8.4\% -7.6\%} $ & $^{+27.4\%}_{-25.7\%} $ & $^{+25.1\%}_{-27.5\%}$ &
$260$ & $13.52^{+6.8\% +8.0\% +3.3\%}_{-8.1\% -7.8\% -3.3\%} $ & $^{+19.4\%}_{-18.2\%} $ & $^{+18.1\%}_{-19.2\%}$ \\ \hline
$130$ & $48.09^{+8.9\% +8.5\% +7.5\%}_{-11.4\% -8.3\% -7.5\%} $ & $^{+27.2\%}_{-25.5\%} $ & $^{+25.0\%}_{-27.3\%}$ &
$270$ & $12.79^{+6.7\% +8.0\% +3.1\%}_{-8.1\% -7.8\% -3.1\%} $ & $^{+19.1\%}_{-18.1\%} $ & $^{+17.9\%}_{-19.1\%}$ \\ \hline
$135$ & $45.06^{+8.7\% +8.4\% +7.5\%}_{-11.1\% -8.2\% -7.5\%} $ & $^{+26.8\%}_{-25.1\%} $ & $^{+24.6\%}_{-26.8\%}$ &
$280$ & $12.17^{+6.7\% +8.0\% +2.9\%}_{-8.1\% -7.8\% -2.9\%} $ & $^{+18.9\%}_{-17.8\%} $ & $^{+17.7\%}_{-18.8\%}$ \\ \hline
$140$ & $42.30^{+8.5\% +8.4\% +7.4\%}_{-10.8\% -8.2\% -7.4\%} $ & $^{+26.4\%}_{-24.7\%} $ & $^{+24.3\%}_{-26.4\%}$ &
$290$ & $11.65^{+6.6\% +8.0\% +2.8\%}_{-8.0\% -7.8\% -2.8\%} $ & $^{+18.6\%}_{-17.6\%} $ & $^{+17.4\%}_{-18.6\%}$ \\ \hline
$145$ & $39.80^{+8.4\% +8.3\% +7.3\%}_{-10.6\% -8.1\% -7.3\%} $ & $^{+26.0\%}_{-24.4\%} $ & $^{+24.0\%}_{-26.1\%}$ &
$300$ & $11.22^{+6.5\% +8.0\% +3.4\%}_{-8.0\% -7.8\% -3.4\%} $ & $^{+19.2\%}_{-18.4\%} $ & $^{+18.0\%}_{-19.3\%}$ \\ \hline
$150$ & $37.50^{+8.3\% +8.3\% +7.2\%}_{-10.4\% -8.1\% -7.2\%} $ & $^{+25.8\%}_{-24.1\%} $ & $^{+23.8\%}_{-25.7\%}$ &
$320$ & $10.70^{+6.5\% +8.1\% +3.1\%}_{-8.0\% -7.9\% -3.1\%} $ & $^{+18.8\%}_{-18.1\%} $ & $^{+17.8\%}_{-19.1\%}$ \\ \hline
$155$ & $35.32^{+8.1\% +8.3\% +7.0\%}_{-10.2\% -8.1\% -7.0\%} $ & $^{+25.3\%}_{-23.7\%} $ & $^{+23.4\%}_{-25.3\%}$ &
$340$ & $10.83^{+6.5\% +8.1\% +2.8\%}_{-8.0\% -7.9\% -2.8\%} $ & $^{+18.4\%}_{-17.8\%} $ & $^{+17.4\%}_{-18.7\%}$ \\ \hline
$160$ & $33.08^{+8.0\% +8.2\% +6.3\%}_{-10.0\% -8.0\% -6.3\%} $ & $^{+24.4\%}_{-22.9\%} $ & $^{+22.6\%}_{-24.4\%}$ &
$360$ & $11.77^{+6.4\% +8.1\% +3.5\%}_{-8.0\% -8.0\% -3.5\%} $ & $^{+19.0\%}_{-18.6\%} $ & $^{+18.1\%}_{-19.5\%}$ \\ \hline
$165$ & $30.77^{+7.9\% +8.2\% +5.3\%}_{-9.8\% -8.0\% -5.3\%} $ & $^{+23.1\%}_{-21.8\%} $ & $^{+21.4\%}_{-23.1\%}$ &
$380$ & $11.46^{+6.0\% +8.2\% +4.4\%}_{-7.7\% -8.1\% -4.4\%} $ & $^{+19.5\%}_{-19.2\%} $ & $^{+18.6\%}_{-20.2\%}$ \\ \hline
$170$ & $28.89^{+7.8\% +8.2\% +4.5\%}_{-9.7\% -8.0\% -4.5\%} $ & $^{+22.3\%}_{-21.0\%} $ & $^{+20.6\%}_{-22.3\%}$ &
$400$ & $10.46^{+5.6\% +8.2\% +5.0\%}_{-7.4\% -8.1\% -5.0\%} $ & $^{+19.7\%}_{-19.7\%} $ & $^{+18.9\%}_{-20.6\%}$ \\ \hline
$175$ & $27.24^{+7.8\% +8.2\% +4.0\%}_{-9.5\% -7.9\% -4.0\%} $ & $^{+21.6\%}_{-20.2\%} $ & $^{+20.0\%}_{-21.4\%}$ &
$450$ & $7.42^{+5.0\% +8.4\% +6.0\%}_{-7.0\% -8.3\% -6.0\%} $ & $^{+20.0\%}_{-20.4\%} $ & $^{+19.3\%}_{-21.3\%}$ \\ \hline
$180$ & $25.71^{+7.7\% +8.2\% +3.5\%}_{-9.4\% -7.9\% -3.5\%} $ & $^{+20.9\%}_{-19.6\%} $ & $^{+19.3\%}_{-20.8\%}$ &
$500$ & $4.97^{+4.6\% +8.6\% +6.4\%}_{-6.7\% -8.6\% -6.4\%} $ & $^{+20.4\%}_{-20.9\%} $ & $^{+19.7\%}_{-21.7\%}$ \\ \hline
$185$ & $24.28^{+7.6\% +8.1\% +3.3\%}_{-9.1\% -7.9\% -3.3\%} $ & $^{+20.6\%}_{-19.2\%} $ & $^{+19.0\%}_{-20.4\%}$ &
$550$ & $3.32^{+4.3\% +8.9\% +7.4\%}_{-6.5\% -8.8\% -7.4\%} $ & $^{+21.3\%}_{-21.9\%} $ & $^{+20.7\%}_{-22.7\%}$ \\ \hline
$190$ & $22.97^{+7.6\% +8.1\% +3.8\%}_{-9.1\% -7.9\% -3.8\%} $ & $^{+21.0\%}_{-19.6\%} $ & $^{+19.5\%}_{-20.7\%}$ &
$600$ & $2.24^{+4.1\% +9.2\% +8.3\%}_{-6.3\% -9.1\% -8.3\%} $ & $^{+22.1\%}_{-22.9\%} $ & $^{+21.6\%}_{-23.7\%}$ \\ \hline
$195$ & $21.83^{+7.5\% +8.1\% +4.0\%}_{-9.0\% -7.9\% -4.0\%} $ & $^{+21.1\%}_{-19.7\%} $ & $^{+19.6\%}_{-20.8\%}$ &
$650$ & $1.53^{+3.9\% +9.5\% +9.0\%}_{-6.2\% -9.4\% -9.0\%} $ & $^{+22.7\%}_{-23.9\%} $ & $^{+22.4\%}_{-24.6\%}$ \\ \hline
$200$ & $20.83^{+7.4\% +8.1\% +4.1\%}_{-8.8\% -7.9\% -4.1\%} $ & $^{+21.0\%}_{-19.6\%} $ & $^{+19.5\%}_{-20.7\%}$ &
$700$ & $1.05^{+3.8\% +9.8\% +9.6\%}_{-6.1\% -9.6\% -9.6\%} $ & $^{+23.4\%}_{-24.6\%} $ & $^{+23.2\%}_{-25.3\%}$ \\ \hline
$210$ & $19.10^{+7.3\% +8.1\% +4.1\%}_{-8.6\% -7.8\% -4.1\%} $ & $^{+20.9\%}_{-19.4\%} $ & $^{+19.5\%}_{-20.5\%}$ &
$800$ & $0.52^{+3.5\% +10.4\% +10.5\%}_{-6.0\% -10.2\% -10.5\%} $ & $^{+24.6\%}_{-26.1\%} $ & $^{+24.4\%}_{-26.7\%}$ \\ \hline
$220$ & $17.64^{+7.2\% +8.1\% +4.0\%}_{-8.4\% -7.8\% -4.0\%} $ & $^{+20.6\%}_{-19.2\%} $ & $^{+19.3\%}_{-20.3\%}$ &
$900$ & $0.27^{+3.4\% +11.0\% +11.3\%}_{-5.9\% -10.7\% -11.3\%} $ & $^{+25.6\%}_{-27.5\%} $ & $^{+25.7\%}_{-28.0\%}$ \\ \hline
$230$ & $16.38^{+7.1\% +8.0\% +3.8\%}_{-8.3\% -7.8\% -3.8\%} $ & $^{+20.4\%}_{-18.9\%} $ & $^{+19.0\%}_{-20.0\%}$ &
$1000$ & $0.15^{+3.3\% +11.5\% +12.0\%}_{-5.8\% -11.3\% -12.0\%} $ & $^{+26.4\%}_{-28.8\%} $ & $^{+26.9\%}_{-29.1\%}$ \\ \hline
\end{tabular}
\end{bigcenter} 
\vspace*{-3mm}
\caption{The NNLO total Higgs production cross sections in the $\protect{gg\to
H}$ process at the LHC with $\sqrt s=14$ TeV  (in pb) for given Higgs mass
values (in GeV) at a central scale $\mu_F=\mu_R=\frac12 M_H$. Shown also are 
the corresponding shifts due to the theoretical uncertainties from the various
sources discussed (first from scale, then from PDF+$\Delta^{\rm \exp+th}\alpha_s$ 
at 90\%CL and from EFT), as well as the total uncertainty when all errors are added 
using the procedures A and B  described in the text. } 
\label{table_lhc14}
\vspace*{-1mm}
}
\end{table}

\clearpage

\section{The Higgs decay branching fractions} 

\subsection{The parametric uncertainties}

There is an additional source of theoretical errors which has not been
considered in  Ref.~\cite{Hpaper} and not included in the Tevatron and  LHC
experimental analyses \cite{Tevatron,ATLAS,CMS}:  the one due to the  branching
ratios in the Higgs boson decays\footnote{A discussion of QCD uncertainties in
Higgs decays has also been made recently  in Ref.~\cite{LHCXS}.}.  Indeed, 
contrary to the partial widths of Higgs decays into leptons and gauge bosons
where only electroweak corrections appear at the lowest orders and  are thus
known precisely, the partial decays widths into quark pairs and gluons are
plagued with various uncertainties due to strong interactions
\cite{DSZ}. 

As it is well known, the QCD corrections to the $H \to Q\bar Q$ partial widths,
with $Q=b$ and $c$,  are extremely large; for reviews, see 
Refs.~\cite{Review,MichaelR}.  The bulk of these corrections can be absorbed
into running quarks masses by defining an effective $H Q\bar Q$ Yukawa
interaction  in which the pole quark masses $M_Q(M_Q)$  are replaced by 
running quark masses in the $\overline{\rm MS}$ scheme, $\overline{m}_Q$, with
$\alpha_s$  evaluated at the scale of the pole mass $\mu=M_Q$
\cite{melnikov-ritbergen}. One then evolves $\overline{m}_Q$ from $M_{Q}$
upward to a renormalization scale $\mu$, which should be  chosen to be close to
the Higgs mass, $\mu \approx M_H$. Using as starting points the pole masses
$M_Q$,  the values of the running $b,c$ masses at  the  scale $\mu \sim 100$
GeV  are, respectively, $\approx 1.5$ and $\approx 2$ times smaller than the
pole masses. The partial widths, $\Gamma(H \to Q\bar Q) \propto 
\overline{m}_Q^2$ (but where additional QCD corrections that are known up to
three--loops need to be implemented \cite{MichaelR,HqqQCD-2loop+runmass}), are
thus reduced by factors of $\approx 2$ and $\approx 4$ for Higgs decays into
$b\bar b$ and $c\bar c$, respectively. 

Nevertheless, residual uncertainties in the $H \to Q\bar Q$ partial decay
widths will remain. These uncertainties are mainly due to three sources:  $i)$
the  imperfect knowledge of the input bottom and charm quark masses, $ii)$ the
error on $\alpha_s$ which migrates into uncertainties on the running quarks
masses at the scale $\mu$ and on the residual QCD corrections to the partial
widths and, finally,  $iii)$ the variation  of the scale $\mu$ which enters as
an argument in $\alpha_s$, in the running quarks masses $\overline{m}_Q$ and in
the residual QCD corrections. This last uncertainty, which is expected to
account for the missing (not yet calculated) higher order contributions to the
partial widths, turns out to be very small (as both the quark mass evolution
and the residual corrections are known up to three loops) and can be safely
neglected \cite{MichaelT}.

In the case of the decay $H\to gg$, the leading order partial width, that is 
proportional to $\alpha_s^2$, has to be corrected by large QCD  corrections
which nearly double the width \cite{SDGZ,Hgg-HO}. It is affected by
uncertainties due to the error on  $\alpha_s$ and to the residual scale
dependence  which is about $\approx 10\%$ at NNLO \cite{Hgg-HO}. However,  as the
branching ratio of this decay is of order a few percent at most in the
entire Higgs mass range,  these uncertainties do not affect the total Higgs
width and thus, the dominant branching ratios, in a substantial way. 

We will assume that the uncertainties in the decays widths  $H \to \tau^+
\tau^-$ and $H\to  VV^*$ are negligible\footnote{This might not be entirely true
in the case of the $H\!\to\! WW$ and  $H\!\to\! ZZ$ decay channels near the
$2M_W$ and $2M_Z$ thresholds where the  regularization of the spurious spikes
that  appear when including  one--loop electroweak corrections \cite{HWWEW},
might introduce some ambiguities. Furthermore, for heavy Higgs bosons $M_H \gsim
500$ GeV, additional corrections to the partial $H\!\to\! WW,ZZ$  widths from
the Higgs self--coupling, $\lambda_{HHH}\!\propto\!M_H^2$  become very large
\cite{HVV-HO}; these uncertainties  mostly cancel in the branching ratios,
though (but they will substantially affect the cross sections for  Higgs
production in  the vector boson  fusion  processes which are related to the
$H\to WW/ZZ$ partial decay widths).} as QCD effects will enter the partial
widths only at higher orders, $\geq {\cal O} (\alpha_s^2)$. In the case of $H\to
\gamma \gamma$, the decay is mediated mainly by a $W$ loop and to a lesser
extent by a top--quark loop; in the low to moderate $M_H$ range where the decay
is important, the uncertainty due to the latter contribution is small, less than
1\%, as the QCD corrections are tiny and the large $m_t$ limit a good
approximation \cite{Hgamma}.

\subsection{Uncertainties on the Higgs branching ratios} 

An analysis of the uncertainties affecting the Higgs branching ratios has been 
performed long ago in Ref.~\cite{DSZ}. Here, we update this analysis in the
following way:
\vspace{-2mm}
\begin{enumerate}[{\it i)}]
\itemsep-3pt
\item{For the strong coupling constant $\alpha_s$, we adopt for consistency
reasons the same value that we use for determining the central predictions for
the production cross section, that is, the MSTW value with its associated 
$1\sigma$ error, $\alpha_s(M_Z^2)=0.1171 \pm 0.0014$ at NNLO.} 

\item{As starting points for the $b,c$  masses we use the $\overline{\rm MS}$
masses $\overline{m}_Q$ evaluated at the mass itself with  
values\footnote{While the central values of these masses are close to those 
obtained in Ref.~\cite{mb-Kuhn} directly from QCD sum  rules in a consistent
${\cal O}(\alpha_s)$ expansion, the corresponding uncertainties are larger. We
thus conservatively adopt the PDG values \cite{PDG}, which could also account 
for possible theory errors; in any case the total uncertainties obtained when
using the errors of Ref.~\cite{mb-Kuhn} would nearly be equal to the
uncertainty due to $\alpha_s$ error alone. Note that with these starting inputs
and the value of $\alpha_s(M_Z^2)$ given above, the central values of  the
$b,c$ pole masses are $m_b=4.71$ GeV and $M_c=1.54$ GeV with variation ranges
of, respectively,  4.64--4.90 GeV and 1.42--1.63 GeV. For the pole
bottom--quark mass, the central value  is rather close to the one adopted  in
the MSTW scheme for parameterizing the parton densities \cite{PDF-MSTW},
$m_b=4.75$ GeV, and we expect that this difference will have no practical impact. }
$\overline{m}_b(\overline{m}_b)=4.19^{+0.18}_{-0.06}$ GeV and  
$\overline{m}_c(\overline{m}_c)=1.27^{+0.07}_{-0.09}$ GeV \cite{PDG}. When we
will quote separately the impact of these uncertainties on the Higgs  branching
ratio, we will assume the  central value $\alpha_s(M_Z^2)=0.1171$ at NNLO for
the strong coupling constant.}

\item{For the $H\to gg$ channel we will consider only the parametric
uncertainties that is, the error on $\alpha_s$ and quark masses (the latter
being negligible). At NNLO, the scale uncertainty is of the order 10\% but since
the branching ratio is small it migrates into an error of less than  1\%. In
addition, the N$^3$LO contribution would reduce this scale variation to the
level of a few percent at most and this uncertainty can be safely neglected.}
\end{enumerate}

\begin{figure}[!h]
\vspace*{.1cm}
\begin{center}
\hspace*{-.2cm}
\epsfig{file=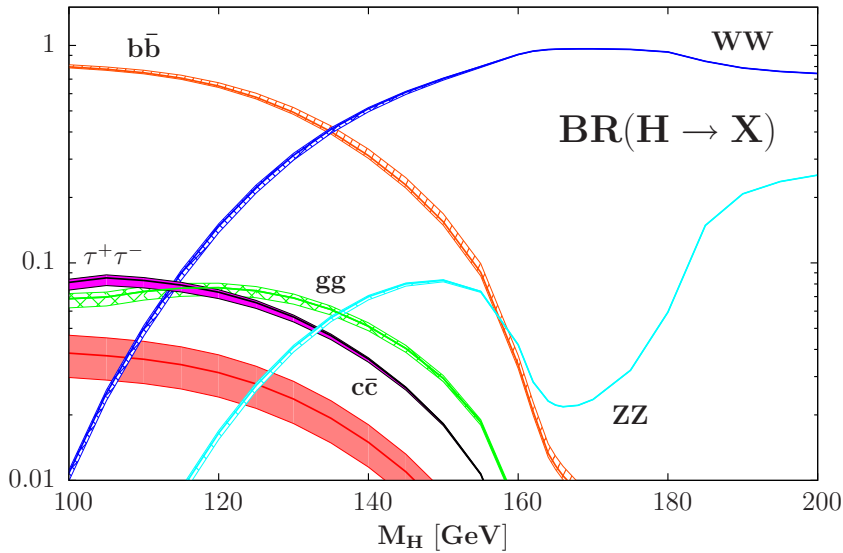,scale=0.85}
\end{center}
\vspace*{-6mm}
\caption[]{The Higgs decay branching ratios as a function of $M_H$ including 
the total uncertainty bands from the 1$\sigma$ errors on the input quark 
masses and the coupling $\alpha_s$ (the individual errors have been added in 
quadrature).} 
\vspace*{-2mm}
\label{Hbrs}
\end{figure}

Using an adapted version of the program {\tt HDECAY} \cite{HDECAY} which
calculates the partial and total Higgs decay widths including the relevant QCD
and electroweak higher order corrections, we have
evaluated\footnote{We thank M. Spira and collaborators for pointing to
  us a numerical problem in the evaluation of $H\to gg$ with {\tt
    HDECAY} in an earlier version of the paper.} the uncertainties
discussed above and added in quadrature the errors due the input masses
$\overline{m}_b, \overline{m}_c$  and the coupling $\alpha_s (M_Z^2)$. We
obtain the branching ratios (BR) shown in Fig.~\ref{Hbrs} as a  function of the
Higgs mass, including the total uncertainty bands.  In
Table~\ref{table_decay1}, the branching ratios for the main decay modes $H\to
b\bar b$  and $H\to WW$  are given for three selected values of the Higgs
mass, $M_H=120,135$ and 150 GeV, together with the individual and total
uncertainties. The uncertainties (in percentage) on the branching ratios for
the $H\to \tau^+\tau^-$ and $H\to ZZ,\gamma \gamma$ channels are the same as
those affecting the $H\to WW$ mode.  In Table~\ref{table_decay2},  displayed
are the branching ratios for the various decays as well as their corresponding 
total uncertainties for a selection of Higgs masses that are relevant for the
$\lhc$ (and the Tevatron as well). 

The largest total errors are by far the ones that affect BR($H \to c\bar c)$ 
which are of the order of 20\%, while the errors on BR$(H\to gg)$ are at the
level of a few to 10\% at most. In the case of the $H\to c\bar c$
channel, it is mainly due to the uncertainty on the input charm--quark mass
$\overline{m}_c$ and, to a lesser extent, to the uncertainty on $\alpha_s$ ; their
combination leads to a very strong variation of the charm--quark mass at the
high scale $\mu$, $\overline{m}_c(\mu) \propto [\alpha_s
(\mu)]^{12/13}$. For the $H\to g g$ channel the uncertainty is mainly
due to the uncertainty on $\alpha_s$, and in both cases, the error 
on the input $b$--quark mass leads to a 2 to 5\% uncertainty. 
Nevertheless, since the branching ratios for these two
decays are small, at most a few percent in the Higgs mass range of interest,
the associated uncertainties will affect the Higgs total width, and hence the
branching ratios for the other decay channels, in a less significant
way.

\begin{table}[!h]{\small%
\let\lbr\{\def\{{\char'173}%
\let\rbr\}\def\}{\char'175}%
\renewcommand{\arraystretch}{1.4}
\vspace*{-1mm}
\begin{center}
\begin{tabular}{|c|c|c|ccc|c|}\hline
channel & $M_H$ & BR(\%) & $\Delta m_c$ & $\Delta m_b$
& $\Delta\alpha_s$ & $\Delta$BR    \\ \hline
& $120$ & 65.40 & $^{+0.7\%}_{-0.6\%}$ & $^{+3.4\%}_{-1.2\%}$ & 
$^{+0.7\%}_{-0.8\%}$ & $^{+3.6\%}_{-1.6\%}$ 
\\ 
$H\to b\bar b$ & $135$ & 41.01 & $^{+0.4\%}_{-0.4\%}$ & $^{+6.0\%}_{-2.1\%}$ & 
$^{+1.3\%}_{-1.3\%}$ & $^{+6.2\%}_{-2.5\%}$ 
\\ 
& $150$ & 16.07 & $^{+0.2\%}_{-0.1\%}$ & $^{+8.7\%}_{-3.0\%}$ & 
$^{+1.9\%}_{-1.9\%}$ & $^{+8.9\%}_{-3.6\%}$ 
\\ \hline
& $120$ & 14.06 & $^{+0.7\%}_{-0.6\%}$ & $^{+2.3\%}_{-6.3\%}$ & 
$^{+1.4\%}_{-1.4\%}$ & $^{+2.8\%}_{-6.5\%}$ 
\\ 
$H\to WW$ & $135$ & 39.86 & $^{+0.4\%}_{-0.4\%}$ & $^{+1.4\%}_{-4.0\%}$ & 
$^{+0.9\%}_{-0.9\%}$ & $^{+1.7\%}_{-4.1\%}$ 
\\ 
& $150$ & 69.45 & $^{+0.2\%}_{-0.1\%}$ & $^{+0.5\%}_{-1.6\%}$ & 
$^{+0.3\%}_{-0.3\%}$ & $^{+0.7\%}_{-1.6\%}$ 
\\ \hline
\end{tabular}
\end{center}
\vspace*{-3mm}
\caption{The Higgs decay branching ratio  BR($H\rightarrow b\bar b)$ 
and BR($H\rightarrow WW)$ (in \%) for given Higgs mass values (in GeV) 
with the corresponding uncertainties from the various sources discussed in 
the text; the total uncertainties (adding the individual ones in quadrature) 
are also shown.}
\label{table_decay1}
\vspace*{-2mm}
}
\end{table}

The uncertainty on the  $H\to b\bar b$ decay channel (which is important at
the Tevatron) depends strongly on the considered  Higgs mass range. For $M_H 
\lsim 120$ GeV where the branching ratio is by far the dominant one, BR($H \to
b\bar b) \gsim 65\%$, the uncertainty is less than a few percent ($\approx 
-1\%, +2\%$ at $M_H\!=\!100$ GeV and  $\approx -2\%,+4\%$ at $M_H\!=\!120$
GeV). This is mainly due to the fact that as the channel is dominant, the
$b\bar b$ partial width is the major component of the total width and the
errors partly cancel in the branching ratio. There is, however, a residual
error stemming mainly from the input bottom mass and  to a lesser extent,
$\alpha_s$ and the charm mass.

In the mass range $M_H\! \approx\! 120$--150 GeV, the partial widths for  $H
\!\to\! b\bar b$ and $WW$ decays have the same  magnitude and the two
branching ratios  have larger uncertainties. The uncertainty on BR$(H\!\to\!
b\bar b)$ increases from a few percent at $M_H\! \approx\! 120$ GeV to $\approx
\!-\!4\%, \!+\!9\%$ at $M_H\!\approx \!150$ GeV. At the same time, the 
uncertainty on BR$(H\!\to \!WW)$, which is the same as the one on
BR($H\!\to\! ZZ^*)$ and BR($H\! \to \!\gamma\gamma)$, drops from $\approx\! -
\!7\%, \!+\!3\%$ at $M_H \! \approx \! 120$ GeV to $\approx\! -\!2\%, +\!1\%$
at $M_H\! \approx \!150$ GeV. Here,  $\overline{m}_c$ has little impact as 
BR$(H\!\to \!c\bar c)$ is small. 

For higher Higgs masses, $M_H \gsim 160$ GeV, the Higgs width is mainly
controlled by the $H\to WW$ decay (and $H \to ZZ$ for $M_H \gsim 2M_Z$) and
the uncertainty on the branching ratio  BR$(H\to WW)$ drops to a level below
1\%. The uncertainty on BR$(H\to b\bar b)$ stays relatively large, $ \approx
-4\%, +11\%$, but the branching ratio itself is too small to be relevant. 

For Higgs masses beyond $M_H \gsim 350$ GeV, the decay channel $H\to t\bar t$
opens up and has a branching ratio of $\approx 20\%$ at $M_H \approx 500$ GeV  
dropping to less than $10\%$ for $M_H \lsim 400$ GeV or $M_H \gsim 700$ GeV. It
will be affected by the uncertainties on $m_t, \alpha_s$ as well as by some
electroweak contributions from the Higgs self--couplings. This leads to a 
total uncertainty that is estimated  to be below the 5\% level \cite{LHCXS},
which translates to an uncertainty of less than 1\% in the branching ratios for
the important decays $H \to WW,ZZ$. 

Thus, the uncertainties on the important Higgs branching ratios  BR($H\to
WW,ZZ,\gamma \gamma)$ as well as BR($H\to b\bar b)$  can be
significant in the intermediate  mass range $M_H \approx 120$--150 GeV where
the $b\bar b$ and $WW$ decays are competing with each other. These
uncertainties should therefore be included in the experimental analyses.

\begin{table}[!h]{\small%
\let\lbr\{\def\{{\char'173}%
\let\rbr\}\def\}{\char'175}%
\renewcommand{\arraystretch}{1.3}
\vspace*{-.1mm}
\begin{bigcenter}
\begin{tabular}{|c||cc|cc|cc|ccccc|}\hline
$M_H$ & $b\bar b $ & $\Delta^{\rm tot}$
& $c\bar c$ & $\Delta^{\rm tot}$ & gg & $\Delta^{\rm tot}$
&WW & ZZ & $\tau \tau$ & $\gamma \gamma$ & $\Delta^{\rm tot}$\\ \hline
$100$ & 79.48 & $^{+2.2\%}_{-1.1\%}$ & 
3.90& $^{+20.9\%}_{-22.7\%}$ & 6.94& $^{+5.6\%}_{-9.2\%}$ & 
1.09 & 0.11 & 8.22 & 0.16 & $^{+3.4\%}_{-7.8\%}$ \\ \hline
$105$ & 77.69 & $^{+2.4\%}_{-1.1\%}$ & 
3.81& $^{+20.9\%}_{-22.7\%}$ & 7.50& $^{+5.4\%}_{-8.9\%}$ & 
2.39 & 0.21 & 8.11 & 0.18 & $^{+3.3\%}_{-7.8\%}$ \\ \hline
$110$ & 74.96 & $^{+2.6\%}_{-1.2\%}$ & 
3.68& $^{+20.8\%}_{-22.7\%}$ & 7.97& $^{+5.3\%}_{-8.7\%}$ & 
4.75 & 0.43 & 7.89 & 0.19 & $^{+3.2\%}_{-7.5\%}$ \\ \hline
$115$ & 70.95 & $^{+3.1\%}_{-1.4\%}$ & 
3.48& $^{+20.8\%}_{-22.7\%}$ & 8.27& $^{+5.1\%}_{-8.3\%}$ & 
8.54 & 0.86 & 7.53 & 0.21 & $^{+3.0\%}_{-7.1\%}$ \\ \hline
$120$ & 65.40 & $^{+3.6\%}_{-1.6\%}$ & 
3.21& $^{+20.7\%}_{-22.8\%}$ & 8.34& $^{+4.9\%}_{-7.8\%}$ & 
14.06 & 1.57 & 7.00 & 0.22 & $^{+2.7\%}_{-6.6\%}$ \\ \hline
$125$ & 58.32 & $^{+4.4\%}_{-1.8\%}$ & 
2.86& $^{+20.7\%}_{-22.8\%}$ & 8.11& $^{+4.6\%}_{-7.1\%}$ & 
21.35 & 2.62 & 6.29 & 0.23 & $^{+2.4\%}_{-5.9\%}$ \\ \hline
$130$ & 50.00 & $^{+5.3\%}_{-2.1\%}$ & 
2.45& $^{+20.6\%}_{-22.9\%}$ & 7.56& $^{+4.2\%}_{-6.4\%}$ & 
30.13 & 3.95 & 5.43 & 0.22 & $^{+2.1\%}_{-5.0\%}$ \\ \hline
$135$ & 41.01 & $^{+6.2\%}_{-2.5\%}$ & 
2.01& $^{+20.5\%}_{-23.0\%}$ & 6.72& $^{+3.9\%}_{-5.5\%}$ & 
39.86 & 5.42 & 4.49 & 0.21 & $^{+1.7\%}_{-4.1\%}$ \\ \hline
$140$ & 32.02 & $^{+7.2\%}_{-2.8\%}$ & 
1.57& $^{+20.5\%}_{-23.1\%}$ & 5.68& $^{+3.6\%}_{-4.7\%}$ & 
49.92 & 6.82 & 3.53 & 0.19 & $^{+1.3\%}_{-3.2\%}$ \\ \hline
$145$ & 23.60 & $^{+8.1\%}_{-3.2\%}$ & 
1.16& $^{+20.6\%}_{-23.1\%}$ & 4.51& $^{+3.2\%}_{-4.0\%}$ & 
59.82 & 7.86 & 2.62 & 0.17 & $^{+1.0\%}_{-2.4\%}$ \\ \hline
$150$ & 16.07 & $^{+8.9\%}_{-3.5\%}$ & 
0.79& $^{+20.6\%}_{-23.2\%}$ & 3.31& $^{+3.0\%}_{-3.4\%}$ & 
69.45 & 8.21 & 1.79 & 0.14 & $^{+0.6\%}_{-1.6\%}$ \\ \hline
$155$ & 9.47 & $^{+9.7\%}_{-3.7\%}$ & 
0.46& $^{+20.7\%}_{-23.3\%}$ & 2.09& $^{+2.8\%}_{-3.0\%}$ & 
79.28 & 7.33 & 1.06 & 0.10 & $^{+0.4\%}_{-1.0\%}$ \\ \hline
$160$ & 3.58 & $^{+10.4\%}_{-3.9\%}$ & 
0.18& $^{+20.7\%}_{-23.4\%}$ & 0.84& $^{+2.7\%}_{-2.7\%}$ & 
90.63 & 4.19 & 0.40 & 0.05 & $^{+0.1\%}_{-0.4\%}$ \\ \hline
$165$ & 1.23 & $^{+10.7\%}_{-4.0\%}$ & 
0.06& $^{+20.8\%}_{-23.4\%}$ & 0.30& $^{+2.6\%}_{-2.6\%}$ & 
95.97 & 2.22 & 0.14 & 0.02 & $^{+0.0\%}_{-0.1\%}$ \\ \hline
$170$ & 0.82 & $^{+10.7\%}_{-4.0\%}$ &
0.04& $^{+20.8\%}_{-23.4\%}$ & 0.21& $^{+2.6\%}_{-2.5\%}$ & 
96.43 & 2.35 & 0.09 & 0.02 & $^{+0.0\%}_{-0.1\%}$ \\ \hline
$175$ & 0.63 & $^{+10.7\%}_{-4.1\%}$ & 
0.03& $^{+20.8\%}_{-23.5\%}$ & 0.17& $^{+2.5\%}_{-2.6\%}$ & 
95.84 & 3.21 & 0.07 & 0.01 & $^{+0.0\%}_{-0.1\%}$ \\ \hline
$180$ & 0.51 & $^{+10.7\%}_{-4.1\%}$ & 
0.02& $^{+20.8\%}_{-23.5\%}$ & 0.15& $^{+2.5\%}_{-2.5\%}$ & 
93.28 & 5.94 & 0.06 & 0.01 & $^{+0.0\%}_{-0.0\%}$ \\ \hline
$185$ & 0.40 & $^{+10.7\%}_{-4.0\%}$ & 
0.02& $^{+20.8\%}_{-23.4\%}$ & 0.12& $^{+2.6\%}_{-2.4\%}$ & 
84.52 & 14.86 & 0.05 & 0.01 & $^{+0.0\%}_{-0.0\%}$ \\ \hline
$190$ & 0.32 & $^{+10.7\%}_{-4.1\%}$ & 
0.02& $^{+20.8\%}_{-23.4\%}$ & 0.10& $^{+2.5\%}_{-2.5\%}$ & 
78.72 & 20.77 & 0.04 & 0.01 & $^{+0.0\%}_{-0.0\%}$ \\ \hline
$195$ & 0.28 & $^{+10.8\%}_{-4.1\%}$ & 
0.01& $^{+20.8\%}_{-23.4\%}$ & 0.10& $^{+2.5\%}_{-2.5\%}$ & 
75.89 & 23.66 & 0.03 & 0.01 & $^{+0.0\%}_{-0.0\%}$ \\ \hline
$200$ & 0.25 & $^{+10.8\%}_{-4.1\%}$ & 
0.01& $^{+20.8\%}_{-23.5\%}$ & 0.09& $^{+2.5\%}_{-2.5\%}$ & 
74.26 & 25.34 & 0.03 & 0.01 & $^{+0.0\%}_{-0.0\%}$ \\ \hline
\end{tabular}
\end{bigcenter}
\vspace*{-2mm}
\caption{The Higgs decay branching ratios (in \% ) for given Higgs mass values 
(in GeV) with the corresponding total uncertainties from the various sources 
discussed in the text.}
\vspace*{-1mm}
\label{table_decay2}
}
\end{table}

\subsection{Combination with the cross section uncertainty} 

In the previous section we have discussed the overall uncertainty on the Higgs
production cross section. To this overall uncertainty, one has to add the 
uncertainty on the Higgs branching ratios obtained  when considering a specific
Higgs decay channel. This combination\footnote{Here, we will not address  two
(connected) issues which become important for heavy Higgs bosons: $i)$  the
effect of the total Higgs decay width which becomes large and $ii)$ the
interference between the $gg \to H \to VV$ signal and the $gg \to VV$
background. A study of these two effects is in progress \cite{DKK}.} is in
general rather complicated to perform when several production and decay channels
are involved as is the case for the Tevatron where both $q\bar q \to HV$ and
$gg\to H$ production and $H \to b\bar b$ and $H\to WW$ decays channel have to be
considered, the main reason being that the rates for the $H\to b\bar b$ and
$H\to WW$ decays are anti-correlated, the sum of all branching ratios being
equal to unity. At the $\lhc$, the situation is much simpler as, in practice,
only the  $gg\to H$ production channel is to be  considered and only the decays
$H\to WW, ZZ$ and to a lesser extent $H\to \gamma \gamma$ are relevant. Since we
have considered only the dominant QCD uncertainties, these decays are affected
by the same uncertainties as discussed previously (see the last column of Table
\ref{table_decay2}).

Nevertheless, combining the  uncertainty in the production cross section and the
uncertainty in the Higgs decay branching ratios  is still not obvious. Indeed,
while the uncertainties on the Higgs  production cross  section are purely
theoretical, one should in principle  consider the uncertainties on the Higgs
branching ratios as experimental errors, since they are mainly due to the 
``experimental" determination of the quark masses and $\alpha_s$ (and we have
not included the corresponding theoretical errors). In addition, there is one
parameter which is common in the calculation of the branching ratios and the
production cross sections:  the  coupling $\alpha_s$. The uncertainty on
$\alpha_s$ will affect at the same time $\sigma(gg\to H)$ and BR($H\to VV)$ and
it occurs that,   in both the Higgs production and in Higgs decays, the minimal
(maximal) values are obtained with the minimal (maximal) value of $\alpha_s$ when
the error $\Delta^{\rm exp} \alpha_s$ is included (we have assumed that no
theoretical uncertainty on $\alpha_s$ is present in the decays).

\begin{figure}[!h]
\begin{center}
\mbox{
\epsfig{file=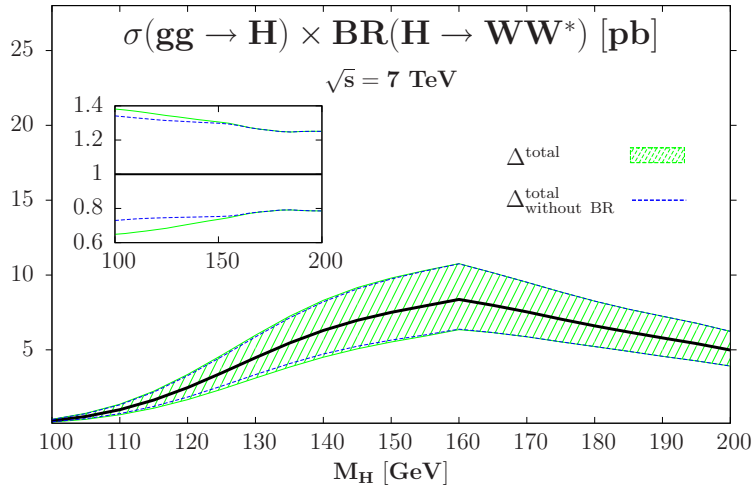, scale=.75} 
}
\end{center}
\vspace*{-4mm}
\caption[]{The production cross section times branching ratio for the  process
$gg\to H \to WW$ at the $\lhc$ with $\sqrt s=7$ TeV,  including the total
theoretical uncertainty band  when the errors in the decay branching ratios
are  taken into account. In the inserts the relative deviations are shown.}
\label{finalerror}
\vspace*{-1mm}
\end{figure}

Here, we will thus adopt the simple  attitude of
adding linearly the errors on the Higgs branching ratios to the
theoretical uncertainty on the production cross section. In the case of the
procedure A for the combination of the cross sections uncertainties,  for
consistency reasons, one should also use the $90\%$ CL error on the branching 
ratios due to $\Delta ^{\rm exp} \alpha_s$ together  with the $90\%$ CL
PDF+$\Delta^{\rm exp} \alpha_s$ uncertainties in the cross section.  In this
case, one obtains the combined uncertainty of $\sigma^{\rm NNLO}(gg\to H)
\times {\rm BR}(H \to VV^*)$ at the $\lhc$ that is displayed  in 
Fig.~\ref{finalerror} in the case of the $H \to WW$  decay as a function of
$M_H$. Besides the uncertainty on the cross section which is shown by  the
dashed lines, we display the effect of adding the error on the $H \to WW$ 
branching ratio as shown by the full lines which slightly increase the
overall uncertainty.   This is only visible  in the region  $M_H \approx
100$--150 GeV where the errors on BR$(H\to WW)$, dominated in practice by the
errors in the decay channel $H\to b\bar b$, is significant. Above the $M_H
\gsim 2M_W$ threshold, the  branching ratio  uncertainty is small, below 
$\approx 1\%$, and can be safely neglected\footnote{For instance, the branching
ratio uncertainties have no effect in the Higgs mass range $M_H\approx
150$--180 GeV to which both the Tevatron and the $\lhc$ are most sensitive.
However, at the Tevatron these errors should be taken into account for $M_H
\lsim 150$ GeV  in both the $H \to b\bar b$ and $WW$ channels.}.  Note,
however, that for large $M_H$ values, the total Higgs width becomes
significant and should be taken into account in both the production and in
the  decay \cite{LHCXS}.

In the case of the $H\to ZZ, \gamma \gamma$ (and $\tau\tau)$ decays, since the
errors on the branching ratios are the same  as those affecting BR($H\to WW)$,
the overall uncertainties in $\sigma \times {\rm BR}$ can also be seen from
Fig.~\ref{finalerror}: only the normalization of the branching ratio  is
different and can be obtained for a specific decay mode from  Table
\ref{table_decay2}.

\section{The MSSM neutral Higgs particles} 

\subsection{The MSSM Higgs sector}

In supersymmetric extensions of the SM, at least two Higgs doublet fields are
required for a consistent electroweak symmetry breaking and in the minimal
model, the MSSM,  the Higgs sector is extended to contain five Higgs bosons: two
CP--even $h$ and $H$, a CP-odd $A$ and two charged Higgs $H^\pm$ particles
\cite{HHG,Review2,MichaelR,SvenPR}.  Besides the  four masses, the properties of
the MSSM Higgs sector are in principle determined by  two more parameters: a
mixing angle $\alpha$ in the neutral CP--even sector  and the ratio of the 
vacuum expectation values of the two Higgs fields $\tb$; the value of the 
latter is expected to lie in the range $1 \lsim \tb \lsim m_t/m_b$. In fact, in
the MSSM, the scalar  potential  does not involve arbitrary self couplings as
is the case with the SM, but involves only the gauge couplings and, as a result,
only two free parameters are needed at tree--level to describe the Higgs sector;
these are usually taken to be  the mass $M_A$ of the pseudoscalar $A$ boson
and   $\tan\beta$.  In addition,  while the  masses of the heavy neutral and
charged $H,A,H^\pm$ particles are expected to range from $M_Z$ to the SUSY
breaking scale $M_S={\cal O}(1$ TeV),  the mass of the lightest Higgs boson $h$
is bounded from above, $M_h \le M_Z$ at the tree level. 

It is well known that this relation is altered by the large radiative
corrections which enter in the Higgs sector,  the leading part of which grow as
the fourth power of $m_t$ and logarithmically with the SUSY scale  or common
squark mass $M_S$; the mixing (or trilinear coupling) in the stop sector $A_t$
and, to a lesser extent a Higgs mass parameter $\mu$, play an important role.
When higher order corrections are included, the upper bound on the mass of the
lightest $h$ boson is shifted from the tree level value $M_Z$ to the value
$M_h^{\rm max} \sim 110$--135 GeV \cite{SvenPR} in the maximal mixing scenario
where $X_t= A_t -\mu/\tb \sim \sqrt 6 M_S$.  The maximal mixing scenario, together
with the no--mixing scenario with $X_t\approx 0$ which leads to a lower
$M_h^{\rm max}$ value, are often used as benchmarks  to describe MSSM Higgs
phenomenology \cite{S-benchmark}.

For a heavy enough  $A$ boson, $M_A \gg M_Z$,  the $h$ boson reaches its 
maximal mass value $M_h\simeq  M_h^{\rm max}$ and has SM--like couplings to
fermions and gauge bosons. In this decoupling regime which, for large $\tb$
values occurs already for $M_A \gsim  M_h^{\rm max}$, the three other Higgs
particles are  almost degenerate in mass, $M_H \!\approx\! M_A\! \approx\!
M_H^{\pm}$. The couplings of the CP--even $H$ boson (as well as those of the
charged $H^\pm$  bosons) become similar to that of the pseudoscalar $A$ boson:
no tree--level couplings to the gauge bosons and couplings to isospin down (up)
type fermions  that are (inversely) proportional to $\tb$. In particular, for
high  $\tb$ values, $\tb \gsim 10$, the $H, A$ Yukawa couplings to bottom quarks
and tau leptons are strongly  enhanced, while those to top quarks are strongly
suppressed. 

For a light pseudoscalar boson,  $M_A \lsim M_h^{\rm max}$ at high $\tb$, the
roles of the CP--even $h$ and $H$ states are reversed: it is the $H$ boson
which has a mass $M_H\simeq  M_h^{\rm max}$ and SM--like couplings, while  the
$h$ particle behaves exactly like the pseudoscalar $A$ state, i.e.  $M_h \simeq
M_A$, no couplings to gauge bosons and enhanced (suppressed) couplings to $b,
\tau$ ($t$) states.  We are thus always in a scenario where one has a SM--like
Higgs boson $H_{\rm SM}= h\;(H)$ and two CP--odd like Higgs particles $\Phi=A$
or $H\;(h)$ when we are in  the decoupling (anti--decoupling)
regime\footnote{Note that there is an intermediate scenario for $\tan \beta
\gsim 10$: the intense coupling regime \cite{S-intense} in which the three
neutral Higgs bosons have comparable masses, $M_h \approx M_H \approx M_A
\approx M_h^{\rm max}$. The couplings of the three states to isospin down type
fermions are  enhanced and the squares of the CP--even Higgs couplings
approximately add to the square of the CP--odd Higgs coupling. This property,
together with the fact that  $M_H \approx M_h$, makes that for our main purpose
in this paper, one has the same phenomenology as in the decoupling or
anti--decoupling regimes.} which, for  values $\tb \gsim 10$, occurs  already
for $M_A \gsim M_h^{\rm max}$  ($M_A \lsim M_h^{\rm max}$). 

Let us now discuss the Higgs Yukawa couplings to bottom quarks, which play  a
major role in the analysis that will be presented in this section. As mentioned
earlier, in almost the entire parameter space for large enough $\tb$ values, the
couplings of one of the CP--even Higgs particles are SM--like, thus  leading to
the phenomenology described in the previous sections for the SM Higgs boson, 
while the couplings of the other CP--even particle are the same as those of the
pseudoscalar $A$ boson, on   which we will focus in the rest of our discussion.
At the tree level, this  coupling is given in terms of the $b$--quark mass, the
SM vacuum expectation value $v$ and $\tb$, by
\beq  
\lambda_{\Phi bb} = \frac{\sqrt 2 m_b}{v \cos\beta}
\stackrel{\small \tb \gg 1} \longrightarrow \frac{\sqrt 2 m_b}{v} \tb \ , \ 
\Phi=A, H\;(h) \label{abb} 
\eeq 
First of all, in the MSSM, one usually uses the modified dimensional reduction 
$\overline{\rm DR}$ scheme which, contrary to the  $\overline{\rm MS}$ scheme, 
preserves Supersymmetry. In the case of the $b$--quark mass, the relation 
between the $\overline{\rm DR}$ and $\overline{\rm MS}$ running masses at a  
given scale $\mu$ reads \cite{S-mbdr}
\beq  
{\overline{m}}_{b}^{\overline{\rm DR}} (\mu) = {\overline{m}}_{b}^{\overline{\rm
MS}} (\mu) \, \bigg[ 1- \frac{1}{3} \frac{\alpha_{s} (\mu^2)}{\pi} - 
\frac{\alpha_s^2(\mu^2)}{\pi^2} +  \cdots \bigg]  
\eeq 
where the strong coupling constant $\alpha_s$ is also evaluated at the  scale
$\mu$ and additional but small electroweak contributions are present. Since the
difference between the quark masses in the two schemes is not very large,
$\Delta m_b/m_b \sim 1\%$, to be compared  with an ``experimental" error  on
$\overline{m}_b(\overline{m}_b)$ of the order of a few percent, it is common practice to  neglect this
difference, at least in unconstrained SUSY models where one does not have to
evolve the parameters up to very high scales. We will thus adopt this
approximation at least when we quote the central values of the cross sections
(see later however, when the theoretical uncertainties will be  considered).  

A second important fact is that the Yukawa coupling eq.~(\ref{abb}), besides 
the standard QCD corrections, receives  additional one--loop  vertex corrections
which  involve squarks and gluinos in the loops and which can be very large as
they grow with $\tan\beta$ \cite{S-deltar},
\beq \Delta_b \approx  \frac{2\alpha_s}{3\pi} \mu m_{\tilde{g}} \tan \beta 
/{\rm max}(m_{\tilde{g}}^2,  m_{\tilde{b}_1}^2,m_{\tilde{b}_2}^2) 
\eeq 
with $m_{\tilde{b}_{1,2}}$ and $m_{{\tilde g}}$ the sbottom and gluino masses;
large corrections, proportional to $\mu A_t$ also appear when the electroweak
stop and chargino loops are included. This correction is very important for
large values of $\tb$ and $\mu$ and significantly increase or decrease
(depending on the sign of $\mu$) the Yukawa coupling. It can be properly taken
into account using a resumation procedure where one writes the modified Yukawa
coupling, normalized to the coupling of the SM Higgs boson, as  \cite{S-mbdr}
\beq
\lambda_{\Phi bb}/\lambda_{H_{\rm SM} bb} =   
\frac{\tan \beta}{1+\Delta_b} \bigg[1-\frac{\Delta_b}  {\tan^2\beta} \bigg] 
\stackrel{\small \tb \gg 1} \longrightarrow \frac{\tan \beta}{1+\Delta_b}
\label{ghff:threshold} 
\eeq 
It has been shown that in this case  all terms of ${\cal O}(\alpha_s \tb)$ in 
perturbation theory are resummed \cite{S-mbdr}. The inclusion of the NNLO
corrections in $\Delta_b$  allows, in addition,  to have a very precise
description of the coupling, with a scale variation at the percent level
\cite{S-Michaelmb}. 

The resummation procedure which allows to include properly the SUSY radiative
corrections  in the $\Phi bb$ Yukawa couplings is implemented in the major 
codes which  calculate the MSSM Higgs spectra such as {\tt HDECAY}
\cite{HDECAY}, {\tt FeynHiggs} \cite{Feynhiggs} or the RGE codes that calculate 
the MSSM particle spectrum such as {\tt SuSpect} \cite{RGEcodes}. Since in our
analysis we will be mainly concerned with the standard QCD effects and we would
like to have  a model independent approach (see next subsection), we will assume
that these SUSY QCD and electroweak corrections are handled by the previously
mentioned  programs that provide the radiatively corrected Higgs couplings.
Nevertheless, we will see later that in the  main channels
discussed in this paper, the major part of these corrections cancels out.

Before closing this survey of the MSSM Higgs sector, we  note that there are
experimental constraints on the Higgs masses, which mainly come from the
negative LEP2 searches \cite{LEP-Higgs}. In the decoupling limit where the $h$ boson
is SM--like, the constraint $M_h \gsim 114$ GeV holds and  rules out $\tb$
values smaller than $\sim 3$. Outside this regime, the most conservative and
model independent limits on the Higgs masses from LEP2 searches are  $M_h, M_A
\gsim M_Z$ \cite{LEP-Higgs}.  Some exclusion limits in the [$M_A, \tb$]
parameter space are also available from Tevatron  searches
\cite{Tevatron-MSSM} and rule out extremely large $\tb$ and low $M_A$ values.  


\subsection{The production cross sections at the lHC}

In the MSSM, the dominant production processes for the CP--even neutral $h$ and
$H$ bosons are essentially  the same as those for the SM Higgs particle
discussed in section 2.1. In fact, for the lighter $h$ boson in the decoupling
and the heavier $H$ boson in the anti--decoupling regimes, the cross sections
are almost exactly the same as for  the SM Higgs particle with a mass $\approx
M_h^{\rm max}$ shown in Fig.~\ref{pp-H-TeV} with the numerical values
of Table \ref{lhc7_allchannels}. The only difference is that some SUSY particles might
contribute to the $gg$ fusion processes and, eventually, to the SUSY
radiative corrections in the other processes. However, for heavy
enough squarks and gluinos, as is the case for the benchmark scenarios
that are in general discussed  and which have $M_S \gg M_h^{\rm max}$,
these contributions are rather small as the SUSY states decouple from
the amplitudes (their couplings to the Higgs bosons are not
proportional to their masses). These SUSY contributions, including
even the known QCD  corrections \cite{SUSY-QCD},  can be nevertheless
evaluated exactly using the program {\tt HIGLU} for instance.

In the case of the CP--odd like particles, the situation is completely
different. Because of CP invariance which forbids $A$ couplings to gauge
bosons  at tree--level, the pseudoscalar $A$ boson cannot be produced in the
Higgs-strahlung and vector boson fusion processes;  only the $gg\to A$ fusion 
as well as associated production with  heavy quark pairs, $q\bar q, gg \to
Q\bar Q A$, will be in practice relevant (additional processes, such as
associated production of CP--even and CP--odd Higgs particles, have too  small
cross sections). This will therefore be also the case of the CP--even $H$ and
$h$ particles in, respectively, the decoupling and anti--decoupling  scenario.
In addition, if one concentrates on the high $\tb$ regime (that is the only
relevant one in this context, both at the Tevatron and the $\lhc$),  the
$b$--quark will play the major  role as its couplings to the CP--odd like 
bosons are enhanced.  

Therefore, for $\Phi=A$ or $H(h)$,   one first has to take into account in the
$gg \to \Phi$  processes the $b$--quark loop  which provides the dominant
contribution  in the MSSM, as the one of the top--quark loop is suppressed, 
$\lambda_{\Phi tt} \propto 1/\tb$. Moreover, in associated Higgs production with
heavy quarks, the rates  for the $pp \to t\bar t  \Phi$ processes are also
suppressed  and,  instead, associated  Higgs production with $b\bar{b}$ final
states must be considered. In fact, the $pp \to b \bar b \Phi$  production
processes become the dominant ones in the MSSM.  

In the $gg \to \Phi$ processes with only the $b$--quark loop included,  as the
$b$--quark mass is very small compared to the Higgs masses, chiral symmetry
approximately holds and the cross sections are approximately the same for the
CP--even $H\;(h)$ and CP--odd $A$ bosons\footnote{This is only true if the SUSY
particle loop contributions are not included. In the case of the CP--even
particles, their relative contribution  are suppressed at high enough $\tb$ and
we will ignore them here. In the case of the $A$ boson,   the SUSY
contributions appear only at two--loops and they can be safely neglected.}. The
QCD corrections are known only to  NLO for which the exact calculation with
finite loop quark masses is available \cite{SDGZ}. Contrary to the SM case, 
they increase only moderately the production cross sections. The calculation of
the higher order corrections that have been made available  for the SM Higgs
boson,  the NNLO QCD corrections (performed in the infinite quark mass limit)
and the NLO electroweak corrections (the dominant part of which arises because
of the large Higgs--$t \bar t$ Yukawa coupling)  do not apply here and will be
thus ignored. However, in the case of  $h$ and $H$ production,  to approach
properly the decoupling and anti--decoupling limits and to reproduce the SM
Higgs results discussed in section 2 when these higher order corrections are
included, we will adopt the central value $\mu_0=\frac12 M_\Phi$ for the
renormalization and factorization scales, i.e. as in the SM case.

In the case of the $pp \to b\bar b \Phi$ processes, the NLO QCD corrections
have been calculated in Ref.~\cite{gg-bbH-QCD} and turn out to be rather
large,  in contrast to $pp \to t\bar t$+Higgs production.   Because of the
small $m_b$ value, the cross sections  develop large logarithms
$\log(Q^2/m_b^2)$ with the scale $Q$ being typically of the order of the
factorization scale, $\mu_F \sim M_\Phi \gg m_b$. These logarithms can be
resummed via the  Altarelli--Parisi equations by considering the $b$--quark as
a massless parton and  using heavy quark distribution functions at a scale 
$\mu_F \sim Q$ in a five active flavor scheme. In this scheme, the inclusive
process where one does not require to observe the $b$ quarks is simply the $2
\to 1$ process $b \bar b \to \Phi$ at leading order \cite{bbH-LO}. If one
requires the observation of a high--$p_T$ final $b$--quark, one has to consider
its NLO corrections \cite{bbH-NLO} and in particular the $2\to 2$ process
$gb\to \Phi b$, which indeed generates the $p_T$ of the $b$--quark. Requiring
the observation of two $b$ quarks in the final state, one has to consider the
$2 \to 3$ process $gg \to b\bar b \Phi$ discussed above, which is the leading
mechanism at NNLO \cite{bbH-NNLO}. Thus, instead of  $q\bar q, gg \to b\bar b
\Phi$,   we will consider the process $b\bar b \to \Phi$ for which  the cross
section is known up to NNLO in QCD \cite{bbH-NLO,bbH-NNLO}, with corrections
that are of moderate size if $i$) the bottom quark mass in the Yukawa coupling
is defined at the scale $M_\Phi$ to absorb large logarithms
$\log(\mu_R^2/m_b^2)$ and $ii)$  if the  factorization  scale, that we will set
here equal to the   renormalization  scale,  is chosen to be small,
$\mu_F=\mu_R=\mu_0= \frac14 M_\Phi$.

In order to evaluate the Higgs production cross sections at $\lhc$ energies in
these two main processes, $gg \to \Phi$ and $b\bar b \to \Phi$, we will proceed
as follows. We only evaluate the cross sections  for the pseudoscalar $A$ 
boson: in the $gg \to A$ process at NLO using the program {\tt HIGLU}
\cite{Michael} with a central scale  $\mu_R=\mu_F=\mu_0= \frac12 M_A$ (and 
where only the bottom quark loop contribution is included by setting
$\lambda_{A tt}=0$), and in the $b\bar b\to A$ process up to NNLO using the
program {\tt bbh@nnlo}  \cite{bbH-NNLO,Robert} with a central scale
$\mu_R=\mu_F=\mu_0= \frac14 M_A$. In both cases, we work in the $\overline{\rm
MS}$ scheme for the renormalization  of the bottom--quark mass. However, while
$\overline{m}_b(\overline{m}_b)$ is  used in the $gg$ fusion process, 
$\overline{m}_b(\mu_R)$ is adopted in the $b\bar b$ fusion channel. In both
processes, we assume the $Ab\bar b$ coupling to be SM--like, that is, we will
not include the $\tb$ term and the SUSY corrections to $\Delta m_b$. To obtain
the true cross sections for $A$ production, one will have therefore to rescale
the numbers that we provide by the factor of eq.~(\ref{ghff:threshold})
squared. Furthermore, to obtain the cross section for both CP--even and CP--odd
Higgs production, an additional factor of two has to be included. As a
consequence of  chiral symmetry for $M_\Phi\! \gg\!\overline{m}_b$ and since
the $H\;(h)$ masses and couplings are very close to those of $A$, this turns
out to be an excellent approximation. 

\begin{figure}[!h]
\begin{center}
\vspace*{-.2mm}
\mbox{
\epsfig{file=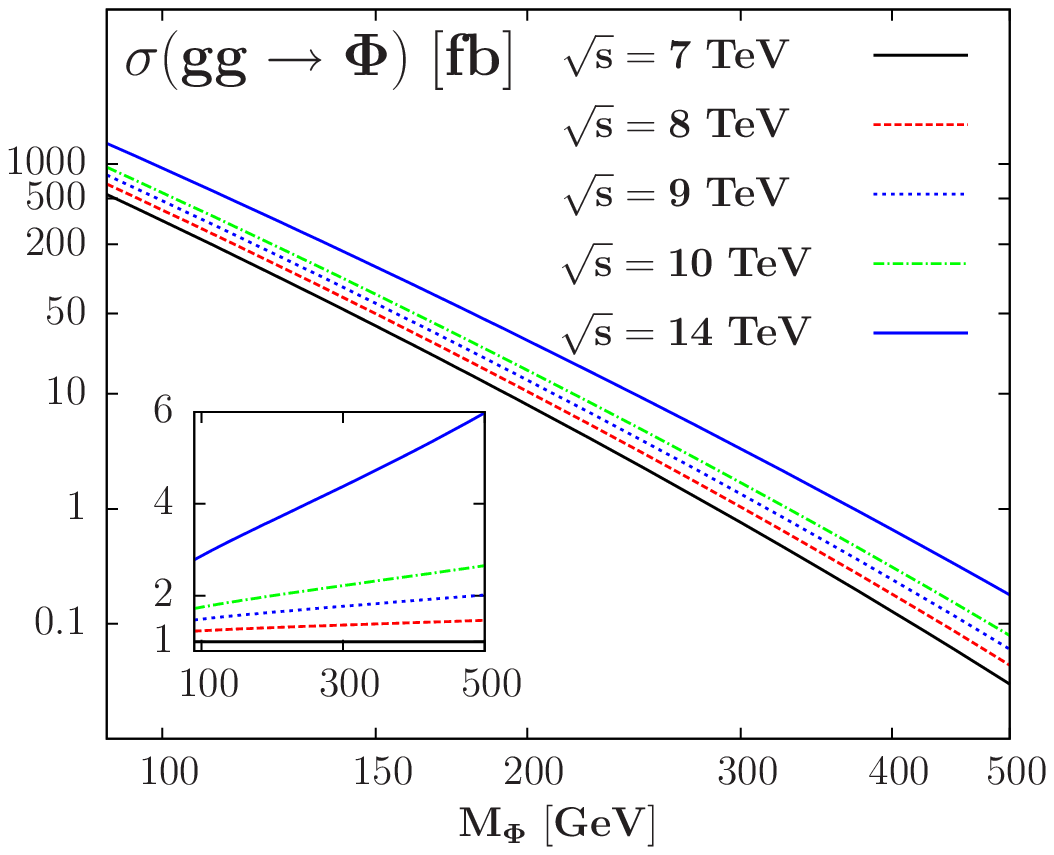,height=6.5cm, width=7.4cm}
\epsfig{file=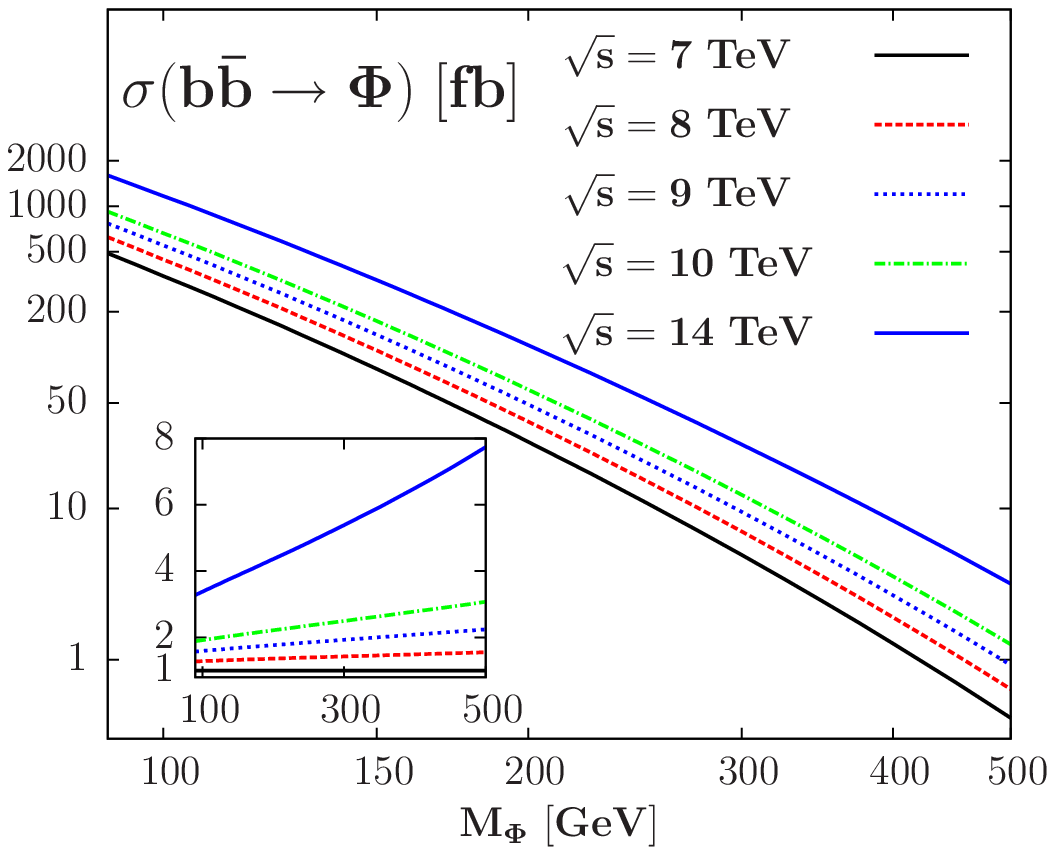,height=6.5cm, width=7.4cm}
}
\end{center}
\vspace*{-4mm}
\caption[]{The production cross sections in the processes $gg \to \Phi$ (left) 
at a central scale $\mu_R=\mu_F=\mu_0= \frac12 M_\Phi$ and $b\bar b  \to \Phi$
(right) at a central scale $\mu_R=\mu_F=\mu_0= \frac14 M_\Phi$ as a function
of $M_\Phi$ for several center of mass energies relevant for the LHC. The MSTW 
sets of PDFs at the required perturbative order have been used.  Only the cross
section in the pseudoscalar $A$ case but with a SM-like Yukawa coupling  is 
included.}
\label{MSSM-xs}
\vspace*{-.6mm}
\end{figure}

Within this set--up, the best values of the  cross sections for $gg \to \Phi$
and  $b\bar b \to \Phi$  are shown in  Fig.~\ref{MSSM-xs}  as a function of the
Higgs mass $M_\Phi$ when the MSTW sets of (NLO for the former and NNLO for the
latter process) PDFs are used to parametrize the gluon and bottom--quark
densities. Center of mass energies in a range between $\sqrt s=7$ TeV and 14
TeV relevant for the LHC  are considered. One can first notice that the cross
sections for  $gg \to \Phi$ and  $b\bar b \to \Phi$ are comparable at a given
energy and they significantly increase with increasing center of mass  energies
or decreasing Higgs mass. If, for example, the value $\tb=10$ is adopted, the
numbers in Fig.~\ref{MSSM-xs} have to  be multiplied by a factor $\simeq 200$
to obtain the true cross section for  both $A$ and $H(h)$ production. For low
to moderate $M_\Phi$ values, the expected event rates are thus simply huge  at
the $\lhc$,   despite of the  relatively low luminosities that are expected. 
This explains why the chances for observing a Higgs boson at the $\lhc$ are
much higher in the MSSM than in the SM, as in the former case the production
rates can be  two to three orders of magnitude larger.

\subsection{The theoretical uncertainties}

Left to evaluate are then the theoretical uncertainties on the production cross
sections  and, for this purpose, we will follow very closely the procedure (A) 
discussed  in section 2.2 for the SM Higgs boson, at least for the scale and
PDF+$\alpha_s$ uncertainties. For the numerical analysis, we will only consider
the case of the $\lhc$ at $\sqrt s=7$ TeV: the uncertainties at center of mass 
energies slightly above this value, $\sqrt s=8$--10 TeV, and even for the LHC
energy $\sqrt s=14$ TeV are expected to be comparable. The procedure and the
main results that we obtain are briefly summarized in the
following\footnote{Note that since in our analysis we are not considering  the
SUSY particle contributions and focus only on the standard QCD effects, 
additional uncertainties from the SUSY sector should, in principle, also be
present.  Nevertheless,  at high $\tb$, the genuine SUSY contributions should 
not be significant and can be ignored, while the corrections  entering in
$\Delta_b$ will almost cancel out when the $\tau^+ \tau^-$ decays are
considered.}.

In the case of the $gg\to \Phi$ process,  as for the SM Higgs boson,  the scale
uncertainty is evaluated by allowing for a variation of the renormalization and
factorization scales within a factor of two  around the central scale, $\frac12
\mu_0 \le \mu_R,\mu_F  \le 2 \mu_0$ with  $\mu_0= \frac12 M_\Phi$.  For the
$b\bar b  \to \Phi$ case, we will extend the domain of scale variation to a
factor of three around the central scale $\mu_0= \frac14 M_\Phi$, $\frac13
\mu_0 \le \mu_R,\mu_F  \le 3 \mu_0$. The reason for this non  democratic
treatment is that, first, in  the $gg \to \Phi$ process, there is another
uncertainty due to the scheme dependence in the renormalization of the
bottom--quark which will add to the scale uncertainty, as will be discussed
later.  A second reason is that it is well  known that when the same final
states are considered, the cross sections in the $b\bar b \to \Phi$ process in
the five--flavor scheme and in the $q\bar q, gg \to b\bar b \Phi$ channel in
the four--flavor scheme differ significantly \cite{bbH-diff} and only by
allowing a wider domain for scale variation  and, hence, a larger scale
uncertainty that the two results become consistent with each other.   In our
final numbers for the uncertainties, we  will assume a  variation of $\mu_R$
and $\mu_F$ that is taken to be independent  but with the additional
restriction $1/\kappa \le \mu_R/\mu_F \le \kappa$ imposed. However, to
illustrate the much larger scale uncertainty that is possible in the $b \bar b
\to \Phi$ case, we will also show results when this  constraint is relaxed, in
much the same way as in Ref.~\cite{LHCXS}. 

\begin{figure}[!h]
\begin{center}
\vspace*{-2mm}
\mbox{
\epsfig{file=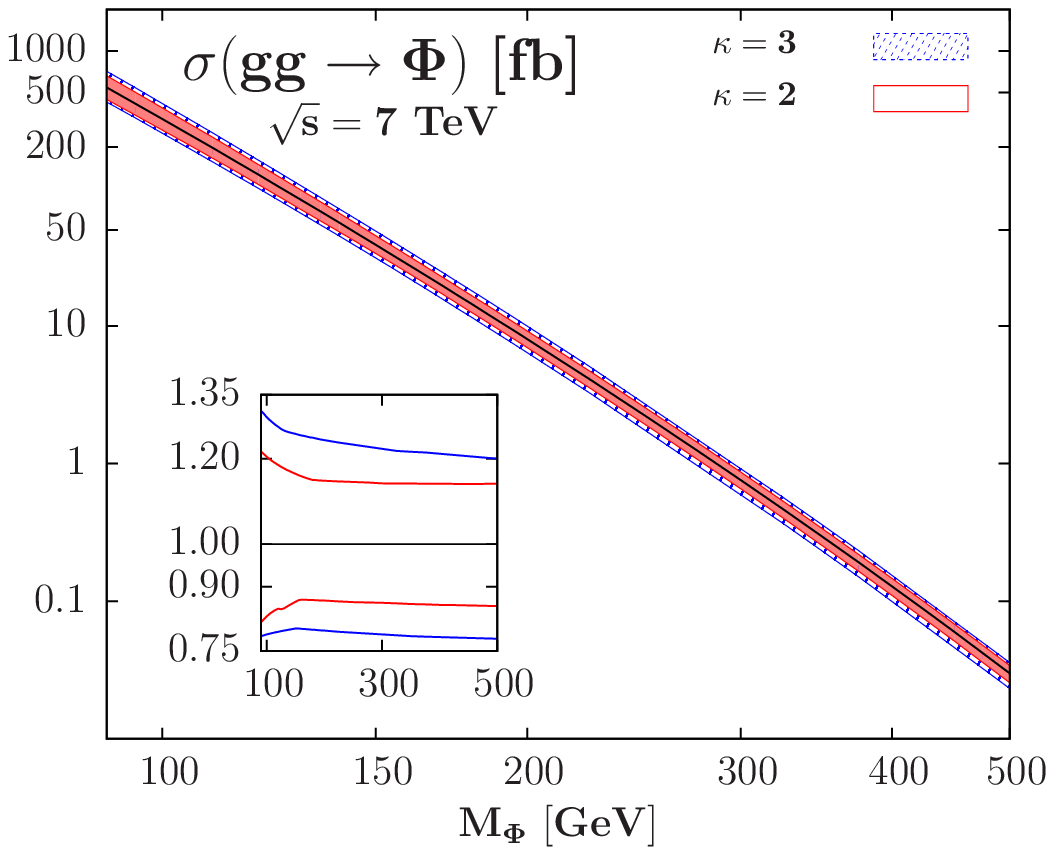,scale=0.7}
\epsfig{file=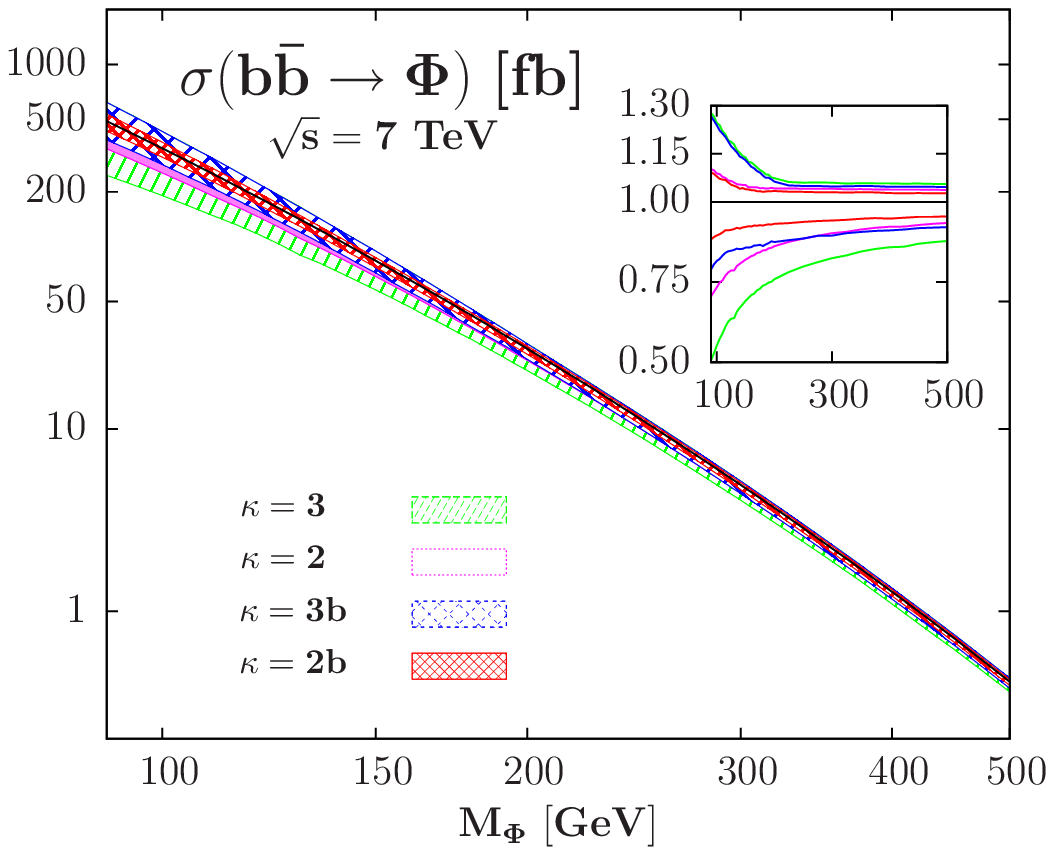,scale=0.7}
}
\end{center}
\vspace*{-4mm}
\caption[]{The scale uncertainty bands of the NLO $gg \to \Phi$ (left) and
the NNLO $b\bar b  \to \Phi$ (right) cross sections  at the $\lhc$ at 7 TeV 
as a function of $M_\Phi$; different values $\kappa=2,3$ are used and the 
results are shown when the additional constraint $1/\kappa \le \mu_R/\mu_F 
\le \kappa$ is imposed or not (marked as $\kappa$b). In the inserts, the relative deviations 
(compared to the central cross section values) are shown.}
\label{MSSM-scale}
\vspace*{-2mm}
\end{figure}

The results for $\sigma(gg\to \Phi)$ and $\sigma(b\bar b\to \Phi)$ at the  lHC,
for scale variations in the domains $ \mu_0/\kappa \le \mu_R,\mu_F  \le \kappa
\mu_0$ with $\kappa=2$ and 3 are shown in Fig.~\ref{MSSM-scale}  as a function
of $M_\Phi$. One can see that the scale variation is moderate for $gg \to \Phi$
with $\kappa=2$, despite of the fact that the process is known only at NLO,
leading to an uncertainty  of order $\pm 10\%$ in the entire  Higgs mass range.
Extending the variation domain to $\kappa=3$ will increase the uncertainty by
another $\approx 10\%$. As in the SM Higgs case, the maximal and minimal values
of the cross sections are approximately obtained for $\mu_R \approx \mu_F$ and
thus, varying independently the two scale does not affect the uncertainty.  
For $\kappa=2$ when the restriction $\frac12 \le \mu_R/\mu_F  \le 2$ is
imposed,  the uncertainty is also small in the case of  $b \bar b \to \Phi$,
$\approx \pm 10\%$ at low Higgs masses and less for higher masses; this was to
be expected as the process is evaluated at NNLO. However, the  extension to
$\kappa=3$,  while keeping the restriction $\frac13 \le \mu_R/\mu_F  \le 3$,
will significantly increase the uncertainty: one would have $-18\%,+24\%$ at
$M_\Phi=100$ GeV and $-13\%,+6\%$ at $M_\Phi=200$ GeV. If the restriction  on
$\mu_R/\mu_F$ is ignored the uncertainty blows up, especially for very
low Higgs masses:  one would have a variation of $-45\%,+25\%$ at $M_\Phi=100$
GeV, as also noticed in Ref.~\cite{LHCXS}.   

Hence, the cross section for $b \bar b \to \Phi$ is rather unstable against
scale variation and this justifies, a posteriori, the choice of a larger domain
of variation with $\kappa=3$ in this case, a choice that does not  appear  to
be a too extreme  one when looking at Fig.~\ref{MSSM-scale}.

\begin{figure}[!h]
\begin{bigcenter}
\vspace*{-1mm}
\mbox{
\epsfig{file=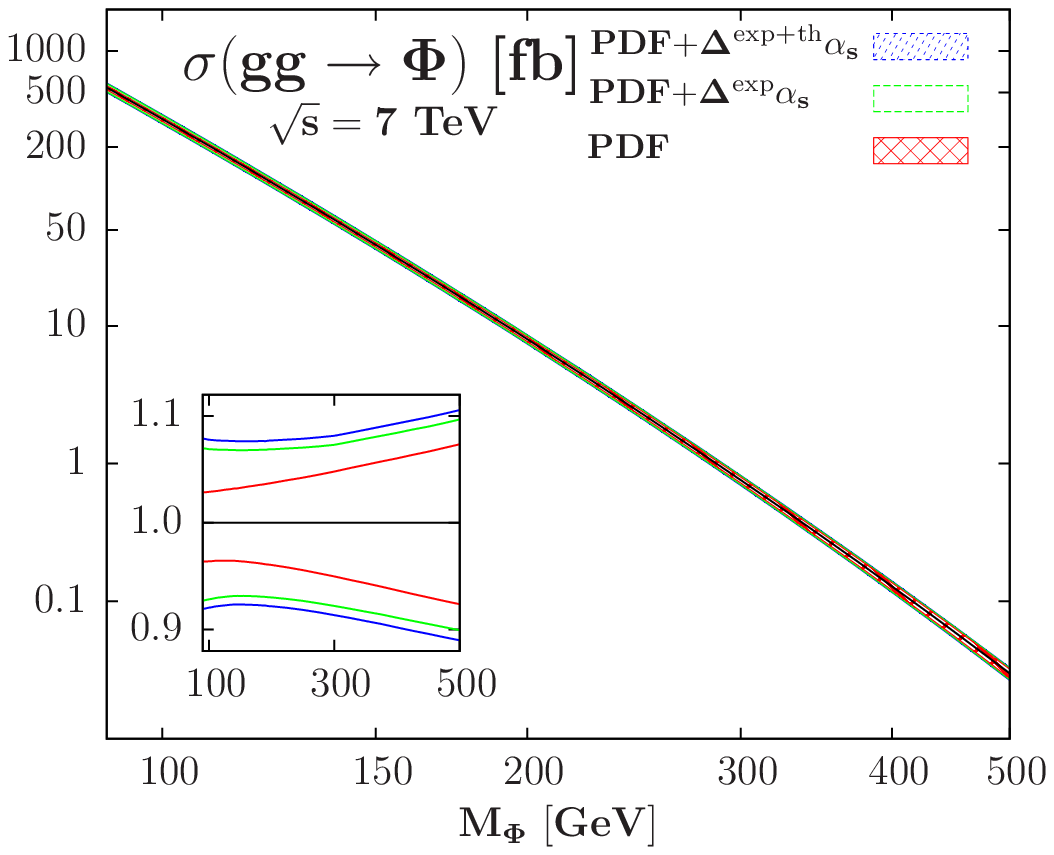,scale=0.7}\hspace*{-0.1cm}
\epsfig{file=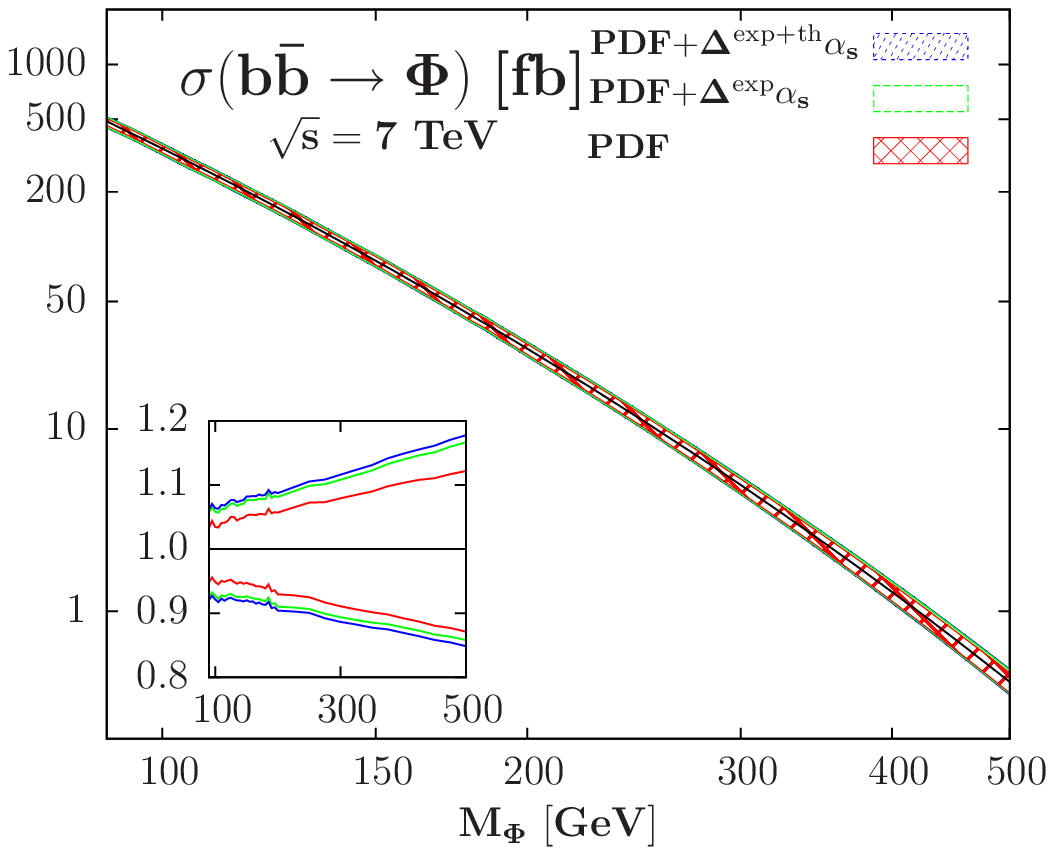,scale=0.7}
}
\end{bigcenter}
\vspace*{-4mm}
\caption[]{The PDF 90\% CL PDF, PDF+$\Delta^{\rm exp}
\alpha_s$ and PDF+$\Delta^{\rm exp}\alpha_s +\Delta^{\rm th}\alpha_s$ 
uncertainties in the MSTW scheme in the $gg \to \Phi$ (left) and
$b\bar b  \to \Phi$ (right) cross sections  at the $\lhc$ at 7 TeV
as a function of $M_\Phi$. In the inserts, the relative deviations are shown.}
\label{MSSM-PDF}
\end{figure}

Let us now turn to the estimation of the uncertainties from the parton 
densities and $\alpha_s$.   The 90\% CL PDF+$\Delta^{\rm exp}\alpha_s$ 
uncertainty, with  $\alpha_s(M_Z^2)=0.120 \pm 0.002$ at NLO for $gg \to \Phi$
and   $\alpha_s(M_Z^2)=0.1171 \pm 0.0014$ at NNLO for $b\bar b \to \Phi$,  is
evaluated within the MSTW parametrization when including the experimental
error on  $\alpha_s$. To that, we add in quadrature the effect of the
theoretical error on $\alpha_s$, estimated by the MSTW collaboration to be
$\Delta^{\rm th} \alpha_s \approx 0.003$ at NLO and $\Delta^{\rm th} \approx
0.002$ at NNLO, using the  MSTW fixed  $\alpha_s$ grid with central PDF sets. 
The 90\%CL PDF, PDF$+\Delta^{\rm exp}\alpha_s$ and the PDF$+\Delta^{\rm
exp+th}\alpha_s$ uncertainties at the $\lhc$ are shown in  Fig.~\ref{MSSM-PDF}
as a function of $M_\Phi$. In both the $gg\to \Phi$ and $b\bar b\to \Phi$
processes,  the total  uncertainty is below  $\pm 10\%$ for $M_\Phi \lsim 200$
GeV but increases at higher masses, in particular in the $b\bar b\to \Phi$ case.

\begin{figure}[!h]
\begin{bigcenter}
\vspace*{-1mm}
\mbox{
\epsfig{file=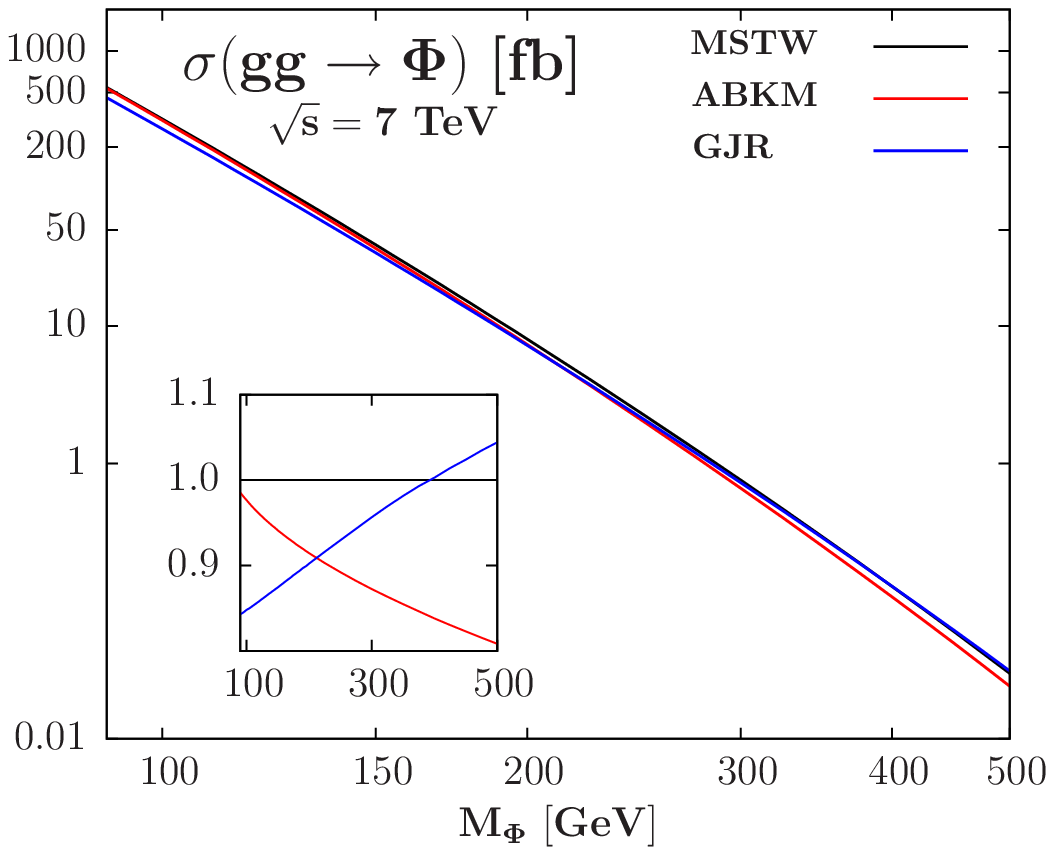,scale=0.7}\hspace*{-0.1cm}
\epsfig{file=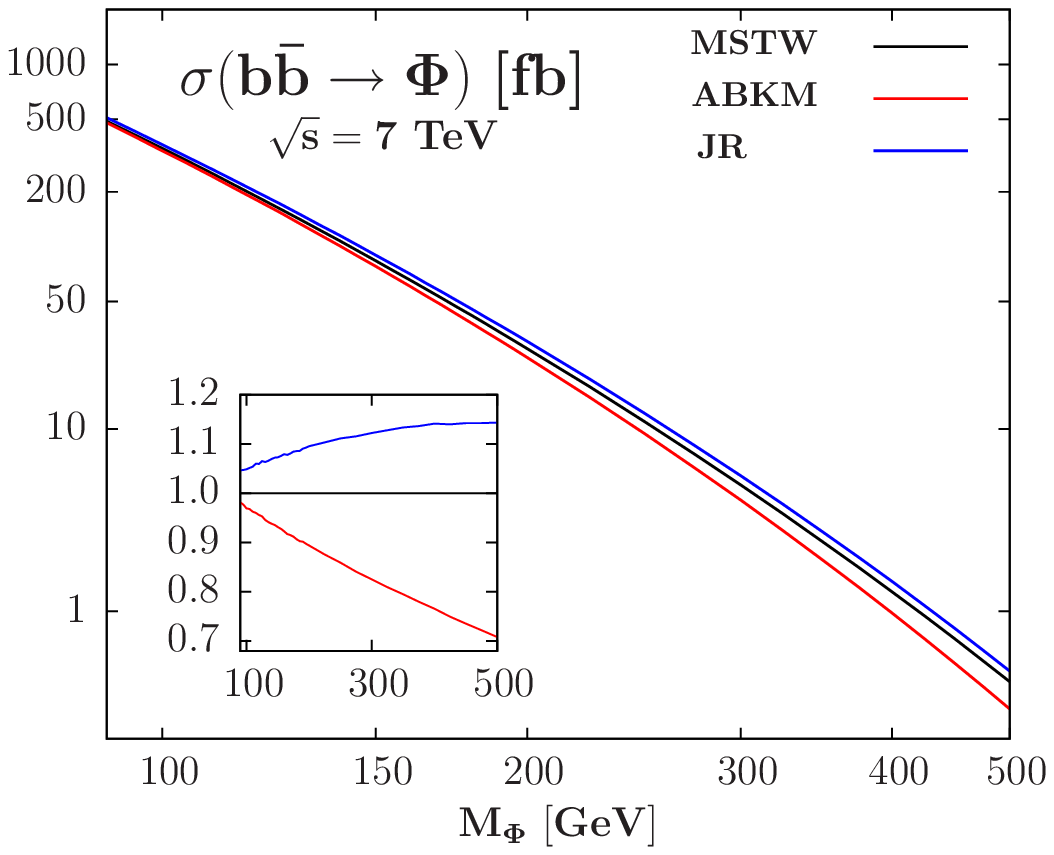,scale=0.7}
}
\end{bigcenter}
\vspace*{-4mm}
\caption[]{The cross sections  in the $gg \to \Phi$ (left) and $b\bar b  \to 
\Phi$ (right)   at the $\lhc$ at 7 TeV as a function of $M_\Phi$ evaluated
when using the ABKM and (G)JR PDF sets. In the inserts, the relative deviations 
from the value in the MSTW scheme are shown.}
\label{MSSM-PDF2}
\vspace*{-3mm}
\end{figure}

For completeness, we also display the two cross sections when ones adopts 
two other PDF sets, ABKM and (G)JR, and compare
the results with that of MSTW. As can be seen in Fig.~\ref{MSSM-PDF2}, the
deviations from the MSTW values are moderate in the case of $b\bar
b\to \Phi$ for the JR scheme: a few percent at $M_\Phi=100$ GeV,
increasing to $\approx 10\%$ at $M_\Phi=500$ GeV. The ABKM scheme
leads to substantial deviations at high masses, $\approx 30\%$ at
$M_\Phi = 500$ GeV. In the $gg$ fusion process, the prediction with the GJR PDF set at
$M_\Phi=100$ GeV is $\approx 20\%$ lower than in the MSTW and ABKM  cases which
give comparable results. The GJR and ABKM parameterizations cross at
$\approx 200$ GeV and at higher masses, the $gg$ cross sections with the
ABKM parametrization are $\approx 20\%$ lower than for MSTW results
while the predictions with the GJR set are $\approx 10\%$ higher than
for MSTW.

Finally, there is the effect of the uncertainty on the $b$--quark mass that had
only a marginal impact in the SM Higgs case but which will be rather important
here. In fact, there are three sources of uncertainties which can be attributed 
to the $b$--quark mass.

The first one  is of purely theoretical nature and is due to the choice of the
renormalization scheme for the $b$--quark mass. In our analysis, we have
adopted the $\overline{\rm  MS}$ renormalization scheme mainly because of the
fact that the calculation of the $b \bar b \to \Phi$ process is available only
in this scheme and we to have chosen to treat the same way both the $b\bar b
\to \Phi$ and  $gg \to \Phi$ channels. Nevertheless, one could choose another
renormalization scheme such as the on--shell scheme as it was discussed in the
case of the $b$--loop contribution in the $gg$ fusion process for SM Higgs
production. In the MSSM case, one could also adopt the $\overline{\rm  DR}$
scheme which appears to be more consistent  theoretically. To estimate this
scheme dependence, one can evaluate the difference of  the $gg\to \Phi$  cross
section  in the cases where the $b$--quark mass is defined in the on--shell and
in the $\overline{\rm  MS}$ schemes.  Using the program {\tt HIGLU}, one 
obtains a $\approx +15\%$ difference. Bearing in mind the fact that the
corrections could have been negative if we had adopted another scheme (as would
have been the case if we have used the $\overline{\rm  DR}$ scheme for
instance, although the difference from the  $\overline{\rm  MS}$ result would
have been at the level of a few percent only), one could assign  an error $
\Delta^{\rm scheme}_{m_b} \approx \pm 15\%$  to the $gg \to \Phi$ cross
section. This is the  procedure that we will adopt here. 

Another way to estimate the renormalization scheme dependence of the 
$b$--quark  mass,  would be to look at the differences that one obtains by using
$\overline{m}_b(\frac12 \overline{m}_b)$ and $\overline{m}_b(2\overline{m}_b)$
as inputs in the $gg\to \Phi$  cross section\footnote{We should note that the
scheme dependence actually appears also as a result  of the truncation of the
perturbative series which, in principle, should already be accounted for by the
scale variation. However,  the scales which enter in the $b$--quark mass, that
is defined  at $m_b$ itself, and in the rest of the $gg \to \Phi$ matrix
element, which is the Higgs mass or the scale $\mu_R$, are different.  We have
checked explictly that the scale uncertainties due to the variation of $\mu_R$
(and $\mu_F$)  in both schemes are comparable.  Adding this scheme dependence to
the scale variation, as we will do here, is similar in practice to increase the
domain of scale variation from the central scale while sticking to a given mass
renormalization scheme.}  \cite{MichaelT}. In this case, one obtains an
uncertainty that is in fact  much larger than the difference between the
on--shell and $\overline{\rm  MS}$ schemes and which goes both ways, 
$\Delta^{\rm scheme}_{m_b}  \approx - 20\%,+40\%$ for $M_\Phi=90$ GeV for
instance. 

In the   $b\bar b  \to \Phi$ process, we cannot perform this exercise  as the 
cross section, using the program {\tt bbh@nnlo} of Ref.~\cite{Robert}, can only
be evaluated  in the $\overline{\rm  MS}$ scheme with the Yukawa couplings
evaluated at the scale $\mu_R$. This is the reason why we extended the domain
of scale variation in this case to $\frac13 \mu_0 \le \mu_R,\mu_F \le 3\mu_0$.
The larger scale uncertainty obtained this way could be seen as indirectly
taking care of the scheme dependence.

The second source of uncertainty is of parametric nature and is the same  as
the one affecting the $H\to b\bar b$ partial decay width of the Higgs boson
discussed in section 3. It is estimated as previously, i.e. by evaluating the
maximal values of the cross sections when one includes the error on the input
$b$--quark mass at the scale $\overline{m}_b$, $\overline{m}_b (\overline{
m}_b)=4.19^{+0.18}_{-0.06}$  GeV,  and  in the case of the $b\bar b \to \Phi$
process where the Yukawa coupling is defined at the high scale, the strong 
coupling constant, $\alpha_s(M_Z^2)=0.1171 \pm 0.0014$ at NNLO, used to run the
mass  $\overline{m}_b (\bar m_b)$ upwards to  $\overline{m}_b (\mu_R)$. In the
considered Higgs mass range, one obtains an uncertainty of  $\Delta_{ m_b}^{\rm
input} \approx -4\%,+14\%$ and $\approx -3\%,+10\%$ in the case of, 
respectively, the $gg\to \Phi$ and $b\bar b\to \Phi$ processes (using the
central MSTW PDF set). The difference is mainly due to the fact that the bottom
quark masses in the two processes are not defined at the  same scale and, 
also, in the case of the $gg \to \Phi$ process, additional  corrections which
involve the $b$--quark mass, $\propto \log(\bar m_b^2/M_\Phi^2)$,  occur at
leading order. 

Finally, a third source of uncertainty originates from the choice of the
$b$--mass value in the $b$--quark densities. The MSTW collaboration has
recently released a set of PDFs with different bottom quark masses
\cite{MSTW-mb}: it  involves six different central PDFs with a range of 
on--shell $m_b$ values between $4.00$ GeV and $5.50$ GeV in $0.25$ GeV steps,
in addition to  the central best--fit with $m_b=4.75$ GeV. In order to
distinguish between the parametric uncertainty in $m_b$ and the one due to the
correlated PDF--$\Delta m_b$ uncertainty, we have chosen to calculate the
latter uncertainty by taking the minimal and maximal values of the production
cross sections when using the central value $m_b=4.75$ GeV and the two closest
ones upwards and downwards,  i.e.  $m_b=4.5$ GeV and $m_b=5$ GeV in the MSTW
PDF set\footnote{ Note that when  including the $1\sigma$ errors on
$\overline{m}_b(\overline{m}_b)$, while the upper value  corresponds to the pole mass
$m_b \approx 5$ GeV, the lower value does not correspond to $m_b=4.5$ GeV; we
will however adopt this smaller value  to estimate the uncertainty as no other
choice is possible within the MSTW set.}. However, we kept in the partonic
calculation the central value of $\overline{m}_b(\overline{m}_b)=4.19$ GeV which,  
approximately corresponds to the pole mass $m_b=4.75$ GeV. One obtains a
$\approx 3$--5\% uncertainty depending on the considered Higgs mass range. Note
that this uncertainty will not only affect the  cross section in the $bb\to
\Phi$ process in  which the $b$--densities play the major role, but  also the
one of the $gg\to \Phi$ channel; however, in this case, the change is below the
percent level and can be safely neglected.

\begin{figure}[!h]
\begin{center}
\vspace*{1mm}
\mbox{
\epsfig{file=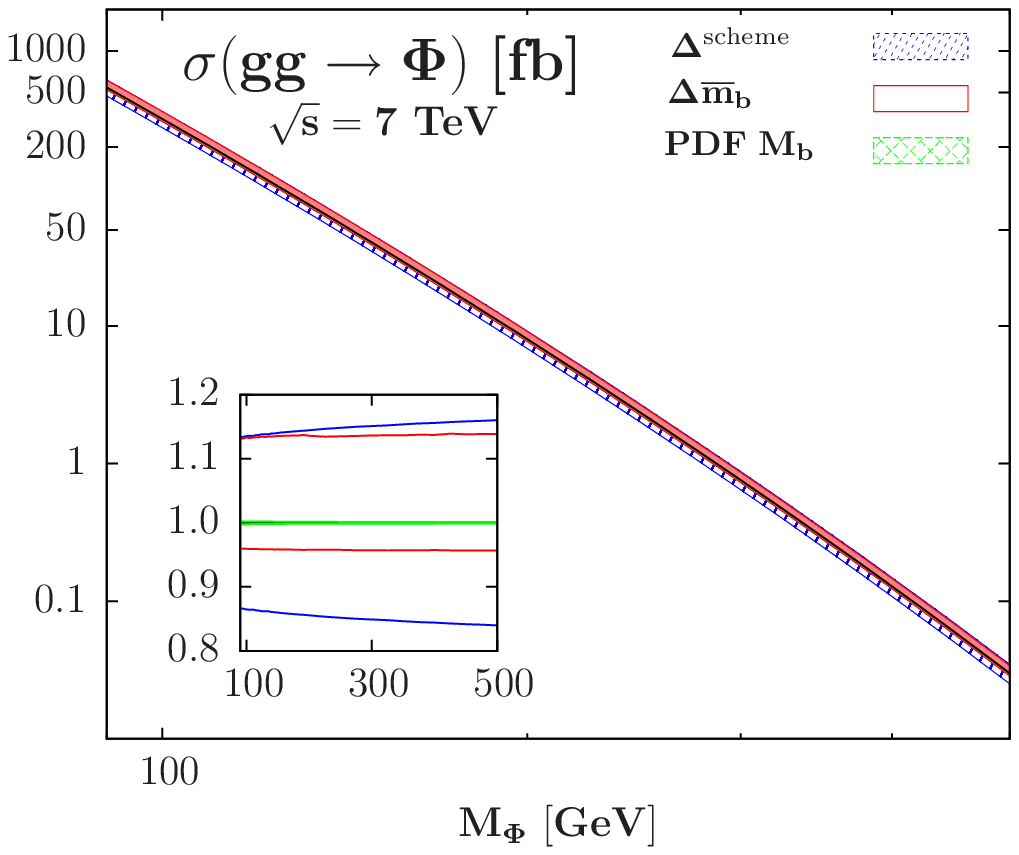,scale=0.7}
\epsfig{file=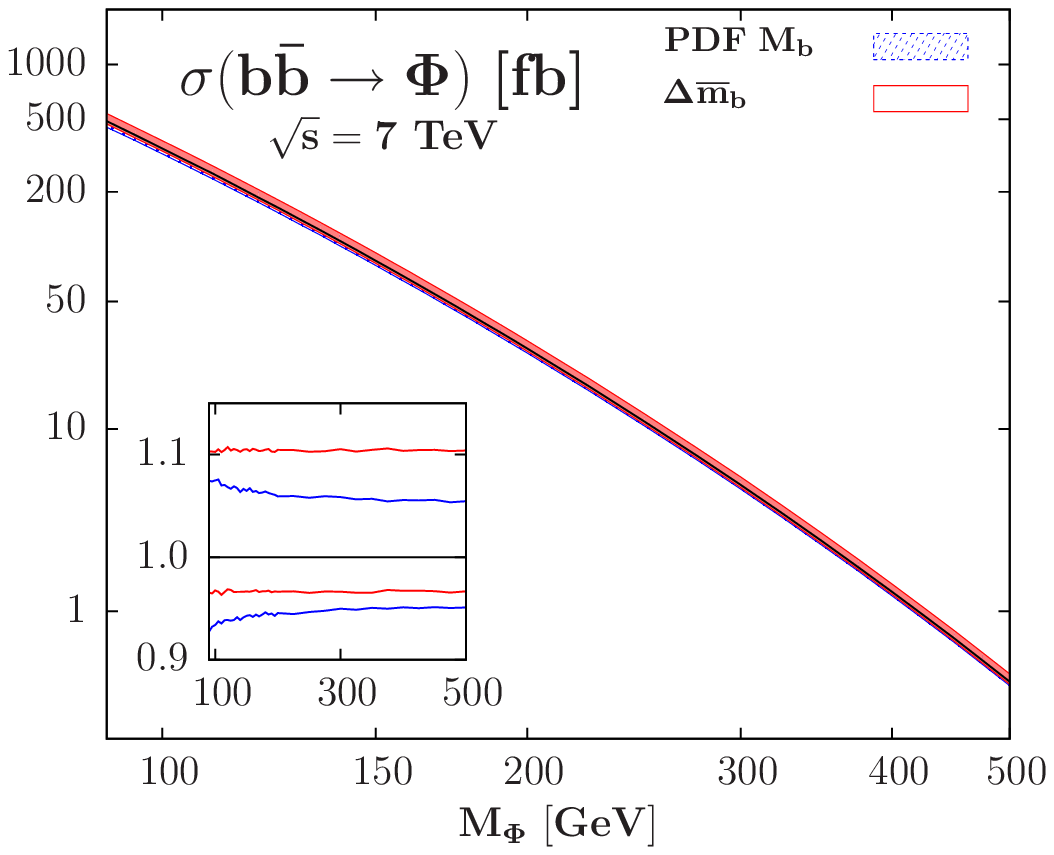,scale=0.7}
}
\end{center}
\vspace*{-4mm}
\caption[]{The scheme, parametric and PDF uncertainties due to the $b$-quark
mass in the $gg \to \Phi$ (left) and $b\bar b  \to \Phi$ (right) cross sections
at the $\lhc$ at 7 TeV as a function of $M_\Phi$. In the inserts, the relative 
deviations are shown.}
\label{MSSM-mb}
\vspace*{-3mm}
\end{figure}

The effect of these three sources of uncertainties is displayed  in
Fig.~\ref{MSSM-mb} for $gg\!\to\! \Phi$ and $b\bar b\!\to\! \Phi$ as a function
of $M_\Phi$. As can be seen, large uncertainties occur, in particular in $gg\!
\to\! \Phi$ where the $\approx 15\%$ scheme uncertainty that is absent in $b
\bar  b\! \to\! \Phi$ dominates.

\subsection{The Higgs decay branching fractions} 

In the most general case, the decay pattern of the MSSM Higgs particles can be
rather complicated, in particular for the heavy states. Indeed, besides the
standard decays into pairs of fermions and gauge bosons, the latter can have
mixed  decays into gauge and Higgs bosons (and the $H$ bosons can decay into
$hh$ states)  and, if some superparticles are light, SUSY decays would also
occur. However,  for the large values of $\tb$ that we are interested in here,
$\tb \gsim 10$, the couplings of the non--SM like Higgs particles to bottom
quarks and $\tau$ leptons are so strongly enhanced and those to top quarks and
gauge bosons    suppressed, that the pattern becomes very simple. To a very
good approximation, the $\Phi=A$ or $H(h)$  bosons  will  decay  almost
exclusively into $b\bar b$ and $\tau^+\tau^-$ pairs with branching ratios of,
respectively, $\approx 90\%$ and $ \approx 10\%$, with the $t\bar t$ decay
channel and the decay involving gauge or Higgs bosons suppressed to a level
where the branching ratios are less than 1\%. The CP--even  $h$ or $H$ boson,
depending on whether we are in the decoupling or anti--decoupling regime, will
have the same decays as the SM Higgs boson  in the mass range below $M_h^{\rm
max} \lsim 135$ GeV,  decays that have been discussed in section 3 (note that
for $M_h^{\rm max}  \sim 135$ GeV, we are in the regime where the decays into
$b \bar b$ and $WW$  are comparable and, thus, the uncertainties in the Higgs
branching ratios are the largest).

\begin{table}[!h]{\small%
\let\lbr\{\def\{{\char'173}%
\let\rbr\}\def\}{\char'175}%
\renewcommand{\arraystretch}{1.27}
\vspace*{1mm}
\begin{center}
\begin{tabular}{|c||cccc|cccc|}\hline
$\!M_\Phi\!$ &$\!b\bar b\!$   & $\!\Delta m_b\!$ &
$\!\Delta\alpha_s\!$ & ~tot~ & $\!\tau\tau\!$ & $\!\Delta m_b\!$ &
$\!\Delta\alpha_s\!$ & ~tot~ \\ \hline
$90$ & 90.40 & $^{+0.9\%}_{-0.3\%}$ &
$^{+0.2\%}_{-0.2\%}$ & $^{+0.9\%}_{-0.4\%}$ &
9.60 & $^{+3.2\%}_{-8.6\%}$ &
$^{+1.9\%}_{-1.8\%}$ & $^{+3.8\%}_{-8.8\%}$ \\ \hline
$100$ & 90.21 & $^{+0.9\%}_{-0.4\%}$ &
$^{+0.2\%}_{-0.2\%}$ & $^{+1.0\%}_{-0.4\%}$ &
9.79 & $^{+3.3\%}_{-8.6\%}$ &
$^{+1.9\%}_{-1.8\%}$ & $^{+3.8\%}_{-8.8\%}$ \\ \hline
$110$ & 90.04 & $^{+0.9\%}_{-0.4\%}$ &
$^{+0.2\%}_{-0.2\%}$ & $^{+1.0\%}_{-0.4\%}$ &
9.96 & $^{+3.2\%}_{-8.6\%}$ &
$^{+2.0\%}_{-1.9\%}$ & $^{+3.8\%}_{-8.8\%}$ \\ \hline
$120$ & 89.88 & $^{+1.0\%}_{-0.4\%}$ &
$^{+0.2\%}_{-0.2\%}$ & $^{+1.0\%}_{-0.4\%}$ &
10.12 & $^{+3.3\%}_{-8.6\%}$ &
$^{+2.0\%}_{-1.9\%}$ & $^{+3.8\%}_{-8.8\%}$ \\ \hline
$130$ & 89.74 & $^{+1.0\%}_{-0.4\%}$ &
$^{+0.2\%}_{-0.2\%}$ & $^{+1.0\%}_{-0.4\%}$ &
10.26 & $^{+3.2\%}_{-8.5\%}$ &
$^{+2.0\%}_{-1.9\%}$ & $^{+3.8\%}_{-8.7\%}$ \\ \hline
$140$ & 89.61 & $^{+1.0\%}_{-0.4\%}$ &
$^{+0.2\%}_{-0.2\%}$ & $^{+1.0\%}_{-0.4\%}$ &
10.39 & $^{+3.3\%}_{-8.5\%}$ &
$^{+2.0\%}_{-1.9\%}$ & $^{+3.8\%}_{-8.7\%}$ \\ \hline
$150$ & 89.48 & $^{+1.0\%}_{-0.4\%}$ &
$^{+0.2\%}_{-0.2\%}$ & $^{+1.0\%}_{-0.4\%}$ &
10.52 & $^{+3.1\%}_{-8.6\%}$ &
$^{+2.0\%}_{-2.0\%}$ & $^{+3.7\%}_{-8.8\%}$ \\ \hline
$160$ & 89.37 & $^{+1.0\%}_{-0.4\%}$ &
$^{+0.2\%}_{-0.2\%}$ & $^{+1.0\%}_{-0.5\%}$ &
10.63 & $^{+3.2\%}_{-8.5\%}$ &
$^{+2.1\%}_{-2.0\%}$ & $^{+3.8\%}_{-8.7\%}$ \\ \hline
$170$ & 89.26 & $^{+1.0\%}_{-0.4\%}$ &
$^{+0.2\%}_{-0.2\%}$ & $^{+1.1\%}_{-0.5\%}$ &
10.74 & $^{+3.2\%}_{-8.5\%}$ &
$^{+2.0\%}_{-2.0\%}$ & $^{+3.8\%}_{-8.8\%}$ \\ \hline
$180$ & 89.16 & $^{+1.0\%}_{-0.4\%}$ &
$^{+0.2\%}_{-0.3\%}$ & $^{+1.1\%}_{-0.5\%}$ &
10.84 & $^{+3.2\%}_{-8.5\%}$ &
$^{+2.1\%}_{-1.9\%}$ & $^{+3.9\%}_{-8.7\%}$ \\ \hline
$190$ & 89.05 & $^{+1.0\%}_{-0.4\%}$ &
$^{+0.2\%}_{-0.2\%}$ & $^{+1.1\%}_{-0.5\%}$ &
10.95 & $^{+3.2\%}_{-8.5\%}$ &
$^{+2.0\%}_{-2.0\%}$ & $^{+3.8\%}_{-8.7\%}$ \\ \hline
$200$ & 88.96 & $^{+1.0\%}_{-0.4\%}$ &
$^{+0.2\%}_{-0.3\%}$ & $^{+1.1\%}_{-0.5\%}$ &
11.04 & $^{+3.3\%}_{-8.4\%}$ &
$^{+2.2\%}_{-2.0\%}$ & $^{+3.9\%}_{-8.7\%}$ \\ \hline
$250$ & 88.53 & $^{+1.1\%}_{-0.4\%}$ &
$^{+0.3\%}_{-0.3\%}$ & $^{+1.1\%}_{-0.5\%}$ &
11.47 & $^{+3.1\%}_{-8.5\%}$ &
$^{+2.1\%}_{-2.1\%}$ & $^{+3.8\%}_{-8.7\%}$ \\ \hline
$300$ & 88.19 & $^{+1.1\%}_{-0.4\%}$ &
$^{+0.3\%}_{-0.3\%}$ & $^{+1.2\%}_{-0.5\%}$ &
11.81 & $^{+3.1\%}_{-8.5\%}$ &
$^{+2.1\%}_{-2.1\%}$ & $^{+3.8\%}_{-8.7\%}$ \\ \hline
$350$ & 87.90 & $^{+1.2\%}_{-0.4\%}$ &
$^{+0.3\%}_{-0.3\%}$ & $^{+1.2\%}_{-0.5\%}$ &
12.10 & $^{+3.1\%}_{-8.4\%}$ &
$^{+2.1\%}_{-2.1\%}$ & $^{+3.8\%}_{-8.7\%}$ \\ \hline
$400$ & 87.66 & $^{+1.2\%}_{-0.4\%}$ &
$^{+0.3\%}_{-0.3\%}$ & $^{+1.2\%}_{-0.5\%}$ &
12.34 & $^{+3.2\%}_{-8.3\%}$ &
$^{+2.2\%}_{-2.1\%}$ & $^{+3.8\%}_{-8.6\%}$ \\ \hline
$450$ & 87.44 & $^{+1.2\%}_{-0.4\%}$ &
$^{+0.3\%}_{-0.3\%}$ & $^{+1.2\%}_{-0.5\%}$ &
12.56 & $^{+3.1\%}_{-8.4\%}$ &
$^{+2.2\%}_{-2.1\%}$ & $^{+3.8\%}_{-8.6\%}$ \\ \hline
$500$ & 87.25 & $^{+1.2\%}_{-0.5\%}$ &
$^{+0.3\%}_{-0.3\%}$ & $^{+1.3\%}_{-0.6\%}$ &
12.75 & $^{+3.1\%}_{-8.3\%}$ &
$^{+2.3\%}_{-2.1\%}$ & $^{+3.9\%}_{-8.6\%}$ \\ \hline
\end{tabular} 
\end{center} 
\vspace*{-3mm}
\caption{The Higgs decay branching ratio into $b\bar b$ and $\tau^+\tau^-$ final
states (in \%) for given Higgs mass values (in GeV) with the corresponding 
individual uncertainties as well as the global uncertainties assuming 
$1\sigma$ uncertainties on the inputs 
$\overline{m}_b(\overline{m}_b)$ and $\alpha_s(M_Z^2)$. }
\label{BR-mssm}
\vspace*{-4mm}
}
\end{table}

For the evaluation of the theoretical uncertainties in the $b\bar b$ and $\tau^+
\tau^-$ decay branching ratios of the $\Phi$ states, the analysis of section 3
for  the SM Higgs boson can be straightforwardly extended to the MSSM case.
Here,  one can ignore the error on the input mass of the charm quark (as the 
decays into $c\bar c$ pairs are strongly suppressed) and consider only the
uncertainties  coming from the two other sources: the inputs
$\overline{m}_b(\overline{m}_b)$ and  $\alpha_s(M_Z^2)$. Again, the impact of a
scale variation  in the range $\frac12 M_\Phi \leq \mu \leq 2M_\Phi$ is
negligibly small. The uncertainties on the two branching ratios are displayed in
Table \ref{BR-mssm} for some values  of the Higgs boson mass, together with the
total uncertainties when  the individual uncertainties resulting from the
``1$\sigma$" errors on the inputs $\overline{m}_b(\overline{m}_b)$ and 
$\alpha_s(M_Z^2)$ are added in quadrature. 

The $b\bar b$ branching ratios of the $\Phi$ states, BR$(\Phi \to b\bar b)
\approx 3 \overline{m}_b^2(M_\Phi)/[3\overline{m}_b^2(M_\Phi)+m_\tau^2]$,
slightly decrease with  increasing $M_\Phi$  as a result of the higher scale
which reduces the $b$--quark mass $\overline{m}_b (M_\Phi)$, but the total
uncertainty  is practically constant and amounts to less than $\approx 3\%$ as a
consequence of the almost complete cancellation of the uncertainties in the
numerator and denominator. In contrast, there is  no such a cancellation in  the
branching fraction for Higgs decays into $\tau^+\tau^-$ pairs, BR$(\Phi \to
\tau^+ \tau^-) \approx m_\tau^2/[3 \overline{m}_b^2(M_\Phi)+m_\tau^2]$, and the
total uncertainty, that is dominated by  the error on the $b$--quark mass, 
reaches the level of 10\% for all Higgs masses\footnote{ We should note for
completeness, that the same results are obtained in the case of the charged
Higgs particles, where for $M_H^\pm \geq m_t+m_b \approx 175$ GeV and
$\tan\beta\gg 1$, one has to  consider only the two decay modes, $H^+ \to t \bar
b$ and $H^+ \to \tau^+ \bar \nu$ with again, BR$(H^+ \to t \bar b) \approx
3\overline{m}_b^2(M_{H^\pm})/[3\overline{m}_b^2 (M_{H^\pm}) +m_\tau^2]$ and
BR$(H^+ \to \tau \nu) \approx m_\tau^2/[3\overline{m}_b^2(M_{H^\pm})+m_\tau^2]$.
The values of the  branching ratios, together with the associated 
uncertainties,  are thus also those given in Table \ref{BR-mssm} when $M_\Phi$
is replaced by  $M_{H^\pm}$.}.  If the error on the input $b$--quark mass is
ignored, the total uncertainty  in the $\tau^+ \tau^-$ branching ratio will
reduce to $\approx 3\%$.

Finally, let us note that in the MSSM at high $\tb$, the total decay widths  of
the $\Phi$ particles should be taken into account. Indeed, they rise as $
\Gamma(\Phi) \propto M_\Phi \tan^2\beta$ and thus, reach the level of ${\cal
O}(10~{\rm GeV})$ for say,  $M_\Phi \approx 200$ GeV and $\tb \approx 50$.  The
total Higgs width  can thus  possibly be larger than  the experimental
resolution on the $\tau^+ \tau^-$ and $b\bar b$ invariant masses when decays
into these final states are analyzed.   If the total width has to be taken into
account in the experimental analyses, the uncertainties that affect it should
also be considered. These uncertainties  are in fact simply those affecting the
$\Phi \to b\bar b$ partial widths and thus, to a good approximation, the $\Phi
\to \tau^+ \tau^-$ branching ratio. The numbers given in Table \ref{BR-mssm} for
the uncertainties of BR$(\Phi \to \tau^+ \tau^-)$ thus correspond (when 
multiplied by a factor $\approx 1.1$) to the uncertainties on the total width
$\Gamma(\Phi)$ with a good accuracy.

\subsection{Combined uncertainties}

Let us come to the delicate issue of combining all the uncertainties
which come from the various sources. Here, we will adopt the procedure A  of
section 2.3 also used in Ref.~\cite{Hpaper}, in which the overall uncertainty  on
the production cross section  is obtained by applying the PDF+$\Delta^{\rm exp} 
\alpha_s$+$\Delta^{\rm th}  \alpha_s$ uncertainty  on the maximal and minimal
cross sections from the scale variation, adding linearly the scheme/EFT 
uncertainty. 

As the difference from the SM case is simply the presence of the additional
sources of uncertainties due to the $b$--quark mass, the combination problem 
reduces in fact to answering the question of  how one should add these three
$m_b$ uncertainties in the production cross sections. The uncertainty from the
scheme dependence in the $gg \to \Phi$ process should be simply, i.e. linearly,
added  to the  scale uncertainty as both emerge as a result of the truncation
of the perturbative series and are thus of pure theoretical nature.  The
uncertainties of the input $b$--quark mass which appears in the parton
densities should be added to the other PDF+$\alpha_s$ uncertainties. As this
uncertainty is of experimental origin, it should be combined in quadrature with
the  PDF+$\Delta^{\rm exp}\alpha_s +\Delta^{\rm th}\alpha_s$ uncertainty and 
since it is much smaller than the latter, it will have practically no impact on
the total PDF+$\alpha_s$ error\footnote{However, one can also view this
uncertainty as being due to the parametrization of the bottom--quark PDF which
is a theoretical problem, and thus add it linearly to the PDF+$\alpha_s$ so
that it increases slightly to the total PDF uncertainty. We will refrain from
doing so, though. Note that one may also consider an uncertainty related to
the charm--quark mass dependence of the PDFs. This can be estimated
again as in footnote 9 and we find at most $\simeq \pm2.6\%$ at $M_\Phi=500$
GeV in both channels. We then neglect this uncertainty added in
quadrature to that of the other PDF--related.}.  Finally, the  parametric  uncertainty, which has a special
status as it appears also in the Higgs  branching ratios, can be  simply
added linearly  to the combined scale+scheme+PDF uncertainty; we will see below
that it will have no impact in practice. 

\begin{figure}[!h]
\begin{center}
\vspace*{-.2mm}
\mbox{
\epsfig{file=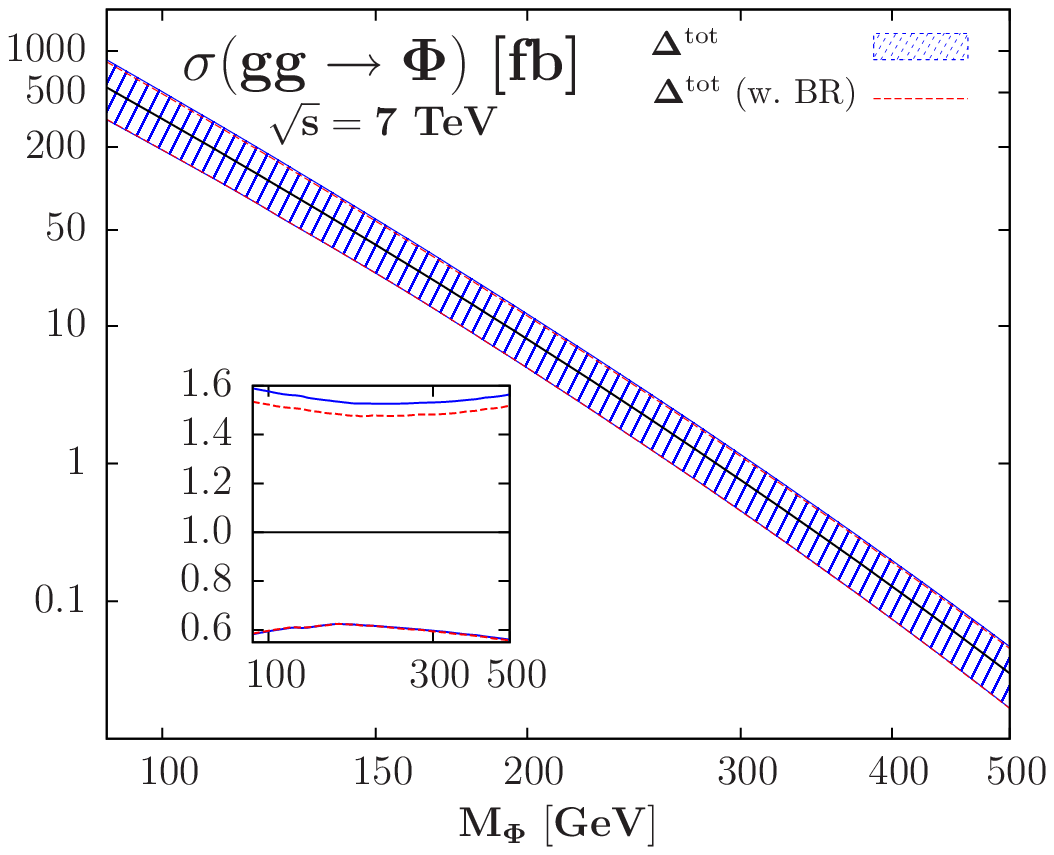,scale=0.7}
\epsfig{file=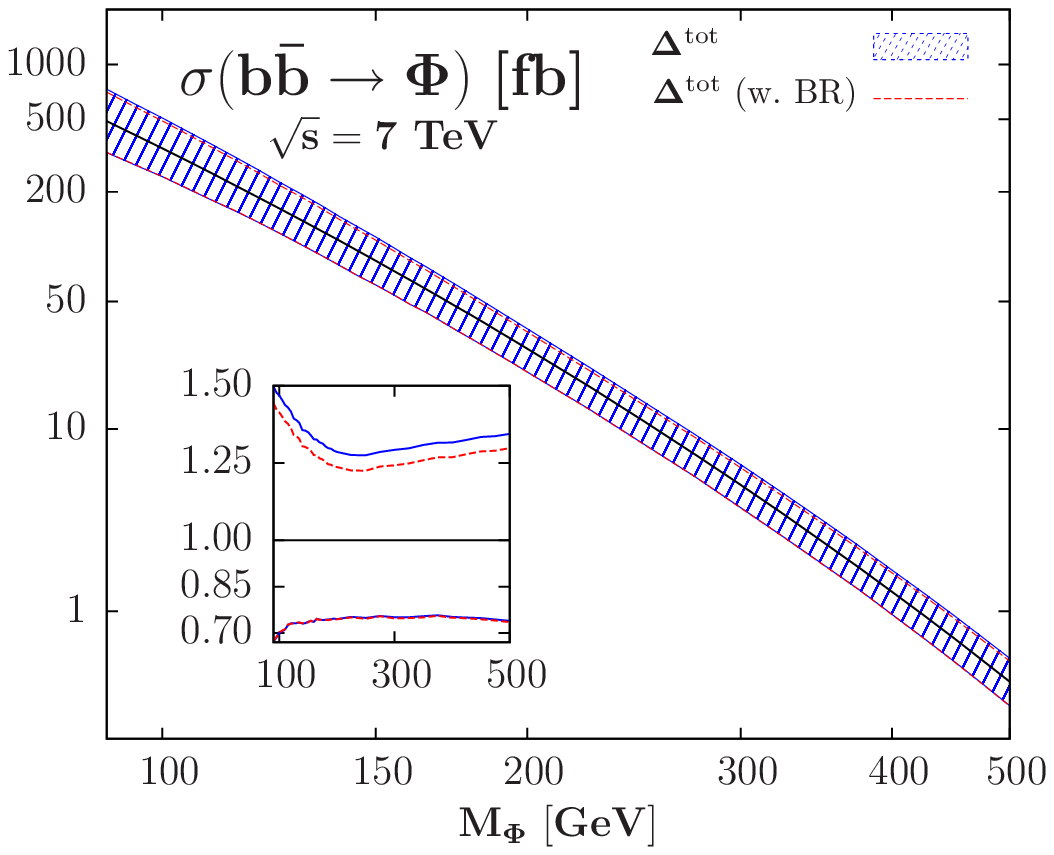,scale=0.7}
}
\end{center}
\vspace*{-4mm}
\caption[]{The total uncertainties due to the scale, PDF and $b$-quark
mass in the $gg \to \Phi$ (left) and $b\bar b  \to \Phi$ (right) cross sections
at the $\lhc$ with 7 TeV as a function of $M_\Phi$. The dotted lines show 
the uncertainties when those on the branching rations for Higgs decays 
into $\tau^+ \tau^-$ final states are added. In the inserts, the relative 
deviations are shown.}
\label{MSSM-all}
\vspace*{-6mm}
\end{figure}

Applying the procedure above for combining the uncertainties,  the results for
the processes $gg \to \Phi$ and $b\bar b\to \Phi$ at the $\lhc$  with  $\sqrt
s= 7$ TeV are displayed in Fig.~\ref{MSSM-all} as a function of $M_\Phi$;  the
numerical values can also be found  in the relevant columns of tables
8 and 9 where
the individual and total uncertainties are given.  One  finds a total
uncertainty of $\approx +60\%,-40\%$ for $\sigma(gg\to \Phi)$ and $\approx
+50\%,-35\%$ for $\sigma(b\bar b \to \Phi)$, in the low Higgs mass range and
slightly less at higher masses.   As mentioned previously, we expect that
these numbers approximately hold at least for slightly higher energies, $\sqrt
s=8$--10 TeV, and probably also at the full fledged LHC with $\sqrt s=14$ TeV.

Finally, the total uncertainty on the cross section times branching ratio, 
$\Delta ({\rm \sigma \times BR})$ is obtained by adding  the total 
uncertainties on the production cross sections and the uncertainties on the
branching fraction in Higgs decays into $\tau^+ \tau^-$ pairs, which is  the
most relevant detection channel.   In this addition, at least in the  $b\bar b
\to \Phi$ process where one defines the $\Phi b\bar b$ coupling at the  scale
$\mu_R$, the uncertainty on the input $b$--mass, which is common  to $\sigma$ 
and to BR, almost cancels out; only $\approx 10\%$ of the error is left 
out\footnote{This is also the case of the $\Delta_b$ SUSY correction which
enters both $\sigma (gg, b\bar b\to  \Phi)$ and BR$(\Phi \to \tau^+ \tau^-)$ and
which thus cancels in the product.   This justifies, a posteriori, why we
ignored this correction.}. In the case of $gg \to \Phi$ where the Yukawa
coupling is defined at the scale $m_b$ in contrast to the Higgs decay widths,
the errors on  $\alpha_s$ used for the running from $m_b$ to $M_\Phi$ will
induce a remaining uncertainty. This $\alpha_s$ uncertainty is correlated in
$\sigma(gg \to \Phi)$ and BR$(\Phi \to \tau^+ \tau^-)$:  smaller $\alpha_s$
values lower $\sigma(gg \to \Phi) \propto \alpha_s^2$ at LO, and also BR$(\Phi
\to \tau^+ \tau^-)$ as the resulting $\bar m_b(M_\Phi)$ is higher, thus
enhancing/reducing the $\Phi \to b \bar b/\tau^+ \tau^-$ rates. This uncertainty
should thus be added linearly to the overall scale+scheme+PDF uncertainty of the
cross section. 

The impact of this additional uncertainty is shown by the dotted lines of
Fig.~\ref{MSSM-all} where the uncertainty $\Delta ({\rm \sigma \times BR})$ for
tau decays is displayed for the two production processes. The impact is
negligible in the case of $b\bar b$ fusion and very modest in $gg$ fusion.


\begin{table}[!h]{\small%
\let\lbr{\def\{{\char'173}%
\let\rbr\}\def\}{\char'175}%
\renewcommand{\arraystretch}{1.2}
\vspace*{1mm}
\begin{bigcenter}
\begin{tabular}{|c|ccc|}\hline
$\!M_\Phi\!$ & $\sigma_{g g\to\Phi}^{\pm \Delta_{\mu}\pm \Delta^{\rm PDF}
\pm \Delta^{ \overline{m}_b}}$ & $\Delta^{\rm tot}$ & $\times\! {\rm BR}$ 
\\  \hline 
$90$ & $542.28^{+21.7\%~+7.9\%~+26.7\%}_{-18.3\%~-8.1\%~-17.6\%} $ & $^{+58.9\%}_{-41.7\%}$ & $^{+53.5\%}_{-41.5\%}$  \\ \hline
$95$ & $414.48^{+21.1\%~+7.8\%~+26.8\%}_{-17.6\%~-8.0\%~-17.7\%} $ & $^{+58.3\%}_{-41.1\%}$ & $^{+52.8\%}_{-40.9\%}$  \\ \hline
$100$ & $320.31^{+20.4\%~+7.8\%~+27.0\%}_{-16.9\%~-7.9\%~-17.8\%} $ & $^{+57.7\%}_{-40.5\%}$ & $^{+52.3\%}_{-40.3\%}$ \\ \hline
$105$ & $250.43^{+19.9\%~+7.8\%~+27.0\%}_{-16.3\%~-7.9\%~-17.9\%} $ & $^{+57.1\%}_{-40.0\%}$ & $^{+51.7\%}_{-39.8\%}$  \\ \hline
$110$ & $197.61^{+19.4\%~+7.7\%~+27.1\%}_{-15.9\%~-7.8\%~-17.9\%} $ & $^{+56.6\%}_{-39.6\%}$ & $^{+51.3\%}_{-39.4\%}$  \\ \hline
$115$ & $157.34^{+18.9\%~+7.7\%~+27.2\%}_{-15.4\%~-7.8\%~-18.0\%} $ & $^{+56.2\%}_{-39.2\%}$ & $^{+51.0\%}_{-39.0\%}$  \\ \hline
$120$ & $126.32^{+18.5\%~+7.7\%~+27.3\%}_{-15.1\%~-7.8\%~-18.0\%} $ & $^{+56.1\%}_{-39.0\%}$ & $^{+50.7\%}_{-38.9\%}$  \\ \hline
$125$ & $102.14^{+18.1\%~+7.7\%~+27.4\%}_{-15.3\%~-7.7\%~-18.1\%} $ & $^{+55.8\%}_{-39.2\%}$ & $^{+50.4\%}_{-39.1\%}$  \\ \hline
$130$ & $83.24^{+17.7\%~+7.7\%~+27.4\%}_{-15.1\%~-7.7\%~-18.1\%} $ & $^{+55.1\%}_{-39.0\%}$ & $^{+50.0\%}_{-38.9\%}$  \\ \hline
$135$ & $68.25^{+17.4\%~+7.7\%~+27.5\%}_{-14.7\%~-7.7\%~-18.2\%} $ & $^{+54.8\%}_{-38.7\%}$ & $^{+49.6\%}_{-38.7\%}$ \\ \hline
$140$ & $56.31^{+17.1\%~+7.7\%~+27.6\%}_{-14.2\%~-7.7\%~-18.2\%} $ & $^{+54.5\%}_{-38.4\%}$ & $^{+49.3\%}_{-38.3\%}$  \\ \hline
$145$ & $46.73^{+16.7\%~+7.6\%~+27.6\%}_{-13.8\%~-7.7\%~-18.3\%} $ & $^{+54.1\%}_{-38.1\%}$ & $^{+48.9\%}_{-38.2\%}$  \\ \hline
$150$ & $38.98^{+16.4\%~+7.6\%~+27.7\%}_{-13.5\%~-7.7\%~-18.4\%} $ & $^{+54.0\%}_{-37.8\%}$ & $^{+48.7\%}_{-37.9\%}$ \\ \hline
$155$ & $32.68^{+16.1\%~+7.6\%~+27.8\%}_{-13.1\%~-7.7\%~-18.4\%} $ & $^{+53.7\%}_{-37.6\%}$ & $^{+48.6\%}_{-37.5\%}$  \\ \hline
$160$ & $27.52^{+15.9\%~+7.6\%~+27.9\%}_{-13.0\%~-7.7\%~-18.5\%} $ & $^{+53.5\%}_{-37.5\%}$ & $^{+48.3\%}_{-37.5\%}$  \\ \hline
$165$ & $23.29^{+15.7\%~+7.6\%~+27.9\%}_{-13.0\%~-7.7\%~-18.5\%} $ & $^{+53.3\%}_{-37.6\%}$ & $^{+48.2\%}_{-37.5\%}$  \\ \hline
$170$ & $19.79^{+15.4\%~+7.6\%~+28.0\%}_{-13.0\%~-7.7\%~-18.6\%} $ & $^{+53.0\%}_{-37.7\%}$ & $^{+47.9\%}_{-37.7\%}$  \\ \hline
$175$ & $16.88^{+15.2\%~+7.6\%~+28.0\%}_{-13.1\%~-7.7\%~-18.6\%} $ & $^{+52.8\%}_{-37.8\%}$ & $^{+47.7\%}_{-37.7\%}$ \\ \hline
$180$ & $14.45^{+15.0\%~+7.7\%~+28.1\%}_{-13.1\%~-7.7\%~-18.6\%} $ & $^{+52.7\%}_{-37.8\%}$ & $^{+47.7\%}_{-37.8\%}$ \\ \hline
$185$ & $12.42^{+15.0\%~+7.7\%~+28.1\%}_{-13.1\%~-7.7\%~-18.7\%} $ & $^{+52.7\%}_{-38.0\%}$ & $^{+47.5\%}_{-38.2\%}$ \\ \hline
$190$ & $10.71^{+14.9\%~+7.6\%~+28.2\%}_{-13.2\%~-7.8\%~-18.8\%} $ & $^{+52.8\%}_{-38.1\%}$ & $^{+47.6\%}_{-38.2\%}$ \\ \hline
$195$ & $9.27^{+14.9\%~+7.7\%~+28.2\%}_{-13.2\%~-7.8\%~-18.8\%} $ & $^{+52.7\%}_{-38.2\%}$ & $^{+47.6\%}_{-38.3\%}$ \\ \hline
$200$ & $8.04^{+14.8\%~+7.7\%~+28.0\%}_{-13.2\%~-7.8\%~-18.7\%} $ & $^{+52.7\%}_{-38.3\%}$ & $^{+47.8\%}_{-38.3\%}$ \\ \hline
$225$ & $4.12^{+14.7\%~+7.8\%~+28.3\%}_{-13.4\%~-8.0\%~-19.0\%} $ & $^{+52.6\%}_{-38.8\%}$ & $^{+47.6\%}_{-39.0\%}$ \\ \hline
$250$ & $2.24^{+14.5\%~+7.9\%~+28.4\%}_{-13.6\%~-8.2\%~-19.2\%} $ & $^{+52.8\%}_{-39.3\%}$ & $^{+47.8\%}_{-39.6\%}$ \\ \hline
$275$ & $1.28^{+14.4\%~+8.0\%~+28.7\%}_{-13.7\%~-8.4\%~-19.4\%} $ & $^{+53.0\%}_{-39.8\%}$ & $^{+48.3\%}_{-39.9\%}$ \\ \hline
$300$ & $0.76^{+14.2\%~+8.2\%~+28.9\%}_{-13.8\%~-8.6\%~-19.5\%} $ & $^{+53.2\%}_{-40.3\%}$ & $^{+48.2\%}_{-40.6\%}$ \\ \hline
$325$ & $0.47^{+14.2\%~+8.4\%~+29.0\%}_{-13.9\%~-8.9\%~-19.6\%} $ & $^{+53.4\%}_{-40.7\%}$ & $^{+48.5\%}_{-41.0\%}$ \\ \hline
$350$ & $0.30^{+14.2\%~+8.7\%~+29.1\%}_{-14.0\%~-9.2\%~-19.8\%} $ & $^{+53.8\%}_{-41.3\%}$ & $^{+48.9\%}_{-41.7\%}$ \\ \hline
$375$ & $0.19^{+14.1\%~+9.0\%~+29.4\%}_{-14.1\%~-9.5\%~-19.9\%} $ & $^{+54.2\%}_{-41.8\%}$ & $^{+49.5\%}_{-41.9\%}$ \\ \hline
$400$ & $0.13^{+14.1\%~+9.3\%~+29.4\%}_{-14.2\%~-9.8\%~-20.0\%} $ & $^{+54.5\%}_{-42.1\%}$ & $^{+49.8\%}_{-42.4\%}$ \\ \hline
$425$ & $0.09^{+14.1\%~+9.6\%~+29.7\%}_{-14.3\%~-10.1\%~-20.2\%} $ & $^{+55.1\%}_{-42.6\%}$ & $^{+50.5\%}_{-42.9\%}$ \\ \hline
$450$ & $0.06^{+14.1\%~+9.9\%~+29.8\%}_{-14.4\%~-10.4\%~-20.3\%} $ & $^{+55.4\%}_{-43.1\%}$ & $^{+50.8\%}_{-43.5\%}$ \\ \hline
$500$ & $0.03^{+14.2\%~+10.6\%~+30.0\%}_{-14.5\%~-11.0\%~-20.5\%} $ & $^{+56.5\%}_{-44.0\%}$ & $^{+51.9\%}_{-44.3\%}$ \\ \hline
\end{tabular} 
\end{bigcenter} 
\caption{The Higgs production cross sections in the $gg\to \Phi$ channel 
(for $\tan\beta=1$) as well as as the individual
uncertainties (first from scale, then from PDF+$\Delta^{\rm \exp+th}\alpha_s$ 
at 90\%CL and from the input mass $\overline{m}_b$ at 1$\sigma$) and the total
uncertainties for selected values of the Higgs mass. The last column
displays the total uncertainty when including the combination with the
total uncertainty on $\Phi \to \tau^+\tau^-$ branching ratio.}
\vspace*{-1mm}
}
\end{table}

\begin{table}[!h]{\small%
\let\lbr{\def\{{\char'173}%
\let\rbr\}\def\}{\char'175}%
\renewcommand{\arraystretch}{1.2}
\vspace*{1mm}
\begin{bigcenter}
\begin{tabular}{|c|ccc|}\hline
$\!M_\Phi\!$ & $\sigma_{b \bar b\to\Phi}^{\pm \Delta_{\mu}\pm \Delta^{\rm PDF}
\pm \Delta^{\overline{m}_b}}$ & $\Delta^{\rm tot}$ & $\times\! {\rm BR}$  \\ \hline 
$90$ & $487.83^{+26.8\%~+6.2\%~+17.8\%}_{-21.1\%~-8.0\%~-10.7\%} $ & $^{+49.8\%}_{-32.8\%}$ & $^{+44.3\%}_{-32.5\%}$  \\ \hline 
$95$ & $409.59^{+25.4\%~+7.1\%~+17.7\%}_{-19.5\%~-7.3\%~-10.3\%} $ & $^{+47.8\%}_{-31.5\%}$ & $^{+42.4\%}_{-31.3\%}$  \\ \hline 
$100$ & $346.50^{+23.9\%~+6.4\%~+17.8\%}_{-18.4\%~-7.8\%~-9.9\%} $ & $^{+46.9\%}_{-29.9\%}$ & $^{+41.4\%}_{-29.8\%}$  \\ \hline 
$105$ & $294.57^{+22.2\%~+6.3\%~+18.1\%}_{-17.8\%~-8.3\%~-9.6\%} $ & $^{+45.4\%}_{-29.5\%}$ & $^{+40.0\%}_{-29.3\%}$ \\ \hline 
$110$ & $252.03^{+21.3\%~+6.9\%~+17.2\%}_{-16.7\%~-7.7\%~-10.1\%} $ & $^{+43.8\%}_{-28.9\%}$ & $^{+38.5\%}_{-28.8\%}$  \\ \hline 
$115$ & $215.89^{+20.3\%~+6.8\%~+17.5\%}_{-15.9\%~-8.0\%~-9.5\%} $ & $^{+43.0\%}_{-27.5\%}$ & $^{+37.7\%}_{-27.3\%}$ \\ \hline 
$120$ & $186.56^{+18.8\%~+7.2\%~+17.6\%}_{-15.4\%~-7.7\%~-9.3\%} $ & $^{+41.7\%}_{-26.9\%}$ & $^{+36.4\%}_{-27.0\%}$  \\ \hline 
$125$ & $161.84^{+17.4\%~+7.7\%~+17.1\%}_{-15.2\%~-7.6\%~-9.4\%} $ & $^{+39.4\%}_{-26.9\%}$ & $^{+34.1\%}_{-26.8\%}$  \\ \hline 
$130$ & $140.97^{+16.4\%~+7.6\%~+17.5\%}_{-14.9\%~-7.8\%~-9.5\%} $ & $^{+38.8\%}_{-26.8\%}$ & $^{+33.7\%}_{-26.7\%}$  \\ \hline 
$135$ & $122.98^{+15.5\%~+7.3\%~+17.2\%}_{-14.6\%~-8.0\%~-9.2\%} $ & $^{+38.0\%}_{-26.6\%}$ & $^{+32.8\%}_{-26.6\%}$  \\ \hline 
$140$ & $108.08^{+14.1\%~+7.5\%~+16.6\%}_{-14.7\%~-8.0\%~-9.5\%} $ & $^{+35.6\%}_{-26.9\%}$ & $^{+30.5\%}_{-26.8\%}$  \\ \hline 
$145$ & $94.85^{+13.7\%~+7.7\%~+17.1\%}_{-14.7\%~-8.2\%~-9.2\%} $ & $^{+35.5\%}_{-26.9\%}$ & $^{+30.2\%}_{-27.0\%}$  \\ \hline 
$150$ & $83.83^{+13.0\%~+8.2\%~+17.0\%}_{-14.3\%~-8.0\%~-9.1\%} $ & $^{+35.0\%}_{-26.4\%}$ & $^{+29.7\%}_{-26.5\%}$ \\ \hline 
$155$ & $74.13^{+12.0\%~+8.2\%~+17.1\%}_{-14.2\%~-8.2\%~-9.0\%} $ & $^{+34.0\%}_{-26.2\%}$ & $^{+28.9\%}_{-26.1\%}$  \\ \hline 
$160$ & $65.84^{+11.0\%~+8.3\%~+16.9\%}_{-14.1\%~-8.2\%~-9.1\%} $ & $^{+32.7\%}_{-26.4\%}$ & $^{+27.6\%}_{-26.5\%}$ \\ \hline 
$165$ & $58.60^{+10.4\%~+8.2\%~+16.9\%}_{-13.6\%~-8.5\%~-8.8\%} $ & $^{+32.6\%}_{-25.4\%}$ & $^{+27.5\%}_{-25.5\%}$  \\ \hline 
$170$ & $52.28^{+9.6\%~+8.5\%~+16.6\%}_{-13.6\%~-8.3\%~-9.2\%} $ & $^{+31.7\%}_{-25.7\%}$ & $^{+26.5\%}_{-25.7\%}$  \\ \hline
$175$ & $46.75^{+9.1\%~+8.4\%~+16.7\%}_{-13.4\%~-8.6\%~-8.9\%} $ & $^{+31.3\%}_{-25.8\%}$ & $^{+26.2\%}_{-25.8\%}$ \\ \hline 
$180$ & $41.97^{+8.3\%~+8.5\%~+16.9\%}_{-13.8\%~-8.7\%~-8.7\%} $ & $^{+30.4\%}_{-25.7\%}$ & $^{+25.4\%}_{-25.7\%}$ \\ \hline 
$185$ & $37.69^{+7.5\%~+9.2\%~+16.8\%}_{-13.3\%~-8.2\%~-8.8\%} $ & $^{+29.8\%}_{-25.6\%}$ & $^{+24.6\%}_{-25.8\%}$ \\ \hline 
$190$ & $33.87^{+7.5\%~+8.6\%~+16.5\%}_{-12.8\%~-9.2\%~-8.9\%} $ & $^{+29.6\%}_{-25.5\%}$ & $^{+24.4\%}_{-25.6\%}$ \\ \hline 
$195$ & $30.55^{+7.0\%~+8.9\%~+16.4\%}_{-12.3\%~-9.1\%~-9.1\%} $ & $^{+29.0\%}_{-25.4\%}$ & $^{+23.9\%}_{-25.5\%}$ \\ \hline 
$200$ & $27.61^{+6.3\%~+8.7\%~+16.4\%}_{-12.5\%~-9.6\%~-8.8\%} $ & $^{+28.6\%}_{-25.5\%}$ & $^{+23.7\%}_{-25.6\%}$ \\ \hline 
$225$ & $17.10^{+5.1\%~+9.6\%~+16.4\%}_{-11.9\%~-9.7\%~-8.9\%} $ & $^{+27.6\%}_{-24.8\%}$ & $^{+22.6\%}_{-25.0\%}$ \\ \hline 
$250$ & $10.97^{+5.0\%~+10.6\%~+16.1\%}_{-11.5\%~-10.0\%~-8.8\%} $ & $^{+27.5\%}_{-25.1\%}$ & $^{+22.5\%}_{-25.4\%}$ \\ \hline 
$275$ & $7.27^{+5.0\%~+10.9\%~+16.3\%}_{-10.6\%~-10.8\%~-8.6\%} $ & $^{+28.6\%}_{-24.5\%}$ & $^{+23.9\%}_{-24.6\%}$ \\ \hline 
$300$ & $4.94^{+4.8\%~+11.6\%~+16.4\%}_{-10.5\%~-11.4\%~-8.4\%} $ & $^{+29.3\%}_{-24.9\%}$ & $^{+24.3\%}_{-25.3\%}$ \\ \hline 
$325$ & $3.43^{+4.8\%~+12.4\%~+15.9\%}_{-9.8\%~-11.8\%~-8.6\%} $ & $^{+29.7\%}_{-24.9\%}$ & $^{+24.9\%}_{-25.2\%}$ \\ \hline 
$350$ & $2.43^{+5.0\%~+13.1\%~+16.2\%}_{-9.4\%~-12.3\%~-8.4\%} $ & $^{+30.8\%}_{-24.6\%}$ & $^{+25.9\%}_{-25.0\%}$ \\ \hline 
$375$ & $1.75^{+4.9\%~+14.1\%~+16.0\%}_{-9.1\%~-12.5\%~-8.2\%} $ & $^{+31.6\%}_{-24.3\%}$ & $^{+26.9\%}_{-24.5\%}$ \\ \hline 
$400$ & $1.28^{+4.7\%~+14.9\%~+15.9\%}_{-8.9\%~-13.1\%~-8.2\%} $ & $^{+31.6\%}_{-24.8\%}$ & $^{+26.9\%}_{-25.1\%}$ \\ \hline 
$425$ & $0.95^{+4.7\%~+15.5\%~+16.0\%}_{-8.6\%~-13.6\%~-8.3\%} $ & $^{+32.5\%}_{-25.0\%}$ & $^{+27.8\%}_{-25.3\%}$ \\ \hline 
$450$ & $0.71^{+4.7\%~+16.1\%~+16.1\%}_{-8.2\%~-14.2\%~-8.2\%} $ & $^{+33.4\%}_{-25.2\%}$ & $^{+28.8\%}_{-25.6\%}$ \\ \hline 
$500$ & $0.41^{+4.7\%~+17.8\%~+15.8\%}_{-7.9\%~-15.1\%~-8.2\%} $ & $^{+34.4\%}_{-26.1\%}$ & $^{+29.8\%}_{-26.5\%}$ \\ \hline

\end{tabular} 
\end{bigcenter} 
\vspace*{-2mm}
\caption{The same as in Table.~8 but for  the $b\bar b\to \Phi$ channel.} 
\vspace*{-1mm}
}
\end{table}

\clearpage


Thus, the theoretical uncertainties in the $b\bar b , gg\to \Phi \to \tau^+
\tau^-$ production and decay channels for the MSSM CP--odd and one of the
CP--even Higgs particles are extremely large. This has important consequences on
the MSSM  $[\tan\beta, M_A]$ parameter space that can be probed at the lHC. In
particular, if no Higgs boson is observed and only exclusion limits on the
parameter space can be imposed, these uncertainties which can lower the
production cross sections by a factor of two will have a significant impact. For
instance,  as the cross sections grow as $\tan^2\beta$, the values of $\tb$
which can be ruled out for a given $A$ boson mass should be multiplied by a
factor $\approx \sqrt 2$ in view of these large uncertainties.

\section{Conclusions}

In the first part of this paper, we have performed a detailed analysis for 
Standard Model Higgs production at the $\lhc$ with $\sqrt s=7$ TeV in the 
dominant  gluon--gluon fusion channel, $g g \to H$. We have first updated the 
production cross sections,  including the relevant higher order corrections in
perturbation theory and then, discussed the  theoretical uncertainties that
affect the predictions:  the scale uncertainties that are usually viewed as an
estimate of the unknown higher order corrections, the uncertainties due to the
parton  distribution functions and the strong coupling constant, and finally,
the uncertainties in using an effective field theory approach in calculating 
the higher order corrections beyond next--to--leading perturbative order. 

We find that the scale uncertainties, estimated by varying the renormalization 
and factorization scales in the interval $\frac12 \mu_0 \leq \mu_R,\mu_F\leq 2
\mu_0$ with $\mu_0=\frac12 M_H$ being the central scale, are moderate, $\approx 
\pm 12\%$ in the low mass range $M_H \approx 120$ GeV, and $\approx  +5\%, -8\%$
in the high mass range, $M_H \approx 500$ GeV. For the PDF uncertainty, we have
considered the four NNLO PDF sets that are available, MSTW, ABKM, JR and
HERAPDF, and shown that there is a sizable spread in the predictions of the
cross section. This spread could be partly accounted for by taking into account
the uncertainties, both experimental and theoretical, on the value of
$\alpha_s$. The evaluation in the MSTW scheme of  the correlated
PDF+$\Delta^{\rm exp + th}\alpha_s$ uncertainties in the $gg \to H$ cross
section leads to a $\approx \pm 10\%$ uncertainty in the entire Higgs  mass
range that is relevant at the lHC, $M_H \lsim 500$ GeV. A third source  of
uncertainties  is due to the use of the approximation of an infinite mass for
the particles running in the  $gg\to H$ loop when calculating the amplitude
beyond NLO, i.e.  for the NNLO QCD and the mixed QCD--electroweak  corrections;
an additional uncertainty arises from the renormalization of the bottom quark
mass. These uncertainties are quite small when taken individually, at
most a few percent each, but they add up to a non--negligible amount, 3 to 7\%
depending on the Higgs mass range and should be thus included.  

We have then addressed the issue of combining these three sources of
uncertainties. Arguing that the PDF uncertainty should be viewed as a pure
theoretical one, we have proposed three procedures for combining it 
with those due the scale and the EFT/scheme, the simplest one being a  linear
addition. All three procedures lead to approximately the same total uncertainty
on the $gg\to H$ cross section at the lHC, $\approx -25\%, +30\%$ in the low
mass range $M_H \lsim 200$ GeV  and $\approx \pm 20\%$ at higher Higgs masses.
The overall uncertainty at the lHC is  significantly smaller than the one
affecting the cross section at Tevatron energies. This a mere consequence of
the smaller scale uncertainty, as the  QCD corrections are more moderate at
$\lhc$ energies, and the fact that, at the Tevatron, one is probing the high
Bjorken--$x$ regime  for the gluon densities which is more uncertain.

We have extended our analysis to other possible energies: $\sqrt s=8$--10 TeV
that are planed to be explored after the $\sqrt s= 7$ TeV run and the LHC
design energy of $\sqrt s=14$ TeV. We provided numerical results for the $gg
\to H$ cross section and in the case of the LHC, we also estimated the
associated theoretical uncertainties. The later turned out to be comparable to
those obtained at 7 TeV and these uncertainties are, thus,  expected to be also
the same at the intermediate energies. 

In a second part of the paper, we have addressed the issue of the Higgs decay 
branching ratios and the uncertainties that affect them, namely, the parametric
ones due the errors in the determination of the bottom and charm quark masses
and the QCD coupling $\alpha_s$. We obtain uncertainties that are significant
in the critical intermediate Higgs mass range $120\; {\rm GeV} \lsim M_H \lsim
150$ GeV where the main Higgs decay channels $H\to b\bar b$ and $H\to WW$ have
comparable rates.  These uncertainties can reach the level of $\approx \pm
3$--10\% in the channels  $H\to WW,ZZ, \gamma \gamma$ used to detect the Higgs
particle  at the lHC; this is also the case for the decays $H\to b\bar b$ and
$\tau^+ \tau^-$  which are also being considered in Tevatron searches. One
should therefore also take into account these uncertainties  which, up to now,
have been overlooked by the experimental collaborations not only at the LHC but
also at the Tevatron.  

Finally, in a third part of the paper, we have extended our analysis of Higgs
production at the lHC to consider the case of the MSSM neutral Higgs
particles.  We have investigated the production of the CP--odd like particles,
$\Phi=A$ as well as one of the CP--even $H$ or $h$ particles (depending on
whether we are in the decoupling or anti--decoupling regimes) that are
degenerate in mass with  $A$ and have the same couplings,  in particular,
enhanced couplings to bottom quarks for the high values of $\tb$ than can be 
probed at the lHC. The two main production processes $gg \to \Phi$ and $b \bar
b \to \Phi$, which could have cross sections that are order of magnitude larger
than in the SM Higgs case, have been considered. Numerical  results at energies
$\sqrt s=7$--14 TeV  have been given and the associated theoretical
uncertainties have been evaluated. The latter are due not only to the scale and
PDF+$\alpha_s$ uncertainties which appear in the SM case, but there are also
uncertainties associated to the $b$--quark which plays a major role in the
MSSM. The error on the input $b$-quark mass, the scheme dependence in  the
renormalization of the mass of the $b$--quark in its contribution to the $gg
\to \Phi$ amplitude and the effect of $m_b$ on the bottom quark densities in 
the process $b\bar b \to \Phi$ will induce additional uncertainties. To these, 
one has to add the parametric uncertainties in the  branching ratio of the
Higgs decay in tau lepton pairs, which is the cleanest detection channel at
hadron colliders, that is found to be at the level of 10\%. 

The overall theoretical uncertainty in these two processes turn out to be 
extremely large at lHC energies, of the order of 50\%. This large uncertainty
will have a significant impact on the MSSM parameter space that can be  probed.
This is particularly true in the case where no MSSM Higgs  signal is observed
and only exclusion bounds can be derived by comparing the experimental data with
the predicted Higgs production cross sections. As the cross sections in both
the  $gg \to \Phi$ and $b\bar b \to \Phi$ processes increase with $\tan^2\beta$,
the  values of $\tb$ which could be excluded in the absence of a signal will be 
smaller by a factor $\approx \sqrt 2$ if these theoretical uncertainties are
taken into account\footnote{This is also the case for the exclusion bounds that
have been  obtained by the CDF and D0 collaborations from negative MSSM Higgs
searches at  the Tevatron \cite{Tevatron-MSSM}. See Ref.~\cite{Julien} for an
analysis of this issue.}.\bigskip

{\bf Acknowledgments:}  Illuminating discussions  with Rohini Godbole on the
PDFs and  with Michael Spira on many aspects of this work are  gratefully
acknowledged.   We  would also like to thank the members of the LHC Higgs cross
section Working Group, in particular C. Mariotti, M. Grazzini and G. Passarino,
for discussions.    This work is supported by the European Network HEPTOOLS. 

\subsection*{A1. Addendum: SM cross sections and uncertainties at
$\mathbf{\sqrt s=8}$ TeV}

\begin{figure}[!t]
\begin{center}
\includegraphics[scale=0.75]{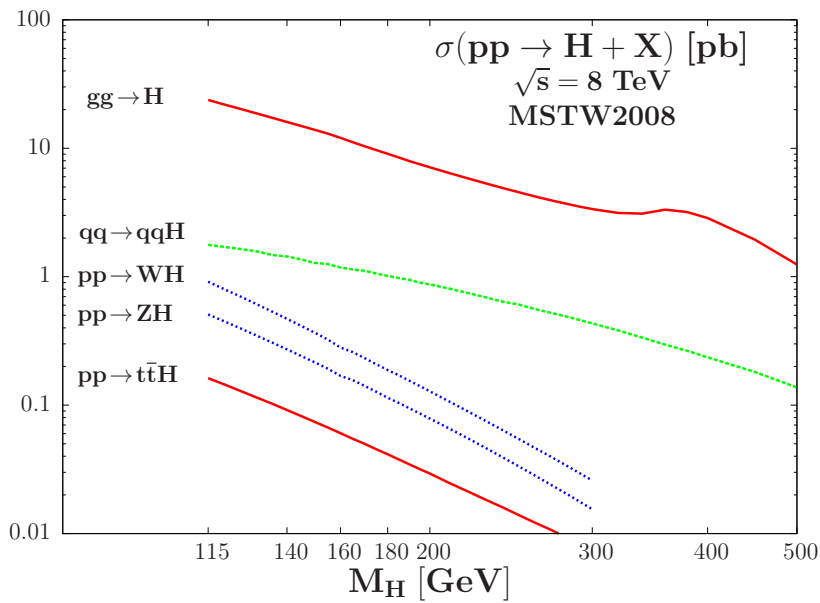}
\end{center}
\vspace*{-4mm}
\caption[]{Same as in Fig.~\ref{pp-H-TeV} for $\sqrt{s}=8$ TeV.}
\vspace*{-2mm}
\label{pp-H-lhc8}
\end{figure}

\begin{figure}[!h]
\begin{bigcenter}
\vspace*{-.1mm}
\mbox{
\epsfig{file=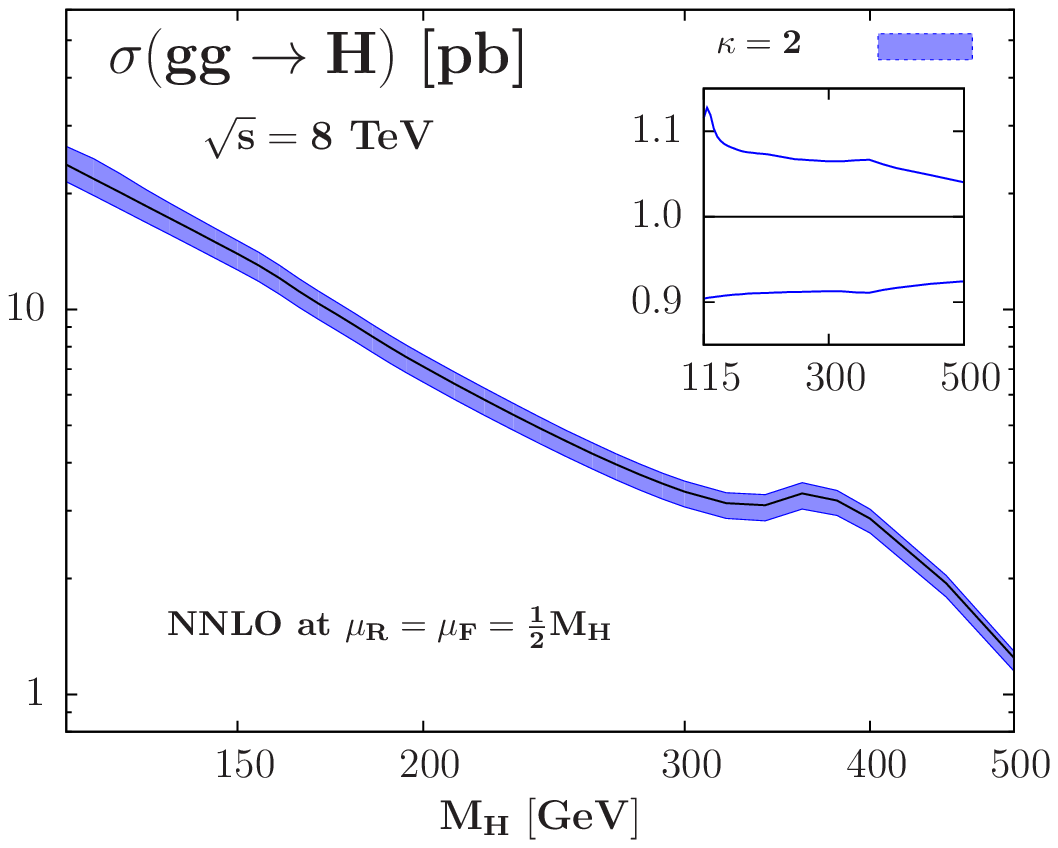,scale=0.67}\hspace*{-0.1cm}
\epsfig{file=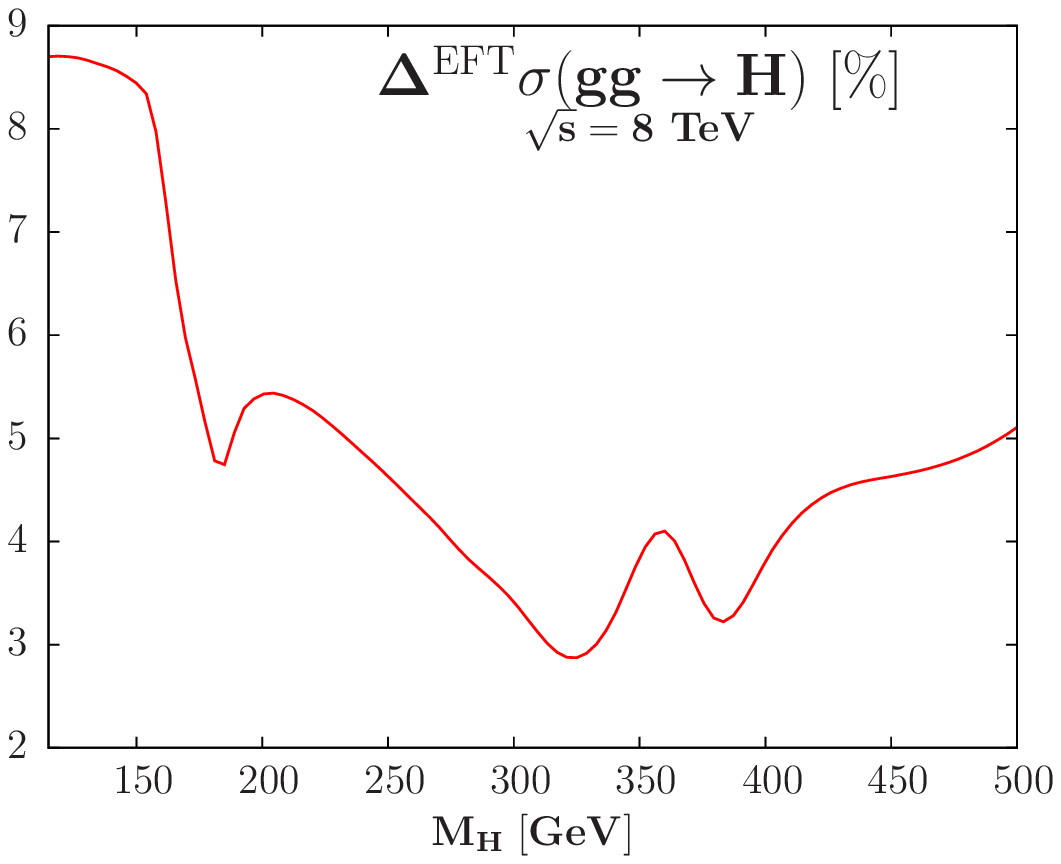,scale=0.67}
}
\bigskip

\mbox{
\epsfig{file=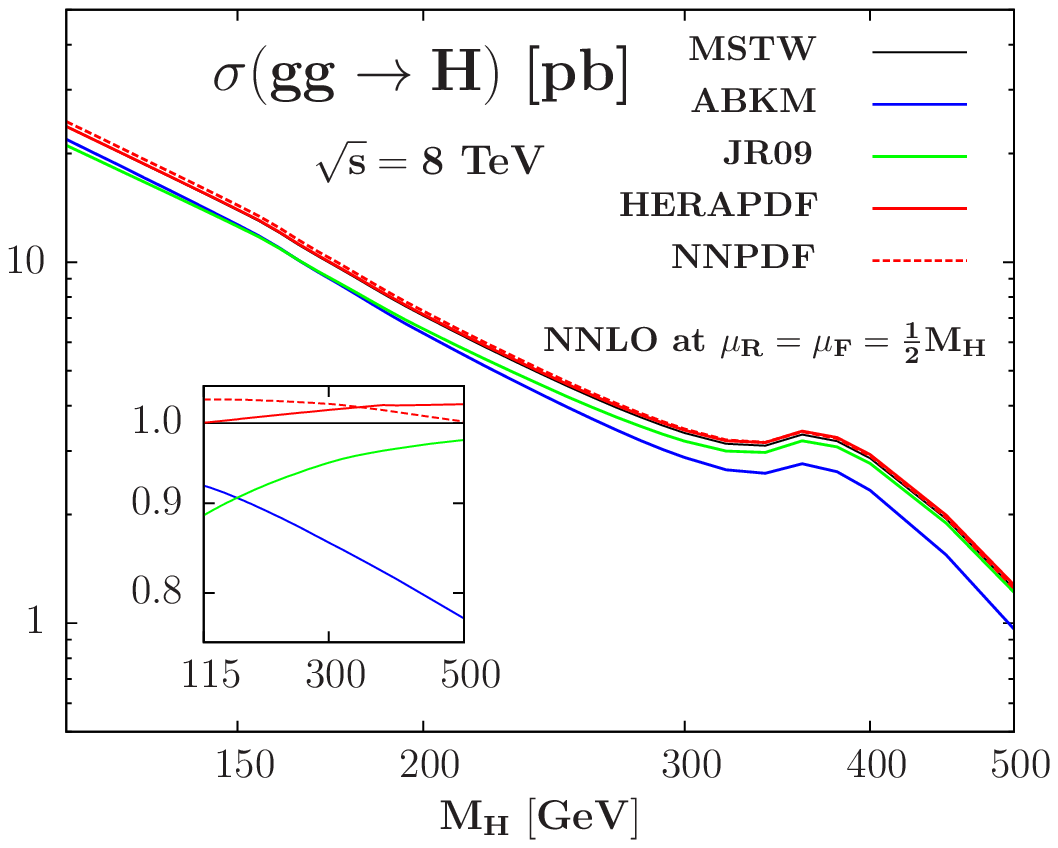,scale=0.67}\hspace*{-0.1cm}
\epsfig{file=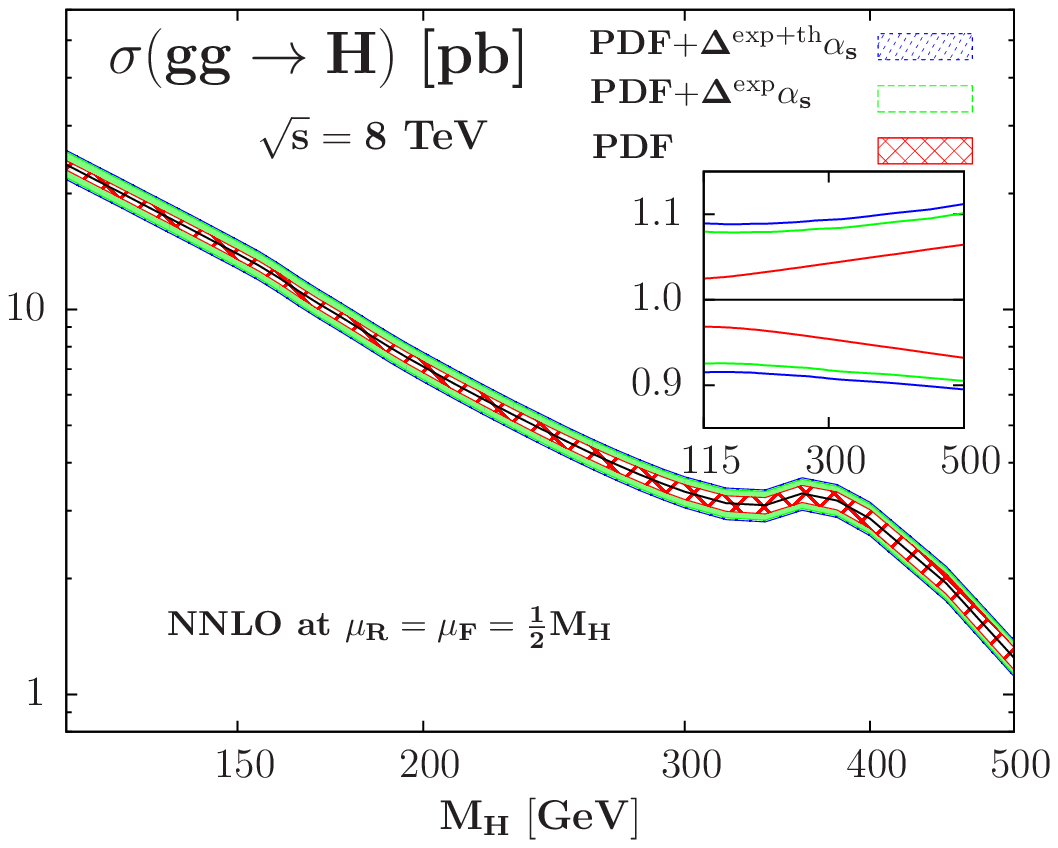,scale=0.67}
}
\bigskip

\mbox{
\epsfig{file=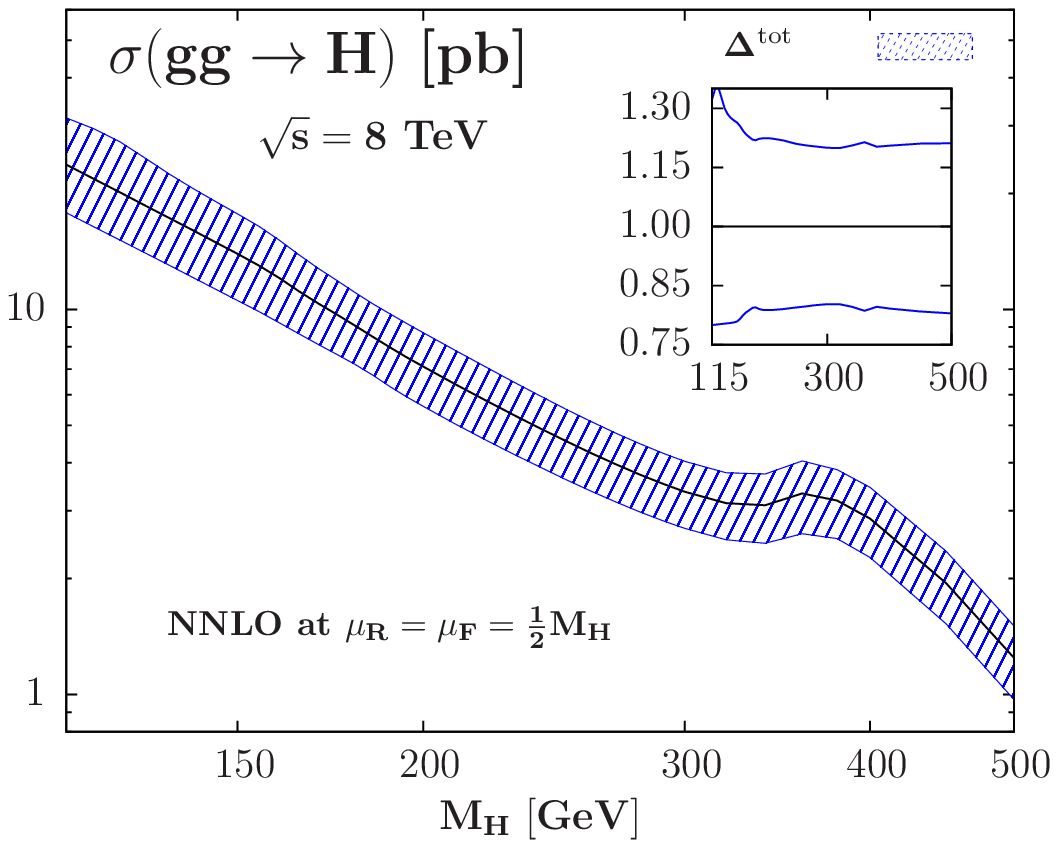,scale=0.67}\hspace*{-0.1cm}
}
\end{bigcenter}
\vspace*{-5mm}
\caption[]{Same as in
  Figs.~\ref{scale_ggH_lhc14}, \ref{pdf_ggH_lhc14}, \ref{ggH_all_lhc14} for
  $\sqrt{s}=8$ TeV.}
\label{uncertainties_ggH_lhc8}
\end{figure}

In this addendum, we summarize the Higgs production cross sections
in the SM and in the case of $gg\to H$ we detail the associated
uncertainties, at a c.m. energy of $\sqrt s=8$ TeV, following exactly
the discussion presented in section 2. In particular we update the
numbers presented in Table \ref{table_lhc8} with more precision and a
full account of the other main processes. The rates for the four
production channels are displayed in Fig.~\ref{pp-H-lhc8} and
Table \ref{lhc8_allchannels}. The scale, PDF (using either only the
MSTW set or including also the other available NNLO PDF sets) and EFT
as well as the total uncertainties are shown in
Fig.~\ref{uncertainties_ggH_lhc8} for the $gg\to H$ process and the
relevant numbers are given in Table \ref{table_lhc8_bis}.

\clearpage

\begin{table}
\renewcommand{\arraystretch}{1.35}
\begin{bigcenter}
\small
\begin{tabular}{|c|ccccc|}\hline
$M_H$ & $\sigma^{\rm NNLO}_{g g\to H}$ & $\sigma^{\rm NLO}_{q q\to H q q}$ &
$\sigma^{\rm NNLO}_{q\bar q\to H W}$ & $\sigma^{\rm NNLO}_{q\bar q\to H Z}$ &
$\sigma^{\rm LO}_{p p\to t \bar{t} H}$ \\ \hline
$115$ & $23757.4$ & $1772.2$ & $913.9$
& $508.7$ & $162.2$  \\  
$120$ & $21832.4$ & $1702.9$ & $794.1$
& $446.0$ & $144.2$  \\  
$125$ & $20126.9$ & $1639.6$ & $694.8$
& $392.7$ & $128.0$  \\  
$130$ & $18608.7$ & $1571.4$ & $607.7$
& $346.1$ & $114.2$  \\  
$135$ & $17252.0$ & $1476.9$ & $533.5$
& $306.6$ & $102.4$  \\  
$140$ & $16032.9$ & $1436.2$ & $469.6$
& $272.1$ & $91.6$  \\  
$145$ & $14933.7$ & $1363.9$ & $414.8$
& $241.9$ & $82.3$  \\  
$150$ & $13934.8$ & $1279.5$ & $366.7$
& $215.8$ & $74.1$  \\  
$155$ & $13009.0$ & $1254.7$ & $324.3$
& $192.5$ & $67.0$  \\  
$160$ & $12063.1$ & $1180.2$ & $280.8$
& $169.0$ & $60.5$  \\  
$165$ & $11143.1$ & $1138.8$ & $258.7$
& $156.6$ & $54.7$  \\  
$170$ & $10360.9$ & $1112.2$ & $232.1$
& $141.1$ & $49.9$  \\  
$175$ & $9688.5$ & $1058.3$ & $208.2$
& $127.0$ & $45.3$  \\  
$180$ & $9065.7$ & $1016.1$ & $187.7$
& $114.1$ & $41.5$  \\  
$185$ & $8483.1$ & $977.8$ & $171.0$
& $104.2$ & $37.9$  \\  
$190$ & $7962.6$ & $944.9$ & $155.0$
& $94.7$ & $34.6$  \\  
$195$ & $7509.0$ & $900.0$ & $141.0$
& $86.2$ & $31.8$  \\  
$200$ & $7105.4$ & $870.3$ & $128.1$
& $78.7$ & $29.4$  \\  
$210$ & $6408.1$ & $807.9$ & $106.6$
& $65.7$ & $24.8$  \\  
$220$ & $5823.0$ & $745.5$ & $89.2$
& $55.0$ & $21.2$  \\  
$230$ & $5326.7$ & $690.4$ & $75.3$
& $46.3$ & $18.3$  \\  
$240$ & $4901.8$ & $638.7$ & $63.7$
& $39.2$ & $16.0$  \\  
$250$ & $4536.1$ & $607.0$ & $54.3$
& $33.3$ & $13.9$  \\  
$260$ & $4221.7$ & $559.7$ & $46.4$
& $28.3$ & $12.2$  \\  
$270$ & $3951.5$ & $525.4$ & $40.0$
& $24.2$ & $10.8$  \\  
$280$ & $3721.0$ & $494.1$ & $34.5$
& $20.8$ & $9.5$  \\  
$290$ & $3524.8$ & $462.8$ & $29.8$
& $17.9$ & $8.5$  \\  
$300$ & $3362.2$ & $433.0$ & $25.9$
& $15.5$ & $7.6$  \\  
\hline  
\end{tabular} 
\caption[]{
Same as in Table \ref{lhc7_allchannels} with $\sqrt{s}=8$ TeV.}
\label{lhc8_allchannels}
\end{bigcenter} 
\end{table}

\clearpage

\clearpage

\begin{table}
\begin{bigcenter}
\small
\begin{tabular}{|c|ccccc|}\hline
$M_H$ & $\sigma$ [pb] & Scale [\%] &
PDF+$\Delta^{\rm exp+th}_{\alpha_s}$  [\%]  & ${\rm EFT}$  [\%] & Total
[\%] \\ \hline
$115$ & $23.76$ & ${+11.6}\;{-9.6}$ & ${+8.9}\;{-8.5}$ & ${\pm 8.7}
$ & ${+32.3}\;{-25.0}$  \\  
$120$ & $21.83$ & ${+12.8}\;{-9.5}$ & ${+8.9}\;{-8.5}$ & ${\pm 8.7}
$ & ${+35.0}\;{-24.9}$  \\  
$125$ & $20.13$ & ${+11.9}\;{-9.5}$ & ${+8.9}\;{-8.5}$ & ${\pm 8.7}
$ & ${+35.3}\;{-24.8}$  \\  
$130$ & $18.61$ & ${+10.3}\;{-9.4}$ & ${+8.9}\;{-8.5}$ & ${\pm 8.7}
$ & ${+32.7}\;{-24.7}$  \\  
$135$ & $17.25$ & ${+9.4}\;{-9.4}$ & ${+8.9}\;{-8.4}$ & ${\pm 8.6}
$ & ${+30.0}\;{-24.6}$  \\  
$140$ & $16.03$ & ${+8.9}\;{-9.3}$ & ${+8.9}\;{-8.4}$ & ${\pm 8.6}
$ & ${+28.5}\;{-24.5}$  \\  
$145$ & $14.93$ & ${+8.5}\;{-9.3}$ & ${+8.9}\;{-8.4}$ & ${\pm 8.5}
$ & ${+27.6}\;{-24.4}$  \\  
$150$ & $13.93$ & ${+8.3}\;{-9.2}$ & ${+8.8}\;{-8.4}$ & ${\pm 8.4}
$ & ${+27.0}\;{-24.3}$  \\  
$155$ & $13.01$ & ${+8.1}\;{-9.2}$ & ${+8.8}\;{-8.5}$ & ${\pm 8.3}
$ & ${+26.4}\;{-24.1}$  \\  
$160$ & $12.06$ & ${+8.0}\;{-9.1}$ & ${+8.8}\;{-8.5}$ & ${\pm 7.6}
$ & ${+25.5}\;{-23.4}$  \\  
$165$ & $11.14$ & ${+7.8}\;{-9.1}$ & ${+8.8}\;{-8.5}$ & ${\pm 6.6}
$ & ${+24.3}\;{-22.5}$  \\  
$170$ & $10.36$ & ${+7.7}\;{-9.1}$ & ${+8.8}\;{-8.5}$ & ${\pm 5.9}
$ & ${+23.3}\;{-21.7}$  \\  
$175$ & $9.69$ & ${+7.6}\;{-9.0}$ & ${+8.9}\;{-8.5}$ & ${\pm 5.4}
$ & ${+22.7}\;{-21.2}$  \\  
$180$ & $9.07$ & ${+7.5}\;{-9.0}$ & ${+8.9}\;{-8.5}$ & ${\pm 4.9}
$ & ${+22.0}\;{-20.6}$  \\  
$185$ & $8.48$ & ${+7.5}\;{-9.0}$ & ${+8.9}\;{-8.5}$ & ${\pm 4.7}
$ & ${+21.9}\;{-20.5}$  \\  
$190$ & $7.96$ & ${+7.5}\;{-9.0}$ & ${+8.9}\;{-8.5}$ & ${\pm 5.1}
$ & ${+22.2}\;{-20.9}$  \\  
$195$ & $7.51$ & ${+7.4}\;{-9.0}$ & ${+8.9}\;{-8.6}$ & ${\pm 5.3}
$ & ${+22.4}\;{-21.1}$  \\  
$200$ & $7.11$ & ${+7.4}\;{-8.9}$ & ${+8.9}\;{-8.6}$ & ${\pm 5.4}
$ & ${+22.4}\;{-21.2}$  \\  
$210$ & $6.41$ & ${+7.3}\;{-8.9}$ & ${+8.9}\;{-8.7}$ & ${\pm 5.4}
$ & ${+22.4}\;{-21.2}$  \\  
$220$ & $5.82$ & ${+7.2}\;{-8.9}$ & ${+8.9}\;{-8.7}$ & ${\pm 5.3}
$ & ${+22.1}\;{-21.1}$  \\  
$230$ & $5.33$ & ${+7.0}\;{-8.9}$ & ${+9.0}\;{-8.8}$ & ${\pm 5.1}
$ & ${+21.8}\;{-20.9}$  \\  
$240$ & $4.90$ & ${+6.9}\;{-8.8}$ & ${+9.0}\;{-8.8}$ & ${\pm 4.9}
$ & ${+21.4}\;{-20.7}$  \\  
$250$ & $4.54$ & ${+6.7}\;{-8.8}$ & ${+9.1}\;{-8.9}$ & ${\pm 4.6}
$ & ${+21.0}\;{-20.5}$  \\  
$260$ & $4.22$ & ${+6.7}\;{-8.8}$ & ${+9.1}\;{-8.9}$ & ${\pm 4.4}
$ & ${+20.7}\;{-20.4}$  \\  
$270$ & $3.95$ & ${+6.6}\;{-8.8}$ & ${+9.2}\;{-8.9}$ & ${\pm 4.1}
$ & ${+20.5}\;{-20.2}$  \\  
$280$ & $3.72$ & ${+6.6}\;{-8.8}$ & ${+9.3}\;{-9.0}$ & ${\pm 3.9}
$ & ${+20.3}\;{-20.0}$  \\  
$290$ & $3.52$ & ${+6.5}\;{-8.7}$ & ${+9.3}\;{-9.1}$ & ${\pm 3.7}
$ & ${+20.1}\;{-19.8}$  \\  
$300$ & $3.36$ & ${+6.5}\;{-8.7}$ & ${+9.3}\;{-9.2}$ & ${\pm 3.4}
$ & ${+20.0}\;{-19.7}$  \\  
$320$ & $3.14$ & ${+6.5}\;{-8.7}$ & ${+9.4}\;{-9.4}$ & ${\pm 2.9}
$ & ${+19.9}\;{-19.7}$  \\  
$340$ & $3.10$ & ${+6.6}\;{-8.9}$ & ${+9.6}\;{-9.5}$ & ${\pm 3.3}
$ & ${+20.5}\;{-20.4}$  \\  
$360$ & $3.33$ & ${+6.7}\;{-8.9}$ & ${+9.8}\;{-9.6}$ & ${\pm 4.1}
$ & ${+21.4}\;{-21.4}$  \\  
$380$ & $3.19$ & ${+6.1}\;{-8.6}$ & ${+10.0}\;{-9.6}$ & ${\pm 3.2}
$ & ${+20.2}\;{-20.4}$  \\  
$400$ & $2.86$ & ${+5.7}\;{-8.3}$ & ${+10.2}\;{-9.8}$ & ${\pm 3.8}
$ & ${+20.5}\;{-20.8}$  \\  
$450$ & $1.94$ & ${+4.9}\;{-7.9}$ & ${+10.6}\;{-10.1}$ & ${\pm 4.6}
$ & ${+21.1}\;{-21.5}$  \\  
$500$ & $1.24$ & ${+4.0}\;{-7.6}$ & ${+11.2}\;{-10.5}$ & ${\pm 5.1}
$ & ${+21.1}\;{-22.0}$  \\ \hline  
\end{tabular} 
\caption[]{Same as in Table \ref{table_lhc14} with $\sqrt{s}=8$ TeV,
  including only procedure A described in the text to calculate the
  total uncertainty.}
\label{table_lhc8_bis}
\end{bigcenter} 
\end{table}

\end{document}